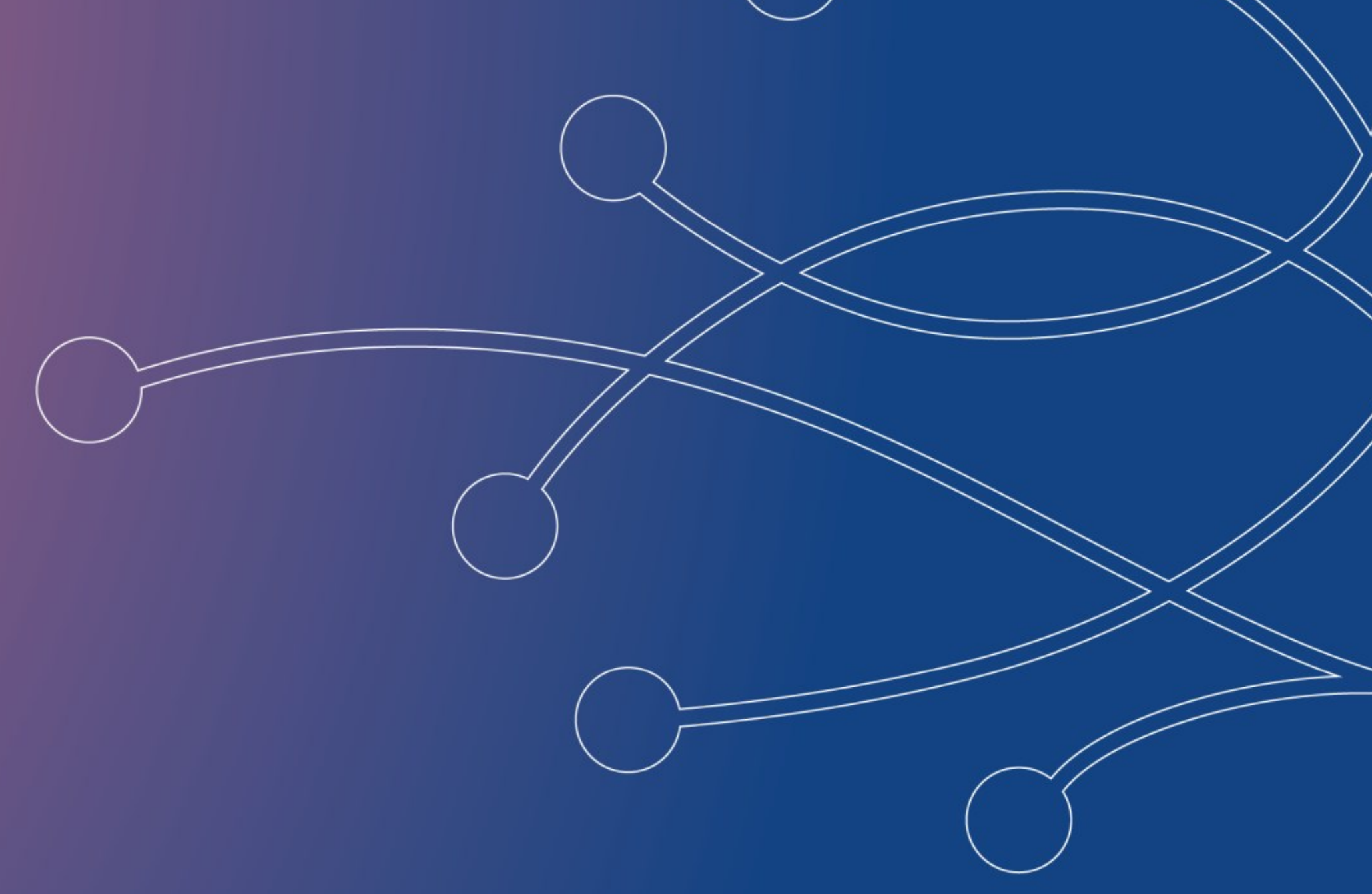

# International Scientific Report on the Safety of Advanced AI

INTERIM REPORT

May 2024

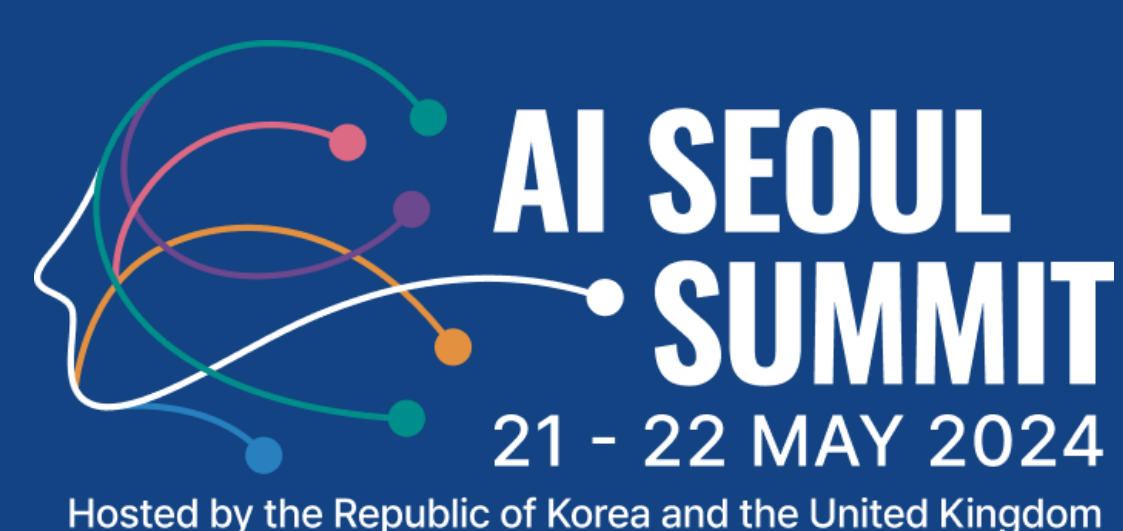

# Contributors

## CHAIR

**Prof. Yoshua Bengio**, Université de Montréal / Mila - Quebec AI Institute

## EXPERT ADVISORY PANEL

**Prof. Bronwyn Fox,** The Commonwealth Scientific and Industrial Research Organisation (CSIRO) (Australia)

**André Carlos Ponce de Leon Ferreira de Carvalho**, Institute of Mathematics and Computer Sciences, University of São Paulo (Brazil)

**Dr. Mona Nemer,** Chief Science Advisor of Canada (Canada)

**Raquel Pezoa Rivera,** Federico Santa María Technical University (Chile)

**Dr. Yi Zeng,** Institute of Automation, Chinese Academy of Sciences (China)

**Juha Heikkilä,** DG Connect (European Union)

**Guillaume Avrin,** General Directorate of Enterprises (France)

**Prof. Antonio Krüger,** German Research Center for Artificial Intelligence (Germany)

**Prof. Balaraman Ravindran,** Indian Institute of Technology, Madras (India)

**Prof. Hammam Riza,** KORIKA (Indonesia)

**Dr. Ciarán Seoighe,** Science Foundation Ireland (Ireland)

**Dr. Ziv Katzir,** Israel Innovation Authority (Israel)

**Dr. Andrea Monti,** University of Chieti-Pescara (Italy)

**Dr. Hiroaki Kitano,** Sony Group (Japan)

[Interim] **Mary Kerema,** Ministry of Information Communications Technology and Digital Economy (Kenya)

**Dr. José Ramón López Portillo,** Q Element (Mexico)

**Prof. Haroon Sheikh,** Netherlands' Scientific Council for Government Policy (Netherlands)

**Dr. Gill Jolly,** Ministry of Business, Innovation and Employment (New Zealand)

**Dr. Olubunmi Ajala,** Innovation and Digital Economy (Nigeria)

**Dominic Ligot,** CirroLytix (Philippines)

**Prof. Kyoung Mu Lee,** Department of Electrical and Computer Engineering, Seoul National University (Republic of Korea)

**Ahmet Halit Hatip,** Turkish Ministry of Industry and Technology (Republic of Turkey)

**Crystal Rugege,** National Center for AI and Innovation Policy (Rwanda)

**Dr. Fahad Albalawi,** Saudi Authority for Data and Artificial Intelligence (Kingdom of Saudi Arabia)

**Denise Wong,** Data Innovation and Protection Group, Infocomm Media Development Authority (IMDA) (Singapore)

**Dr. Nuria Oliver,** ELLIS Alicante (Spain)

**Dr. Christian Busch,** Federal Department of Economic Affairs, Education and Research (Switzerland)

**Oleksii Molchanovskyi,** Expert Committee on the Development of Artificial intelligence in Ukraine (Ukraine)

**Marwan Alserkal,** Ministry of Cabinet Affairs, Prime Minister's Office (United Arab Emirates)

**Saif M. Khan,** U.S. Department of Commerce (United States)

**Dame Angela McLean,** Government Chief Scientific Adviser (United Kingdom)

**Amandeep Gill,** UN Tech Envoy (United Nations)

**SCIENTIFIC LEAD**

**Sören Mindermann,** Mila – Quebec AI Institute

**WRITING GROUP**

**Daniel Privitera** (lead writer), KIRA Center
**Tamay Besiroglu,** Epoch AI
**Rishi Bommasani,** Stanford University
**Stephen Casper,** Massachusetts Institute of Technology
**Yejin Choi,** University of Washington/A12
**Danielle Goldfarb,** Mila – Quebec AI Institute
**Hoda Heidari,** Carnegie Mellon University **Leila Khalatbari,** Hong Kong University of Science and Technology
**Shayne Longpre,** Massachusetts Institute of Technology
**Vasilios Mavroudis,** Alan Turing Institute
**Mantas Mazeika,** University of Illinois at Urbana-Champaign
**Kwan Yee Ng,** Concordia AI
**Chinasa T. Okolo, Ph.D,** The Brookings Institution
**Deborah Raji,** Mozilla
**Theodora Skeadas,** Humane Intelligence
**Florian Tramèr,** ETH Zürich

**SENIOR ADVISERS**

**Bayo Adekanmbi,** Data Science Nigeria
**Paul Christiano,** contributed as a Senior Adviser prior to taking up his role at the US AI Safety Institute
**David Dalrymple,** Advanced Research + Invention Agency (ARIA)
**Thomas G. Dietterich,** Oregon State University **Edward Felten,** Princeton University
**Pascale Fung,** Hong Kong University of Science and Technology, contributed as a Senior Adviser prior to taking up her role at Meta **Pierre-Olivier Gourinchas,** International Monetary Fund (IMF)
**Nick Jennings CB FREng FRS,** University of Loughborough
**Andreas Krause,** ETH Zurich
**Percy Liang,** Stanford University
**Teresa Ludermir,** Federal University of Pernambuco
**Vidushi Marda,** REAL ML
**Helen Margetts OBE FBA,** University of Oxford/Alan Turing Institute
**John A. McDermid OBE FREng,** University of York
**Arvind Narayanan,** Princeton University
**Alondra Nelson,** Institute for Advanced Study
**Alice Oh,** KAIST School of Computing
**Gopal Ramchurn,** RAI UK/UKRI TAS Hub/University of Southampton
**Stuart Russell,** University of California, Berkeley
**Marietje Schaake,** Stanford University
**Dawn Song,** University of California, Berkeley
**Alvaro Soto,** Pontificia Universidad Católica de Chile
**Lee Tiedrich,** Duke University
**Gaël Varoquaux,** The National Institute for Research in Digital Science and Technology (Inria)
**Andrew Yao,** Institute for Interdisciplinary Information Sciences, Tsinghua University
**Ya-Qin Zhang,** Tsinghua University

**SECRETARIAT**

**UK Government Secretariat** hosted by the AI Safety Institute
**Benjamin Prud'homme,** Mila – Quebec AI Institute

**ACKNOWLEDGEMENTS**

The Secretariat appreciate the helpful support, comments, and feedback from the following UK-based organisations: Ada Lovelace Institute, The Alan Turing Institute, The Centre for Long-Term Resilience, Centre for the Governance of AI, and UK AI Safety Institute. Also a special thanks to Dan Hendrycks, Dylan Hadfield-Menell, and Pamela Samuelson.

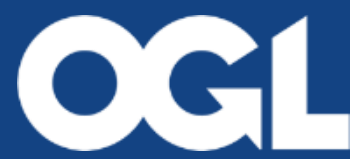

Any enquiries regarding this publication should be sent to us at:
secretariat.AIStateofScience@dsit.gov.uk

**Disclaimer**

The report does not represent the views of the Chair, any particular individual in the writing or advisory groups, nor any of the governments that have supported its development. This report is a synthesis of the existing research on the capabilities and risks of advanced AI. The Chair of the report has ultimate responsibility for it, and has overseen its development from beginning to end.

# Forewords

## This report is the beginning of a journey on AI Safety

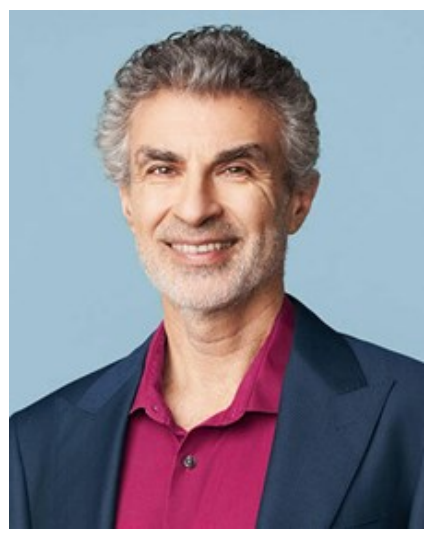

I am honoured to be chairing the delivery of the inaugural International Scientific Report on Advanced AI Safety. I am proud to publish this interim report which is the culmination of huge efforts by many experts over the six months since the work was commissioned at the Bletchley Park AI Safety Summit in November 2023.

We know that advanced AI is developing very rapidly, and that there is considerable uncertainty over how these advanced AI systems might affect how we live and work in the future. AI has tremendous potential to change our lives for the better, but it also poses risks of harm. That is why having this thorough analysis of the available scientific literature and expert opinion is essential. The more we know, the better equipped we are to shape our collective destiny.

Our mission is clear: to drive a shared, science-based, up-to-date understanding of the safety of advanced AI, and to continue to develop that understanding over time. The report rightly highlights that there are areas of consensus among experts and also disagreements over the capabilities and risks of advanced AI, especially those expected to be developed in the future. In order to meet our mission effectively, we have aimed to address disagreement amongst the expert community with intellectual honesty. By dissecting these differences, we pave the way for informed policy-making and stimulate the research needed to help clear the fog and mitigate risks.

I am grateful to our international Expert Advisory Panel for their invaluable comments, initially shaping the report's scope and later providing feedback on the full draft. Their diverse perspectives and careful review have broadened and strengthened this interim report. Equally deserving of recognition are my dedicated team of writers and senior advisers. Their commitment over the past few months has created an interim product that has surpassed my expectations. My thanks also go to the UK Government for starting this process and offering outstanding operational support. It was also important for me that the UK Government agreed that the scientists writing this report should have complete independence.

This interim report is only the beginning of a journey. There are no doubt perspectives and evidence that this report has failed to capture in this first attempt. In a scientific process such as this, feedback is precious. We will incorporate additional evidence and scientific viewpoints as we work toward the final version.

**Professor Yoshua Bengio**

Université de Montréal / Mila - Quebec AI Institute & Chair

## AI Safety is a shared global issue

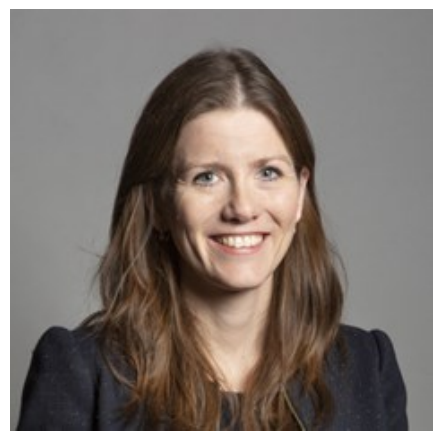

I am delighted to present this interim update on the first International Scientific Report on the Safety of Advanced AI, a key outcome of the groundbreaking AI Safety Summit held at Bletchley Park in November 2023. This landmark report represents an unprecedented global effort to build a shared, science-based understanding of the opportunities and risks posed by rapid advancements in AI, and is a testament to the "Bletchley Effect" - the power of convening brilliant minds to tackle one of humanity's greatest challenges.

We believe that realising the immense potential of AI to benefit humanity will require proactive efforts to ensure these powerful technologies are developed and deployed safely and responsibly. No one country can tackle this challenge alone. That is why I was so passionate about bringing together a diverse group of world-leading experts to contribute their knowledge and perspectives. I want to especially thank Professor Yoshua Bengio for his leadership as Chair in skilfully shepherding this complex international effort.

Crucially the report also shines a light on the significant gaps in our current knowledge and the key uncertainties and debates that urgently require further research and discussion. It is my sincere hope that this report, and the cooperative process behind it, can serve as a catalyst for the research and policy efforts needed to close critical knowledge gaps and a valuable input for the challenging policy choices that lie ahead.

We still have much to learn, but this report marks an important start. The UK looks forward to continuing to work with international partners to promote a responsible, human-centric approach to AI development - one that harnesses these powerful tools to improve lives and livelihoods while vigilantly safeguarding against downside risks and harms. Together, we can work to build a future in which all of humanity can benefit from the wonders of AI.

**The Rt Hon Michelle Donelan MP,** Secretary of State,
Department for Science, Innovation, and Technology

## A critical step forward and a Call to Action on AI Safety

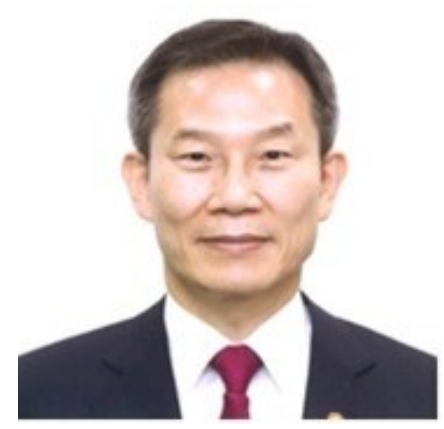

The rapid advancement of AI stands poised to reshape our world in ways both profound and unforeseen. From revolutionising healthcare and transportation to automating complex tasks and unlocking scientific breakthroughs, AI's potential for positive impact is undeniable.

However, alongside these notable possibilities lie significant challenges that necessitate a forward-looking approach. Concerns range from unintended biases embedded in algorithms to the possibility of autonomous systems exceeding human control. These potential risks highlight the urgent need for a global conversation to ensure the safe, and responsible advancement of AI.

In this context, the International AI Safety Report will provide vital groundwork for global collaboration. The report represents a convergence of knowledge from experts across 30 countries, the European Union, and the United Nations, providing a comprehensive analysis of AI safety. By focusing on the early scientific understanding of capabilities and risks from general purpose AI and evaluating technical methods for assessing and mitigating them, the report will spark ongoing dialogue and collaboration among multi-stakeholders.

I hope that based on this report, experts from 30 countries, the EU, and the UN continue to engage in balanced discussions, achieving AI risk mitigation that is acceptable and tailored to the specific context of both developed and developing countries, thereby creating a future where innovation and responsible AI coexist harmoniously.

**Lee Jong-Ho,** Minister of MSIT,
Republic of Korea

# Executive Summary

## About this report

- This is the interim publication of the first 'International Scientific Report on the Safety of Advanced AI'. A diverse group of 75 artificial intelligence (AI) experts contributed to this report, including an international Expert Advisory Panel nominated by 30 countries, the European Union (EU), and the United Nations (UN).
- Led by the Chair of this report, the independent experts writing this report collectively had full discretion over its content.
- At a time of unprecedented progress in AI development, this first publication restricts its focus to a type of AI that has advanced particularly rapidly in recent years: General-purpose AI, or AI that can perform a wide variety of tasks. Amid rapid advancements, research on general-purpose AI is currently in a time of scientific discovery and is not yet settled science.
- People around the world will only be able to enjoy general-purpose AI's many potential benefits safely if its risks are appropriately managed. This report focuses on identifying these risks and evaluating technical methods for assessing and mitigating them. It does not aim to comprehensively assess all possible societal impacts of general-purpose AI, including its many potential benefits.
- For the first time in history, this interim report brought together experts nominated by 30 countries, the EU, and the UN, and other world-leading experts, to provide a shared scientific, evidence-based foundation for discussions and decisions about general-purpose AI safety. We continue to disagree on several questions, minor and major, around general-purpose AI capabilities, risks, and risk mitigations. But we consider this project essential for improving our collective understanding of this technology and its potential risks, and for moving closer towards consensus and effective risk mitigation to ensure people can experience the potential benefits of general-purpose AI safely. The stakes are high. We look forward to continuing this effort.

## Highlights of the executive summary

- If properly governed, general-purpose AI can be applied to advance the public interest, potentially leading to enhanced wellbeing, more prosperity, and new scientific discoveries. However, malfunctioning or maliciously used general-purpose AI can also cause harm, for instance through biased decisions in high-stakes settings or through scams, fake media, or privacy violations.
- As general-purpose AI capabilities continue to advance, risks such as large-scale labour market impacts, AI-enabled hacking or biological attacks, and society losing control over general-purpose AI could emerge, although the likelihood of these scenarios is debated among researchers. Different views on these risks often stem from differing expectations about the steps society will take to limit them, the effectiveness of those steps, and how rapidly general-purpose AI capabilities will be advanced.
- There is considerable uncertainty about the rate of future progress in general-purpose AI capabilities. Some experts think a slowdown of progress is by far most likely, while other experts think that extremely rapid progress is possible or likely.
- There are various technical methods to assess and reduce risks from general-purpose AI that developers can employ and regulators can require, but they all have limitations. For example, current techniques for explaining why general-purpose AI models produce any given output are severely limited.

- The future of general-purpose AI technology is uncertain, with a wide range of trajectories appearing possible even in the near future, including both very positive and very negative outcomes. But nothing about the future of AI is inevitable. It will be the decisions of societies and governments that will determine the future of AI. This interim report aims to facilitate constructive discussion about these decisions.

## This report synthesises the state of scientific understanding of general-purpose AI – AI that can perform a wide variety of tasks – with a focus on understanding and managing its risks

The capabilities of systems using AI have been advancing rapidly. This has highlighted the many opportunities that AI creates for business, research, government, and private life. It has also led to an increased awareness of current harms and potential future risks associated with advanced AI.

The purpose of the International Scientific Report on the Safety of Advanced AI is to take a step towards a shared international understanding of AI risks and how they can be mitigated. This first interim publication of the report restricts its focus to a type of AI whose capabilities have advanced particularly rapidly: general-purpose AI, or AI that can perform a wide variety of tasks.

Amid rapid advancements, research on general-purpose AI is currently in a time of scientific discovery and is not yet settled science. The report provides a snapshot of the current scientific understanding of general-purpose AI and its risks. This includes identifying areas of scientific consensus and areas where there are different views or open research questions.

People around the world will only be able to enjoy the potential benefits of general-purpose AI safely if its risks are appropriately managed. This report focuses on identifying risks from general-purpose AI and evaluating technical methods for assessing and mitigating them, including the beneficial use of general-purpose AI to mitigate risks. It does not aim to comprehensively assess all possible societal impacts of general-purpose AI, including what benefits it may offer.

## General-purpose AI capabilities have grown rapidly in recent years according to many metrics, and there is no consensus on how to predict future progress, making a wide range of scenarios appear possible

According to many metrics, general-purpose AI capabilities are progressing rapidly. Five years ago, the leading general-purpose AI language models could rarely produce a coherent paragraph of text. Today, some general-purpose AI models can engage in multi-turn conversations on a wide range of topics, write short computer programs, or generate videos from a description. However, the capabilities of general-purpose AI are difficult to estimate reliably and define precisely.

The pace of general-purpose AI advancement depends on both the rate of technological advancements and the regulatory environment. This report focuses on the technological aspects and does not provide a discussion of how regulatory efforts might affect the speed of development and deployment of general-purpose AI.

AI developers have rapidly advanced general-purpose AI capabilities in recent years mostly by continuously increasing resources used for training new models (a trend called 'scaling') and refining existing algorithms. For example, state-of-the-art AI models have seen annual increases of approximately 4x in computational resources ('compute') used for training, 2.5x in training dataset size, and 1.5-3x in algorithmic efficiency (performance relative to compute). Whether 'scaling' has resulted in progress on fundamental challenges such as causal reasoning is debated among researchers.

The pace of future progress in general-purpose AI capabilities has substantial implications for managing emerging risks, but experts disagree on what to expect even in the near future. Experts variously support the possibility of general-purpose AI capabilities advancing slowly, rapidly, or extremely rapidly. This disagreement involves a key question: will continued 'scaling' of resources and refining existing techniques be sufficient to yield rapid progress and solve issues such as reliability and factual accuracy, or are new research breakthroughs required to substantially advance general-purpose AI abilities?

Several leading companies that develop general-purpose AI are betting on 'scaling' to continue leading to performance improvements. If recent trends continue, by the end of 2026 some general-purpose AI models will be trained using 40x to 100x more compute than the most compute-intensive models published in 2023, combined with training methods that use this compute 3x to 20x more efficiently. However, there are potential bottlenecks to further increasing both data and compute, including the availability of data, AI chips, capital expenditure, and local energy capacity. Companies developing general-purpose AI are working to navigate these potential bottlenecks.

## Several research efforts aim to understand and evaluate general-purpose AI more reliably, but our overall understanding of how general-purpose AI models and systems work is limited

Approaches to managing risks from general-purpose AI often rest on the assumption that AI developers and policymakers can assess the capabilities and potential impacts of general-purpose AI models and systems. But while technical methods can help with assessment, all existing methods have limitations and cannot provide strong assurances against most harms related to general-purpose AI. Overall, the scientific understanding of the inner workings, capabilities, and societal impacts of general-purpose AI is very limited, and there is broad expert agreement that it should be a priority to improve our understanding of general-purpose AI. Some of the key challenges include:

- Developers still understand little about how their general-purpose AI models operate. This is because general-purpose AI models are not programmed in the traditional sense. Instead, they are trained: AI developers set up a training process that involves a lot of data, and the outcome of that training process is the general-purpose AI model. These models can consist of trillions of components, called parameters, and most of their inner workings are inscrutable, including to the model developers. Model explanation and interpretability techniques can improve researchers' and developers' understanding of how general-purpose AI models operate, but this research is nascent.
- General-purpose AI is mainly assessed through testing the model or system on various inputs. These spot checks are helpful for assessing strengths and weaknesses, including vulnerabilities and potentially harmful capabilities, but do not provide quantitative safety guarantees. The tests often miss hazards and overestimate or underestimate capabilities because general-purpose AI systems may behave differently in different circumstances, with different users, or with additional adjustments to their components.
- Independent actors can, in principle, audit general-purpose AI models or systems developed by a company. However, companies often do not provide independent auditors with the necessary level of direct access to models or the information about data and methods used that are needed for rigorous assessment. Several governments are beginning to build capacity for conducting technical evaluations and audits.
- It is difficult to assess the downstream societal impact of a general-purpose AI system because research into risk assessment has not been sufficient to produce rigorous and comprehensive assessment methodologies. In addition, general-purpose AI has a wide range of use cases, which are often not predefined and only lightly restricted, complicating risk assessment further. Understanding the potential downstream societal impacts of general-purpose AI models and systems requires nuanced and multidisciplinary analysis. Increasing the representation of diverse perspectives in general-purpose AI development and evaluation processes is an ongoing technical and institutional challenge.

## General-purpose AI can pose severe risks to individual and public safety and wellbeing

This report classifies general-purpose AI risks into three categories: malicious use risks, risks from malfunctions, and systemic risks. It also discusses several cross-cutting factors that contribute to many risks.

***Malicious use.*** Like all powerful technologies, general-purpose AI systems can be used maliciously to cause harm. Possible types of malicious use range from relatively well-evidenced ones, such as scams enabled by general-purpose AI, to ones that some experts believe might occur in the coming years, such as malicious use of scientific capabilities of general-purpose AI.

- Harm to individuals through fake content generated by general-purpose AI is a relatively well-documented class of general-purpose AI malicious use. General-purpose AI can be used to increase the scale and sophistication of scams and fraud, for example through 'phishing' attacks enhanced by general-purpose AI. General-purpose AI can also be used to generate fake compromising content featuring individuals without their consent, such as non-consensual deepfake pornography.
- Another area of concern is the malicious use of general-purpose AI for disinformation and manipulation of public opinion. General-purpose AI and other modern technologies make it easier to generate and disseminate disinformation, including in an effort to affect political processes. Technical countermeasures like watermarking content, although useful, can usually be circumvented by moderately sophisticated actors.
- General-purpose AI might also be maliciously used for cyber offence, uplifting the cyber expertise of individuals and making it easier for malicious users to conduct effective cyber-attacks. General-purpose AI systems can be used to scale and partially automate some types of cyber operations, such as social engineering attacks. However, general-purpose AI could also be used in cyber defence. Overall, there is not yet any substantial evidence suggesting that general-purpose AI can automate sophisticated cybersecurity tasks.
- Some experts have also expressed concern that general-purpose AI could be used to support the development and malicious use of weapons, such as biological weapons. There is no strong evidence that current general-purpose AI systems pose this risk. For example, although current general-purpose AI systems demonstrate growing capabilities related to biology, the limited studies available do not provide clear evidence that current systems can 'uplift' malicious actors to obtain biological pathogens more easily than could be done using the internet. However, future large-scale threats have scarcely been assessed and are hard to rule out.

***Risks from malfunctions.*** Even when users have no intention to cause harm, serious risks can arise due to the malfunctioning of general-purpose AI. Such malfunctions can have several possible causes and consequences:

- The functionality of products based on general-purpose AI models and systems might be poorly understood by their users, for example due to miscommunication or misleading advertising. This can cause harm if users then deploy the systems in unsuitable ways or for unsuitable purposes.
- Bias in AI systems generally is a well-evidenced problem and remains unsolved for general-purpose AI, too. General-purpose AI outputs can be biased with respect to protected characteristics like race, gender, culture, age, and disability. This can create risks, including in high-stakes domains such as healthcare, job recruitment, and financial lending. In addition, many widely-used general-purpose AI models are primarily trained on data that disproportionately represents Western cultures, which can increase the potential for harm to individuals not represented well by this data.

- 'Loss of control' scenarios are potential future scenarios in which society can no longer meaningfully constrain general-purpose AI systems, even if it becomes clear that they are causing harm. There is broad consensus that current general-purpose AI lacks the capabilities to pose this risk. Some experts believe that current efforts to develop general-purpose *autonomous* AI – systems that can act, plan, and pursue goals – could lead to a loss of control if successful. Experts disagree about how plausible loss-of-control scenarios are, when they might occur, and how difficult it would be to mitigate them.

***Systemic risks.*** The widespread development and adoption of general-purpose AI technology poses several systemic risks, ranging from potential labour market impacts to privacy risks and environmental effects:

- General-purpose AI, especially if it further advances rapidly, has the potential to automate a very wide range of tasks, which could have a significant effect on the labour market. This could mean many people could lose their current jobs. However, many economists expect that potential job losses could be offset, possibly completely, by the creation of new jobs and by increased demand in non-automated sectors.
- General-purpose AI research and development is currently concentrated in a few Western countries and China. This 'AI Divide' is multicausal, but in part stems from differing levels of access to the compute needed to develop general-purpose AI. Since low-income countries and academic institutions have less access to compute than high-income countries and technology companies do, they are placed at a disadvantage.
- The resulting market concentration in general-purpose AI development makes societies more vulnerable to several systemic risks. For instance, the widespread use of a small number of general-purpose AI systems in critical sectors like finance or healthcare could cause simultaneous failures and disruptions on a broad scale across these interdependent sectors, for instance because of bugs or vulnerabilities.
- Growing compute use in general-purpose AI development and deployment has rapidly increased energy usage associated with general-purpose AI. This trend shows no indications of moderating, potentially leading to further increased $CO_2$ emissions and water consumption.
- General-purpose AI models or systems can pose risks to privacy. For instance, research has shown that by using adversarial inputs, users can extract training data containing information about individuals from a model. For future models trained on sensitive personal data like health or financial data, this may lead to particularly serious privacy leaks.
- Potential copyright infringements in general-purpose AI development pose a challenge to traditional intellectual property laws, as well as to systems of consent, compensation, and control over data. An unclear copyright regime disincentivises general-purpose AI developers from declaring what data they use and makes it unclear what protections are afforded to creators whose work is used without their consent to train general-purpose AI models.

***Cross-cutting risk factors.*** Underpinning the risks associated with general-purpose AI are several cross-cutting risk factors – characteristics of general-purpose AI that increase the probability or severity of not one but several risks:

- Technical cross-cutting risk factors include the difficulty of ensuring that general-purpose AI systems reliably behave as intended, our lack of understanding of their inner workings, and the ongoing development of general-purpose AI 'agents' which can act autonomously with reduced oversight.
- Societal cross-cutting risk factors include the potential disparity between the pace of technological progress and the pace of a regulatory response, as well as competitive incentives for AI developers to release products quickly, potentially at the cost of thorough risk management.

## Several technical approaches can help mitigate risks, but no currently known method provides strong assurances or guarantees against harm associated with general-purpose AI

While this report does not discuss policy interventions for mitigating risks from general-purpose AI, it does discuss technical risk mitigation methods on which researchers are making progress. Despite this progress, current methods have not reliably prevented even overtly harmful general-purpose AI outputs in real-world contexts. Several technical approaches are used to assess and mitigate risks:

- There is some progress in training general-purpose AI models to function more safely. Developers also train models to be more robust to inputs that are designed to make them fail ('adversarial training'). Despite this, adversaries can typically find alternative inputs that reduce the effectiveness of safeguards with low to moderate effort. Limiting a general-purpose AI system's capabilities to a specific use case can help to reduce risks from unforeseen failures or malicious use.
- There are several techniques for identifying risks, inspecting system actions, and evaluating performance once a general-purpose AI system has been deployed. These practices are often referred to as 'monitoring'.
- Mitigation of bias in general-purpose AI systems can be addressed throughout the lifecycle of the system, including design, training, deployment, and usage. However, entirely preventing bias in general-purpose AI systems is challenging because it requires systematic training data collection, ongoing evaluation, and effective identification of bias. It may also require trading off fairness with other objectives such as accuracy and privacy, and deciding what is useful knowledge and what is an undesirable bias that should not be reflected in the outputs.
- Privacy protection is an active area of research and development. Simply minimising the use of sensitive personal data in training is one approach that can substantially reduce privacy risks. However, when sensitive data is either intentionally or unintentionally used, existing technical tools for reducing privacy risks struggle to scale to large general-purpose AI models, and can fail to provide users with meaningful control.

## Conclusion: A wide range of general-purpose AI trajectories are possible, and much will depend on how societies and governments act

The future of general-purpose AI is uncertain, with a wide range of trajectories appearing possible even in the near future, including both very positive and very negative outcomes. But nothing about the future of general-purpose AI is inevitable. How general-purpose AI gets developed and by whom, which problems it gets designed to solve, whether societies will be able to reap general-purpose AI's full economic potential, who benefits from it, the types of risks we expose ourselves to, and how much we invest into research to mitigate risks — these and many other questions depend on the choices that societies and governments make today and in the future to shape the development of general-purpose AI.

To help facilitate constructive discussion about these decisions, this report provides an overview of the current state of scientific research and discussion on managing the risks of general-purpose AI. The stakes are high. We look forward to continuing this effort.

# 1 Introduction

We are in the midst of a technological revolution that will fundamentally alter the way we live, work, and relate to one another. Artificial Intelligence (AI) promises to transform many aspects of our society and economy.

There is broad scientific consensus that the capabilities of AI systems have progressed rapidly on many tasks in the last five years. Large Language Models (LLMs) are a particularly salient example. In 2019, GPT-2, then the most advanced LLM, could not reliably produce a coherent paragraph of text and could not always count to ten. At the time of writing, the most powerful LLMs like Claude 3, GPT-4, and Gemini Ultra can engage consistently in multi-turn conversations, write short computer programs, translate between multiple languages, score highly on university entrance exams, and summarise long documents.

This step-change in capabilities, and the potential for continued progress, could help advance the public interest in many ways. Among the most promising prospects are AI's potential for education, medical applications, research advances in a wide range of fields, and increased innovation leading to increased prosperity. This rapid progress has also increased awareness of the current harms and potential future risks associated with the most capable types of AI.

## This report aims to contribute to an internationally shared scientific understanding of advanced AI safety

To begin forging a shared international understanding of the risks of advanced AI, government representatives and leaders from academia, business, and civil society convened in Bletchley Park in the United Kingdom in November 2023 for the first international AI Safety Summit. At the Summit, the nations present, as well as the EU and the UN, agreed to support the development of an International Scientific Report on Advanced AI Safety. This report aims to contribute to an internationally shared scientific understanding of advanced AI safety. This is the first interim publication of that report: the final version of the first report will be published ahead of the France *AI Summit*.

An international group of 75 AI experts across a breadth of views and, where relevant, a diversity of backgrounds contributed to this interim report. The evidence considered for the report includes relevant scientific, technical, and socio-economic evidence. Since the field of AI is developing at pace, not all sources used for this report are peer-reviewed. However, the report is committed to citing only high-quality sources. Criteria for a source being of high quality include:

- The piece constitutes an original contribution that advances the field.
- The piece engages comprehensively with the existing scientific literature, references the work of others where appropriate, and interprets it accurately.
- The piece discusses possible objections to its claims in good faith.
- The piece clearly describes the methods employed for its analysis. It critically discusses the choice of methods.
- The piece clearly highlights its methodological limitations.
- This piece has been influential in the scientific community.

Because a scientific consensus on the risks from advanced AI is still being forged, in many cases the report does not put forward confident views. Rather, it offers a snapshot of the current state of scientific understanding and consensus, or lack thereof. Where there are gaps in the literature, the report identifies them, in the hope that this will be a spur to further research. Further, this report does not comment on what policy options are appropriate responses to the risks it discusses. Ultimately, policymakers must choose how to balance the opportunities and risks that advanced AI poses.

Policymakers must also judge the appropriate level of prudence and caution to display in response to risks that remain ambiguous.

## This first iteration of the report focuses on 'general-purpose' AI, or AI that can perform a wide range of tasks

Artificial Intelligence (AI) refers to advanced machine-based systems developed with broadly applicable methodologies to achieve given goals or answer given questions. AI is a broad and quickly evolving field of study, and there are many different kinds of AI. This interim report does not address all potential risks from all types of advanced AI. This first iteration of the report focuses on general-purpose AI, or AI that can perform a wide range of tasks. General-purpose AI systems, now known to many through applications like ChatGPT, have generated unprecedented interest in AI both among the public and policymakers in the last 18 months. Its capabilities have been improving particularly rapidly. General-purpose AI is different from so-called 'narrow AI', a kind of AI that is specialised to perform one specific task or a few very similar tasks.

To better understand how we define general-purpose AI for this report, making a distinction between 'AI models' and 'AI systems' is useful. AI models can be thought of as the raw, mathematical essence that is often the 'engine' of AI applications. An AI system is an ensemble of several components, including one or more AI models, that is designed to be particularly useful to humans in some way. For example, the ChatGPT app is an AI system. Its core engine, GPT-4, is an AI model.

This report covers risks from AI models and AI systems if they are 'general-purpose' AI models or systems. We consider an AI model to be general-purpose if it can perform, or can be adapted to perform, a wide variety of tasks. We consider an AI system to be general-purpose if it is based on a general-purpose model, but also if it is based on a specialised model that was derived from a general-purpose model. Within the domain of general-purpose AI, this report focuses on general-purpose AI that is at least as capable as today's most advanced general-purpose AI such as GPT-4 Turbo, Claude 3 and Gemini Ultra. In our definition, a model or system does not need to have multiple modalities, like speech, text, and image, to be considered general-purpose. Instead, AI that can perform a wide variety of tasks within specific domains, like structural biology, also counts as general-purpose in our definition.

Importantly, general-purpose AI is not to be confused with 'Artificial General Intelligence' (AGI), a term sometimes used to refer to a potential future AI system that equals or surpasses human performance on all or almost all cognitive tasks. General-purpose AI is a much weaker concept.

This report does not address risks from 'narrow AI', which is trained to perform a very limited task and captures a correspondingly very limited body of knowledge. The limited timeframe for writing this interim report has led to this focus on advanced general-purpose AI, where progress has been most rapid, and the associated risks are less studied and understood. Narrow AI, however, can also be highly relevant from a risk and safety perspective, and evidence relating to the risks of these systems is used across the report. Narrow AI models and systems are used in a vast range of products and services in fields like medicine, advertising, or banking, and can pose significant risks in many of them. These risks can lead to harms like biased hiring decisions, car crashes, or harmful medical treatment recommendations. Narrow AI also gets used in various military applications. One application, though a very small subset of the application of AI to militaries, (*1*) involves, for instance, Lethal Autonomous Weapon Systems (LAWS). Such topics are covered in other fora and are outside the scope of this interim report.

A large and diverse group of leading international experts contributed to this report, including representatives nominated by 30 nations from all UN Regional Groups, and the EU and the UN. While our individual views sometimes differ, we share the conviction that constructive scientific and public discourse on AI is necessary for people around the world to reap the benefits of this technology safely. We hope that this interim report can contribute to that discourse and be a foundation for future reports that will gradually improve our shared understanding of the capabilities and risks of advanced AI.

The report is organised into six main sections. After this introduction, 2. Capabilities provides information on the current capabilities of general-purpose AI, underlying principles, and potential future trends. 3. Methodology to assess and understand general-purpose AI systems explains how researchers try to understand what general-purpose AI can do and what risks it might pose. 4. Risks discusses specific risks and cross-cutting risk factors. 5. Technical approaches to mitigate risks presents technical approaches to mitigating risk from general-purpose AI and evaluates their strengths and limitations. 6. Conclusion summarises and concludes.

# 2 Capabilities

## 2.1 How does General-Purpose AI gain its capabilities?

### KEY INFORMATION

- General-purpose AI models and systems can produce text, images, video, labels for unlabelled data, and initiate actions.
- The lifecycle of general-purpose AI models and systems typically involves computationally intensive 'pre-training', labour-intensive 'fine-tuning', and continual post-deployment monitoring and updates.

There are various types of general-purpose AI. Examples of general-purpose AI models include:

- Chatbot-style language models, such as GPT-4 (*2**), Gemini-1.5 (*3**), Claude-3 (*4**), Qwen1.5 (*5**), Llama-3 (*6**), and Mistral Large (*7**).
- Image generators (*8*), such as DALLE-3 (*9**), Midjourney-5 (*10**), and Stable Diffusion-3 (*11**).
- Video generators such as SORA (*12**).
- Robotics and navigation systems, such as PaLM-E (*13*).
- Predictors of various structures in molecular biology such as AlphaFold 3 (*14*).

General-purpose AI models rely on *deep learning* (*15*), or the training of artificial neural networks, which are AI models composed of multiple layers of interconnected nodes, loosely inspired by the structure of biological neural networks brains. Most state-of-the-art general-purpose AI models are based on the 'Transformer' neural network architecture (*16*), which has proven particularly efficient at converting increasingly large amounts of training data and computational power into better model performance. General-purpose AI models are, broadly speaking, developed and deployed following the same series of distinct stages: pre-training, fine-tuning, system integration, deployment, and post-deployment updates. Each requires different methods and resources.

Both pre-training and fine-tuning are ways of 'training' a general-purpose AI model. During training, a general-purpose AI model is given some data, which it processes to predict some other data. For example, the model might be given the first 500 words of a Wikipedia article and then predict the 501st word. Initially, it predicts randomly, but as it sees more data it is automatically adapted to learn from its mistakes, and its predictions improve. Each prediction requires some amount of computational resources ('compute'), and so training requires both data and compute. The model architecture, designed by the developers, dictates the broad types of calculations that occur when the model makes a prediction, and the exact numbers used in those calculations are adjusted during training.

**Pre-training:** The goal of pre-training is to build general background knowledge into a general-purpose AI model. During pre-training, general-purpose AI models typically learn from patterns in large amounts of data (usually taken from the Internet). Collecting and preparing pre-training data are large-scale operations, and in most cases, pre-training is the most computationally intensive stage of development. The pre-training of general-purpose AI models today takes weeks or months and uses thousands of Graphics Processing Units (GPUs) - specialised computer chips, designed to rapidly process complex parallelised calculations. For example, the Falcon-180B model used 4,096 GPUs for multiple months, and PaLM (540B) used 6,144 chips for 50 days (*13*). Today, this process uses roughly ten billion times more compute compared to state-of-the-art model training in 2010 (*17*). Some developers conduct pre-training with their own compute, while others use resources provided by specialised cloud compute providers.

**Fine-tuning:** After pre-training, most general-purpose AI models undergo one or more additional fine-tuning stages, to refine their ability to accomplish the intended tasks. Fine-tuning can include various techniques including learning from desirable examples (*18*), pairs of desirable and undesirable examples (*19*), or rewards and penalties (*20, 21**). Fine-tuning usually requires significant human involvement, and tends to be the most labour-intensive part of training, with millions of instances of human feedback needed to fine-tune modern models (*22**). Often, this feedback is provided by thousands of contracted knowledge workers.

**System integration:** After a model is trained, it can be used to build a general-purpose AI system by integrating it with other system components aimed at enhancing both capabilities and safety. In practice, general-purpose AI models are typically integrated with user interfaces, input pre-processors, output postprocessors and content filters.

**Deployment:** Once they are trained, models can be deployed for use. Deployment can be 'internal', where a system is only used by the developers, or 'external', allowing the public or other non-developer entities to use it. External deployments can be 'closed-source' or 'open-source'. Closed-source means that the public can only use the system through a limited interface. Open-source means that the entire system, including all of the model parameters, are made available. Some state-of-the-art general-purpose AI systems, such as GPT-4 (*2**), are closed source, while others like Llama-3 (*6**) are open-source. From a risk mitigation perspective, there are advantages and disadvantages of open-source models which are the subject of ongoing discussions in the scientific community. This interim report does not provide a detailed discussion of the advantages and disadvantages of open-source models.

**Post-deployment monitoring and updates:** Many general-purpose AI systems are continually updated after deployment. This lets developers update capabilities and try to address flaws and vulnerabilities as they are discovered. These changes often amount to a type of 'cat-and-mouse' game where developers continually update high-profile systems in response to newly discovered vulnerabilities (*22**).

## 2.2 What current general-purpose AI systems are capable of

### KEY INFORMATION

- General-purpose AI capabilities are difficult to estimate reliably but most experts agree that current general-purpose AI capabilities include:
  - Assisting programmers and writing short computer programs
  - Engaging in fluent conversation over several turns
  - Solving textbook mathematics and science problems
- Most experts agree that general-purpose AI is currently not capable of tasks including:
  - Performing useful robotic tasks such as household tasks
  - Reliably avoiding false statements
  - Developing entirely novel complex ideas
- A key challenge for assessing general-purpose AI systems' capabilities is that performance is highly context-specific. Methods that elicit improved model capabilities are sometimes discovered only after a model has been deployed, so initial capabilities might be underestimated.

Alternatively, general-purpose AI model and system capabilities may be overestimated due to a lack of robustness across different contexts and using different methods to elicit capabilities.

This section focuses on the capabilities of general-purpose AI models and systems categorised by modality (such as video and language) and by skill (such as reasoning and knowledge). Capabilities can also be categorised by performance on specific benchmarks (see 3. Methodology to assess and understand general-purpose AI systems). While this section covers capabilities generally, 4.4.1. Cross-cutting technical risk factors focuses on 'high-risk' capabilities.

**Difficulty of defining capabilities –** Although general-purpose AI systems are often described in terms of their capabilities, there is no widely-accepted definition of the term 'capability' in the field of AI. Part of the difficulty of defining a capability is that it is not directly observed – AI researchers can only observe an AI system's *behaviour*: the set of outputs or actions that a system actually produces and the context in which it does so (for example, the prompt that leads to the observed behaviour) (*23*). AI researchers can merely summarise the observed system behaviour in many contexts, and thus arrive at an impression of what the system is capable of – the capability. It is difficult to define and measure the full capabilities of a new general-purpose AI model, even after the model is built; researchers and users have often discovered new ways to elicit capabilities after a model is deployed, for example through prompting a model to 'think step-by-step' (*24*, *25*). Another complication in defining a general-purpose AI system's capabilities is that they are shaped by the *affordances* in its environment – the tools and resources it can access. For instance, when a general-purpose AI system is connected to the internet and equipped with a web browser, it gains new affordances for retrieving information and interacting with the real world, effectively expanding its capabilities (*26*).

## 2.2.1 Capabilities by modality

General-purpose AI models can be categorised by the modalities they process (e.g. text, images, video) as input and generate as output. General-purpose AI models exist for 10+ modalities (*27*) such as time series (*28**) and music (*29**), but text-processing models are the source of much of the present attention on general-purpose AI models. Advanced general-purpose AI models are increasingly able to process and generate text, images, video, audio, robotic actions, and proteins and large molecules:

- **Text –** Advanced language models can generate fluent text and can be used for multi-turn conversations across a variety of natural languages, topics, and formats. Text and natural language interfaces are useful for people interacting with general-purpose AI models. Some general-purpose AI models can use text as both input and output, such as OpenAI's GPT-3 (*30*); while others take text as input like Stability AI's Stable Diffusion 3 (*11**) and can process increasingly long textual sequences – for example, Google's Gemini-Pro-1.5 can process 30,000 lines of code (*31**). Text can include many types of data encoded as text, such as mathematical formulae and software code. In the software domain, language models can write short programs and assist programmers (*32*).
  **Images –** Many image-related general-purpose AI models can take images as inputs potentially combined with text, such as Anthropic's Claude 3 (*33**), and can be used to classify (*34*), describe (*2**), encode (*35*), or segment images in order to distinguish different objects inside of them (*36**). General-purpose AI models can also generate images as outputs, such as OpenAI's DALL-E 3 (*9**). Advanced general-purpose AI models can generate increasingly controllable images, with notable improvements for more complex concepts and the rendering of text in images (*9**).
- **Video –** Video-related general-purpose AI models take existing videos as inputs, such as Meta's V-JEPA (*37**), or can generate video from text, like OpenAI's Sora (*38**). Some general-purpose AI models learn to encode object properties that can be tracked across time in videos. Current models can generate realistic videos, but are limited in terms of length (generally less than one minute), fidelity, and consistency.

- **Robotic actions –** General-purpose AI models can be used for planning multi-step robot actions, and to interpret instructions to guide lower-level actions (*39*). Initial efforts are also exploring general-purpose AI models that not only plan or interpret, but also generate, robotic actions such as Google's RT-2-X (*40**), but general-purpose AI model capabilities to generate robotic actions are relatively rudimentary. Part of the reason is that data collection is challenging, although substantial efforts are being made (*41, 42*).
- **Proteins and large molecules –** General-purpose AI models that work with proteins and other large molecules operate on various representations (e.g. residue sequences, 3D structures). These models can predict protein folding, generate useful novel proteins, and perform a range of protein-related tasks. They therefore fall under the definition of general-purpose AI models outlined in 1. Introduction. Protein general-purpose AI models can be increasingly controlled to generate designs of proteins with predictable functions across large protein families (*43, 44*).

### 2.2.2 Capabilities and limitations by skill

To assess general-purpose AI capabilities fully, it can be helpful to categorise them by well-known skills such as displaying knowledge, reasoning, and creativity. Compared to categorising by modality, skills are harder to precisely define, but provide a more intuitive lens into general-purpose AI capabilities.

Viewed through this lens of skills, today's most capable general-purpose AI systems show partial proficiency but are not perfectly reliable. Experts often disagree on whether current general-purpose AI systems can be said to have a specific skill or not. One way to look at this is through capability-limitation pairs.

- **Knowledge (capability) and inconsistency (limitation) –** General-purpose AI models encode an extensive range of facts that are found across the public internet (*45*), yet are limited in identifying subtle factual differences and do not always generate self-consistent text, so the elicitation of knowledge from a general-purpose AI model can be inconsistent (*45, 46*).
- **Creativity (capability) and hallucination (limitation) –** General-purpose AI models can generate novel examples (for example, new images or text). This type of 'creativity' can be useful, but can also lead to 'hallucinations' fabricating content. For example, language models commonly generate non-existent citations, biographies, or facts (*46, 47*, 48, 49, 50*) which pose risks from misinformation (see 4.1. Malicious use risks).
- **Common sense reasoning (capability) and causality (limitation) –** Terms such as 'common sense' or 'reasoning' are often not well defined in the field of artificial intelligence, and are often used in ways that differ from their everyday use to describe human capabilities. In some circumstances, general-purpose AI models demonstrate the ability to emulate broad 'common sense knowledge' (*51*), the ability to work through relatively complex problems step-by-step (*24*), and the ability to learn and apply novel patterns within a given context (known as 'in-context learning') (*30, 52*). However, appropriate forms of 'reasoning' are contextual and task-dependent. The extent to which general-purpose AI systems demonstrate genuine 'reasoning' or 'common sense' is contested. Research has shown (*53*) that some reasoning capabilities are improving even for common sense reasoning problems in unusual contexts (*53*), although there are significant limitations for other forms of basic 'common sense knowledge' (*54*). Even when general-purpose AI models appear to correctly 'reason' about the world, they may not have identified the underlying causal basis for this reasoning (*55*). There is a general consensus that currently, general-purpose AI falls short of human-level reasoning abilities.
- **Formal reasoning (capability) and compositionality (limitation) –** Language models, especially when given additional resources such as tools and multiple attempts, can perform some formal reasoning tasks in domains like mathematics, computer programming, and the natural sciences, noting the caveat above regarding use of the term 'reasoning' in this domain. For example, research shows that the Claude 3 model approaches the performance of graduate-level experts on relevant questions in biology, physics, and chemistry (*4**) – on a benchmark created after Claude 3 was initially trained (for discussion of benchmarking, see 2.4. Capability progress in coming years and 3. Methodology to assess and understand general-purpose AI systems). However, such models use the 'Transformer' architecture, which most of today's general-purpose AI systems are based on, introduced in 2017 (*16*). In principle, this architecture has fundamental limitations in performing arbitrary compositional reasoning, which underpins formal reasoning (*56*) it remains unclear how relevant these theoretical limitations are in practice.

- **Forecasting (capability) and novel concepts (limitation) –** Language models, when integrated into more complex systems, can forecast future events with reasonable predictive accuracy in restricted domains. A recent study (*57*) shows that language model systems using retrieval can match the aggregate performance of expert forecasters on statistical forecasting problems (i.e. predicting the probability of events). However, while current capabilities show models can synthesise information to reason about the likelihood of future events, the extent to which models can synthesise entirely new concepts appears to be limited (*58*).
- **Simulation (capability) and embodiment (limitation) –** General-purpose AI models can simulate the behaviour of virtual agents when integrated into virtual environments (*59*). For example, recent research shows that 25 virtual agents powered by OpenAI's ChatGPT can operate a virtual town in a way that appears to match aspects of human behaviour (*60*). However, while current general-purpose AI models can simulate virtual agents, they are limited in 'embodiment' and cannot yet effectively control physical robots or machines, as the integration of general-purpose AI models with motor control systems remains a challenge (*61*).

## 2.3 Recent trends in capabilities and their drivers

### KEY INFORMATION

- In recent years, general-purpose AI capabilities have advanced rapidly according to many metrics, thanks to both increasing the resources used for training and algorithmic improvements. Per model, these are estimated to have increased:
  - Compute for training: 4x/year
  - Training dataset size: 2.5x/year
  - Algorithmic training efficiency: 1.5x to 3x/year
  - Energy used for powering computer chips during training: 3x/year
  - Hardware efficiency: 1.3x/year
- Using ever more compute and data to train general-purpose AI models in recent years is referred to as 'scaling up' models. Performance on broad metrics improves predictably with scale, and many AI researchers agree that scaling has driven most of the increase in advanced general-purpose AI capabilities in recent years. However, it is debated if this has resulted in progress on fundamental challenges such as causal reasoning.

### 2.3.1 Recent trends in compute, data, and algorithms

Increased investment in computing resources, enhancements in hardware efficiency, the existence of readily accessible datasets online, and incremental innovations in algorithms have contributed to advances in general-purpose AI over the last decade. This section examines recent trends in computing power, data, and algorithms.

## Trends in compute used in training and inference

Computing resources used for training AI models have been increasing fast. Computing resources, often referred to as 'compute', represent the number of operations performed. This has grown exponentially since the early 2010s, with the average amount used to train machine learning models doubling approximately every six months (*17*). In 2010, notable machine learning models (*62, 63, 64*) used an average of approximately 1e15 floating-point operations (FLOP) (*65*), but by 2023 Inflection-2, the largest model with a publicly reported compute budget, used 1e25 FLOP (*66**) - a ten billion-fold increase. This progress is driven by industry labs' willingness to use more data centre capacity for large-scale general-purpose AI training. There is insufficient data to determine if this trend is changing over a shorter period such as the 2020s.

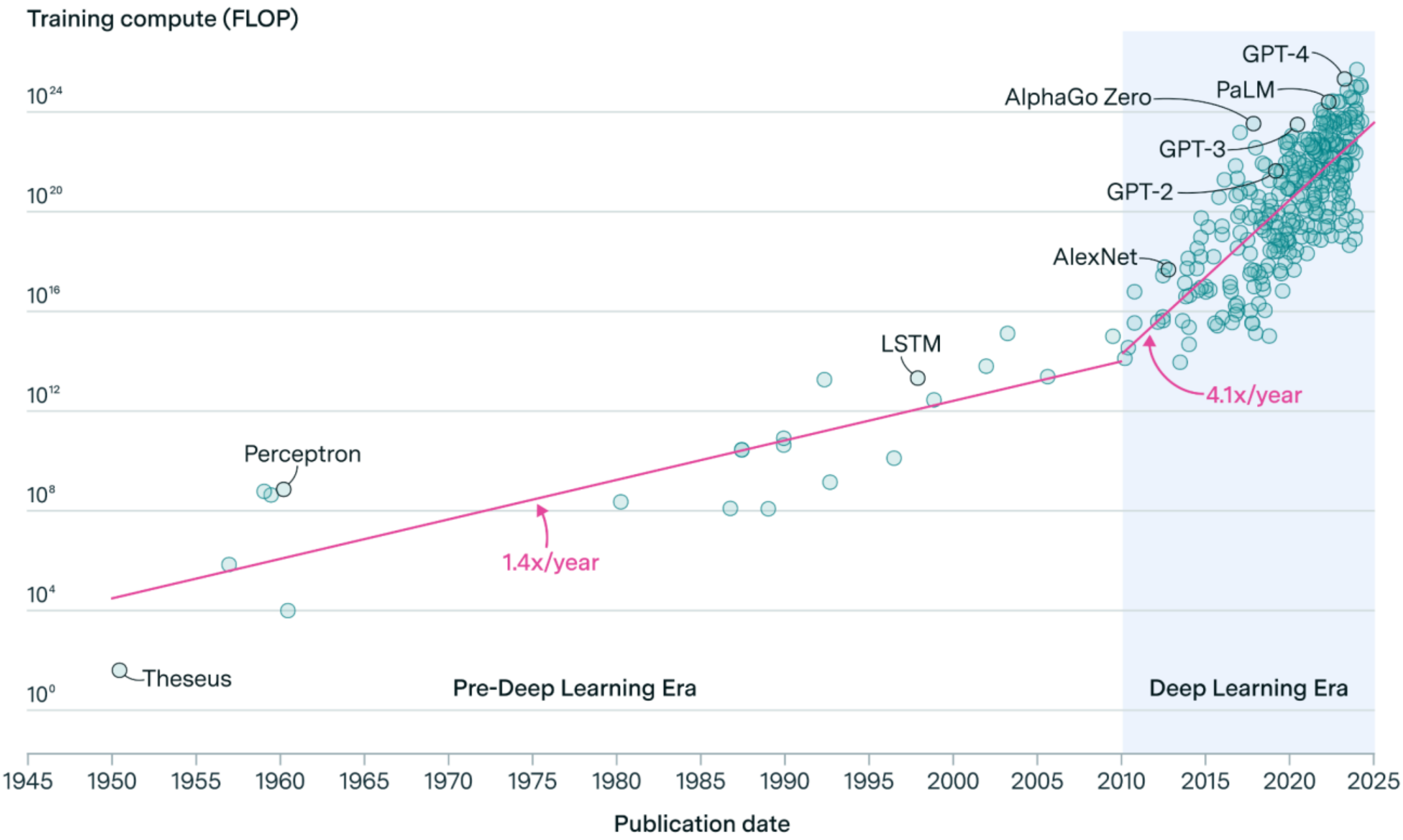

*Figure 1. Training compute of notable machine learning models over time (17, 65). Computation is measured in total floating-point operations (FLOP) estimated from AI literature. Estimates are expected to be accurate within a factor of two, or a factor of five for recent undisclosed models like GPT-4.*

*Reproduced with the kind permission of Epoch AI from 'Parameter, Compute and Data Trends in Machine Learning'. Published online at epochai.org. Retrieved from: 'https://epochai.org/data/epochdb/visualization'.*

Over the last fifteen years, the amount of compute per dollar has increased between around 50- to 200-fold (*67, 68*). However, the total amount of compute used for training general-purpose AI models far outpaced the reduction in computing costs: for example, Google's Word2vec model was trained using around 3e16 FLOP in 2013, around a billion-fold smaller than current frontier models (*65*). While GPU performance improvements have helped, these have been partially limited by data centre GPU shortages and high prices for top-tier GPUs used in AI applications. Supply chain shortages of high-end processors, packaging, high-bandwidth memory, and other components are delaying the technology sector's ability to meet the enormous demand for artificial intelligence hardware like AI servers (*69*). The expansion in general-purpose AI compute usage is mainly the result of industry labs being increasingly willing to allocate data centre resources and engineering staff to large-scale general-purpose AI training runs.

The discovery of neural 'scaling laws', which describe predictable relationships between the amount of compute, the size of the model and data, and performance, has contributed to a compute-centric view of AI development that is prominent at some leading AI labs.[1] The development of flagship general-purpose AI models such as Google Gemini Ultra and OpenAI's GPT-4 was guided by work on scaling laws (*2*, 3**). As a result, there is a greater need for hardware infrastructure expertise, and there are tighter collaborations between AI labs and technology giants such as Microsoft and Google.[2]

Computational resources for deployment have also seen significant growth. Companies are rapidly expanding infrastructure to meet these growing demands. The computational resources required for inference (a key part of serving general-purpose AI systems to users) have experienced significant growth (*76*) because the number of users for deployed general-purpose AI systems has increased rapidly. In April 2023, OpenAI's AI systems were reportedly estimated to incur $700k/day in inference costs (*77*). Some estimates indicate that the total computation spent on general-purpose AI inference already exceeds that devoted to training new models for example, AI inference represented 60% of Google's AI infrastructure emissions as of 2022 (*78*).

Growing compute resources both for training and inference have also rapidly expanded AI's energy usage (see 4.3.4 Risks to the environment).

## Training data trends: larger datasets, multimodal, synthetic data and human preferences

General-purpose AI developers have been able to significantly increase training dataset sizes thanks to the availability of content from the internet including open repositories of web data. These larger datasets contribute to higher performance on a wide range of metrics. Dataset sizes for training general-purpose AI have increased from around two billion tokens (a token is a word, a character, or sometimes part of a word) for the original Transformer model in 2017 to over three trillion tokens in 2023 (*79*, 80**), growing approximately 10x every three years (*65*).

However, general-purpose AI developers only have a limited amount of text data available on the internet to draw on (*81, 82*). While this could be overcome, for example by training on the same data many times, using AI-generated data, or training on other non-text data sources like YouTube videos, some believe that by 2030 shortages of accessible online high-quality text data could slow the rate at which models can be productively scaled (see 2.4.2 Will resources be scaled rapidly?).

Data quality plays a critical role in training high-performing language models. Selecting high-quality data and optimising the overall composition of the dataset can significantly improve model performance, but this process is labour-intensive (*83, 84, 85*). Moreover, measuring and analysing data to identify and mitigate problematic artefacts, such as biases and lack of diversity, is essential for producing high-quality models (*86**).

Training general-purpose AI models on diverse modalities like images, audio, and video alongside text, has recently gained traction. General-purpose AI models such as GPT-4, Claude 3, and Gemini Ultra combine different modalities to perform tasks requiring joint processing of textual, visual, and auditory information, such as analysing documents with text and graphics or creating multimedia presentations (*2*, 3*, 4**).

'Human preference' data captures the types of outputs users prefer and has become crucial for developing general-purpose AI systems. This data cannot be mined from publicly available sources, but must be produced specifically for training; as such it is more expensive than the text data used for pre-training. This data helps fine-tune language models to conform with user and developer needs, adapt to diverse preferences, and ground the models in human judgments of quality and helpfulness (*20, 21*, 87**). AI labs and large companies may have an advantage in producing and accessing large quantities of proprietary human preference data.

---

[1] Neural scaling laws have significantly influenced the compute-centric view of AI development at some frontier AI labs. Anthropic's research principles state that they "develop scaling laws to help us do systematic, empirically-driven research (...) to train networks more efficiently and predictably, and to evaluate our own progress." (70*).

[2] For example, there are partnerships between OpenAI and Microsoft (71*), Anthropic and Google Cloud (72*) as well as Amazon Web Services (73*), Cohere and Google Cloud (74*), and Mistral and Microsoft Azure (75*).

### Techniques and training methods for general-purpose AI have improved consistently

The techniques and training methods underpinning the most capable general-purpose AI models have consistently and reliably improved over time (*88*, 89*). The efficiency of AI techniques and training methods has been increasing 10x approximately every 2 to 5 years in key domains such as image classification, game-playing, and language modelling. For example, the amount of compute required to train a model to perform image classification to achieve a set level of performance decreased by 44x between 2012 and 2019, meaning that efficiency doubled every 16 months. Game-playing AI systems require half as many training examples every 5 to 20 months (*90*). In language modelling, the compute required to reach a fixed performance level has halved approximately every 8 months on average since 2012 (*89*). These advances have enabled general-purpose AI researchers and labs to develop more capable models over time within a limited hardware budget.

There have also been incremental advances in algorithms that are not best understood as increasing compute efficiency. For example, new techniques have significantly increased the size of context windows, allowing general-purpose AI systems to process larger quantities of information (*31*, 91*, 92**), and post-training algorithms allow general-purpose AI systems to use tools and take actions in the world without human assistance (see 2.4.3. Will algorithmic progress lead to rapid progress?).

Despite significant advancements in AI algorithms, general-purpose AI has seen relatively few major conceptual breakthroughs in recent years. The 'Transformer architecture' remains perhaps the most significant innovation, and is used by most advanced general-purpose AI systems (*16*). While many alternative architectures have been proposed, none have yet substantially and consistently outperformed the Transformer. Recent 'selective state space models' (*93*) might prove to be more efficient than the Transformer, once properly tested. These models reinforce the recent trend of allowing language models to analyse longer contexts, such as books and large software projects. If more fundamental conceptual breakthroughs are needed to advance general-purpose AI capabilities, this could be a key barrier to further development even if incremental improvements and scaling continue to drive rapid progress in some areas (see 2.4.1. If resources continue to be scaled rapidly, would this lead to rapid advancements?).

## 2.3.2 Recent trends in capabilities

### General-purpose AI has advanced increasingly rapidly, nearing or surpassing human-level performance on some metrics, with debated implications

The pace of recent general-purpose AI progress has been rapid, often surpassing the expectations of AI experts on some metrics. Over the last decade, AI has achieved or exceeded average human-level performance on some benchmarks in domains such as computer vision, speech recognition, image recognition, and natural language understanding (Figure 2). The latest advances in LLMs build upon this longer-running trend.

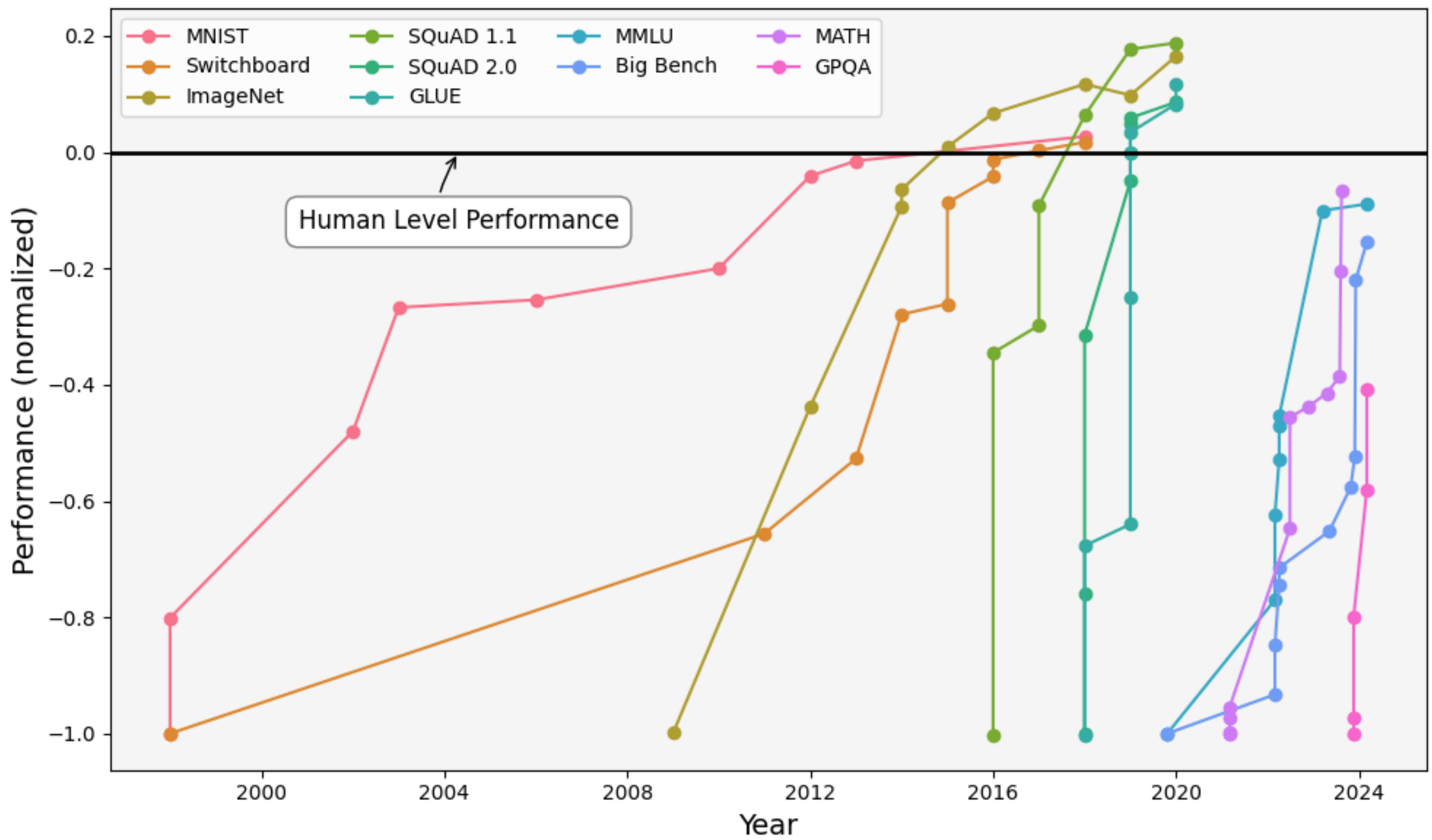

*Figure 2. Performance of AI models on various benchmarks from 1998 to 2024, including computer vision (MNIST, ImageNet), speech recognition (Switchboard), natural language understanding (SQuAD 1.1, MMLU, GLUE), general language model evaluation (MMLU, Big-Bench, and GPQA), and mathematical reasoning (MATH). Many models surpass human-level performance (black solid line) by 2024, demonstrating significant advancements in AI capabilities across different domains over the past two decades. Data are from (94) for MNSIT, Switchboard, ImageNet, SQuAD 1.1, 2 and GLUE. Data for MMLU, Big Bench, GPQA are from the relevant papers (95, 96, 97).*

LLM capabilities have advanced significantly in multiple domains between 2020 and 2024, shown by broad benchmarks such as Massive Multitask Language Understanding (MMLU) (*95*), Big-Bench (*96*), and Graduate-Level Google-Proof Q&A (GPQA) (*97*). In 2020, general-purpose AI models performed substantially worse than average human test subjects on many of these benchmarks; in 2024, advanced general-purpose AI models have approached human-level performance. For example, consider the MATH benchmark (*98*), which tests mathematical problem-solving skills. Initially general-purpose AI systems performed weakly on this benchmark, but two years after its release, GPT-4 seemed to achieve 42.5% accuracy (*99**) and subsequent work pushed state-of-the-art performance using GPT-4 to 84.3% (*100*), which is close to the score obtained by expert human testers.

Despite rapid progress on benchmark metrics, these benchmarks are highly limited compared to real-world tasks, and experts debate whether these metrics effectively evaluate true generalisation and meaningful understanding (*101*). State-of-the-art general-purpose AI models often exhibit unexpected weaknesses on some benchmarks, indicating that they partly or fully rely on memorising patterns rather than employing robust reasoning or abstract thinking (*102, 103*). In some cases, models were accidentally trained on the benchmark solutions, leading to high benchmark performance despite the absence of the actual capability (*104, 105*). Models also struggle to adapt to cultures that are less represented in the training data (*106*). This underscores the significant disparity between benchmark results and the capacity to reliably apply knowledge to practical, real-world scenarios.

## AI and human capabilities have distinct strengths and weaknesses, making comparisons challenging

While it may be tempting to compare the cognitive capabilities of humans to the capabilities of general-purpose AI systems, they have distinct strengths and weaknesses, making these comparisons less meaningful in many cases. While general-purpose AI excels in some domains, it is arguably lacking the deep conceptual understanding and abstract reasoning capabilities of humans (*102*). Current general-purpose AI systems often demonstrate uneven performance, excelling in some narrow domains while struggling in others (*102*).

Current general-purpose AI systems are prone to some failures that humans are not (*107, 108*). General-purpose AI reasoning can be 'brittle' (unable to cope with novel scenarios) and overly influenced by superficial similarities (*102*). LLMs can fail at reasoning in contexts where humans typically excel. For example, a model trained on data including the statement: “Olaf Scholz was the ninth Chancellor of Germany” will not automatically be able to answer the question “Who was the ninth Chancellor of Germany?” (*107*). In addition, LLMs can be exploited by nonsensical input to deviate from their usual safeguards, while humans would recognise these prompts (see 5.2. Training more trustworthy models).

## As general-purpose AI models are scaled up, their capabilities improve overall, but to date, this growth has been hard to predict for specific capabilities

The aggregate performance of language models has improved reliably and predictably with the scale of computing power and data. Researchers have discovered empirical ‘scaling laws’ that quantify the relationship between these inputs and the capabilities of the model on broad performance measures like next-word-prediction (*109*, 110**). Empirical studies across diverse domains have evidenced performance improvements in machine learning systems with increased computational resources, including in vision (*111*, 112*), language modelling (*109*, 110**), and games (*113**).

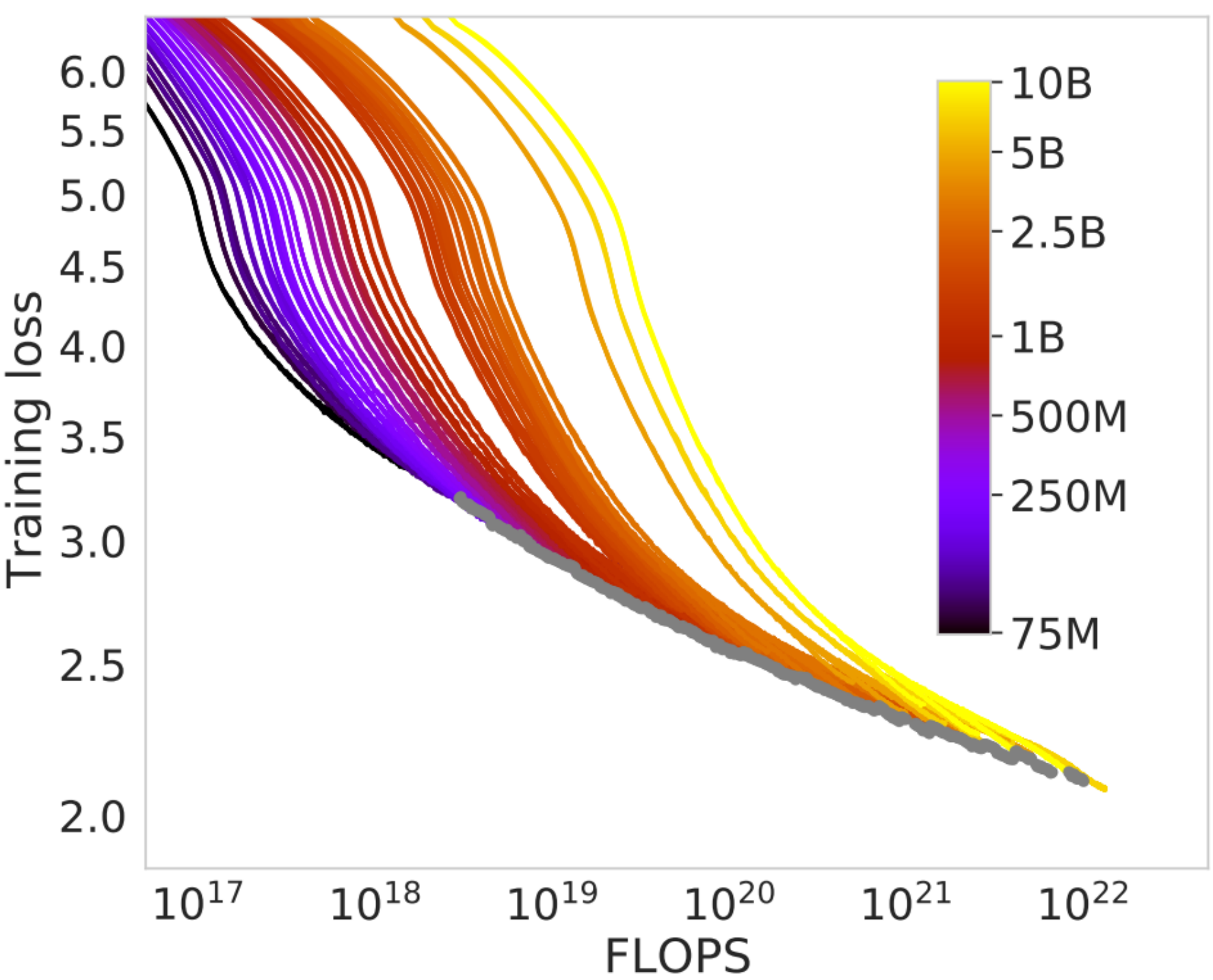

*Figure 3. Cross-entropy loss scales predictably with compute across a broad range of empirically-studied training runs. FLOPS refers to the number of operations performed during training. Figure from (110*). Different colours represent models with different numbers of parameters. Each line shows how loss falls as training FLOP increases for a model. Permissions were sought from the author.*

The best-known scaling law predicts that, as language models are grown in size and are trained on more data, they improve by a predictable amount (*109*, 110**)[3]. Specifically, these models become more accurate at predicting the next 'token' in a sequence, which can be a word, a character, or a number. When this performance improves, the model is effectively getting better at the tasks implicit in the dataset. For example, general-purpose AI model performance has been observed to consistently improve on broad benchmarks that test many capabilities, such as MMLU (*95*), as the models are scaled up.

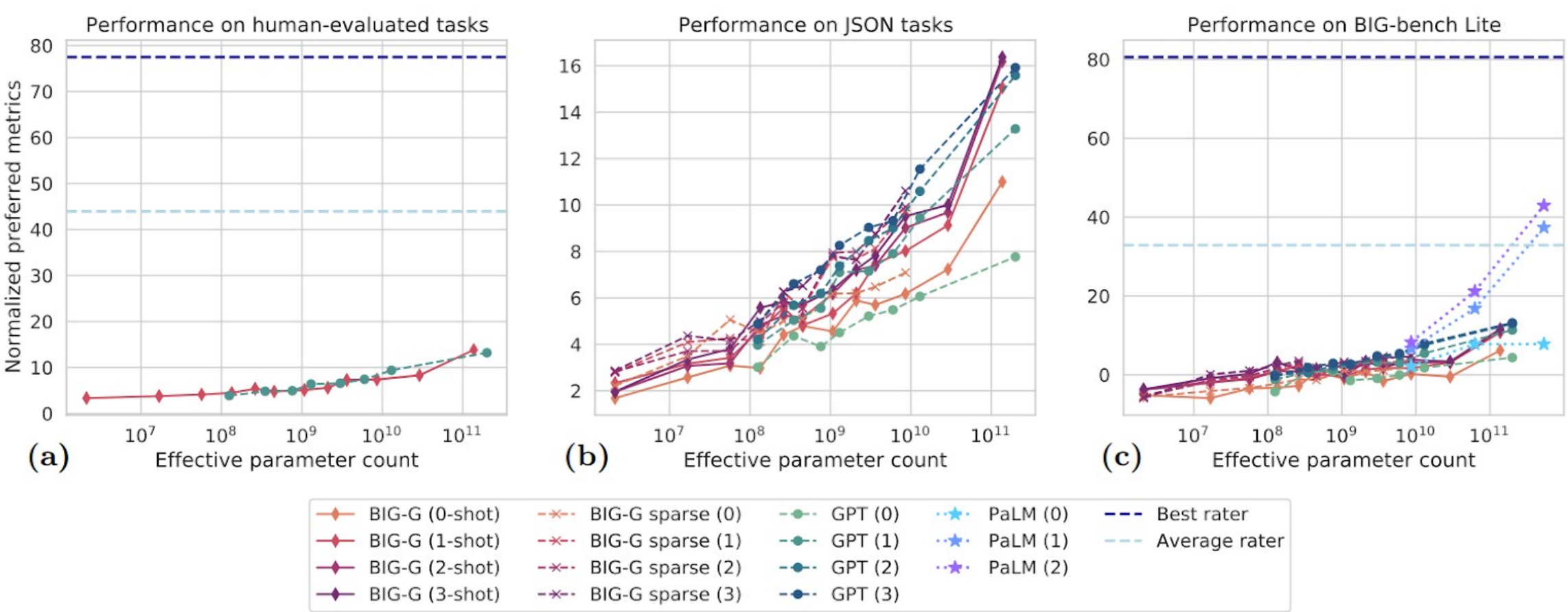

*Figure 4. Performance on broad benchmarks such as Big-Bench has been strongly associated with parameter scaling, and compute scaling more generally. This figure was taken from (96).*

These scaling laws are derived from empirical observations, not from inviolable principles, although theoretical models have been proposed to explain them (*115*, 116*, 117, 118, 119*). As a result, there is no mathematical guarantee that they will continue to hold for scales beyond the range of the empirical data used to establish them.

*Aggregate* performance across many tasks within comprehensive benchmarks can be partially predicted based on the model scale (see Figure 4) However, it is unclear whether we can currently reliably predict far in advance if and when *specific* capabilities will appear. There are many documented examples of capabilities that appear when models reach a certain scale, sometimes suddenly, without being explicitly programmed into the model (*120, 121, 122*, 123*). For example, large language models at a certain scale have gained the ability to perform the addition of large numbers with high accuracy, when prompted to perform the calculation step-by-step. Some researchers sometimes define these as 'emergent' capabilities (*120, 121, 122*, 123*), indicating that they are present in larger models but not in smaller models and so their emergence may be hard to predict in advance. This suggests that new capabilities, including beneficial and potentially harmful ones, may emerge unexpectedly.

It is debated whether these capabilities appear gradually or suddenly, and there is disagreement about how far in advance they can be predicted. Recent research has found that some such capabilities appear more gradually and predictably, if more linear and continuous metrics are used to measure progress (*124*). This has led some researchers to question that capabilities are 'emergent', since some definitions of emergence in AI require that abilities appear suddenly at a certain scale (*124*). This suggests that capabilities that currently seem to appear suddenly may turn out to be predictable further in advance if different progress metrics are used. It is an open question whether, and how far in advance, it will become possible to predict the new capabilities of new models.

[3] This work implicitly treats the data as homogenous; more recent work studies the intersection of scaling laws and data set curation (114).

Recent research has identified examples of 'inverse scaling', where language model performance *worsens* as model size and training compute increase (*125*). For instance, when asked to complete a common phrase with a novel ending, larger models are more likely to fail and simply reproduce the memorised phrase instead (*125*). While model scaling sometimes leads to decreased performance, research on this phenomenon also finds random-seeming fluctuations in performance. Some apparent inverse scaling trends may not persist when extrapolating to much larger models, with performance eventually improving again (*126*). The full implications of inverse scaling remain unclear, and further research is likely needed to better understand this phenomenon.

## 2.4 Capability progress in coming years

### KEY INFORMATION

- The pace of future progress in general-purpose AI capabilities has important implications for managing emerging risks but experts disagree on what to expect, even in the near future. Experts variously support the possibility of general-purpose AI capabilities advancing slowly, rapidly, or extremely rapidly.
- This disagreement involves a key question: Would continued 'scaling up' and refining existing techniques yield rapid progress, or is this approach fundamentally limited, and will unpredictable research breakthroughs be required to substantially advance general-purpose AI abilities? Those who think research breakthroughs are required often think that recent progress hasn't overcome fundamental challenges like common sense reasoning and flexible world models.
- In recent years, three main factors have driven progress in AI: scaling up the computational power ('compute') used in training; scaling up the amount of training data; and improving AI techniques and training methods.
- Leading AI companies are betting on all three factors continuing to drive improvements, particularly increased compute. If recent trends continue, by the end of 2026 some general-purpose AI models will be trained using 40x to 100x the computation of the most compute-intensive models currently published, combined with around 3 to 20x more efficient techniques and training methods.
- However, there are potential bottlenecks to further increasing both data and compute, including the limited availability of data, AI chip production challenges, high overall costs, and limited local energy supply. AI companies are working to overcome these bottlenecks. The pace of scaling also depends on regulations that might place constraints or conditions on AI deployment and development.

This section examines the feasibility and effectiveness of further scaling up compute and training data, and the potential for rapid advancements through algorithm development. Overall, both approaches are likely to add up and lead to further advancements, but no agreed methodology exists to predict the pace of advancements. For a discussion of global disparities in AI capabilities, see 4.3.2 Global AI divide.

### 2.4.1 If resources continue to be scaled rapidly, would this lead to rapid advancements?

A key driver of disagreements about future progress is how to interpret past progress. In recent years, scaling computational resources and data has led to consistent performance gains in general-purpose AI systems as evidenced by higher scores on a range of benchmarks (see As general-purpose AI models are scaled up, their capabilities improve overall, but to date, this growth has been hard to predict for specific capabilities). Some AI researchers judge that this past progress involved significant and meaningful advancements in the understanding and reasoning capabilities of general-purpose AI systems and that, with substantially more compute and perhaps moderate conceptual innovations, continued progress of this kind could lead to the development of general-purpose AI systems that perform at a broadly human level or beyond, for most cognitive tasks (*127*).

Other researchers are sceptical. They argue that current general-purpose AI systems, which are based on deep learning (the currently dominant approach to machine learning relying on deep artificial neural networks (*15*)), fundamentally lack crucial components of intelligence. In particular, current deep learning systems are thought by some to lack causal reasoning abilities (*128*), abstraction from limited data, common sense reasoning (*102, 129, 130*), and flexible predictive world models (*55, 128, 130*). These researchers argue that these shortcomings cannot be simply resolved by scaling alone (*102, 129, 130*). This point is supported by the significant limitations of current systems, which are discussed in 2.2 What current general-purpose AI systems are capable of. Addressing these limitations, they argue, may require significant conceptual breakthroughs and innovations beyond the current deep learning paradigm. This would suggest that achieving human-level performance in general-purpose AI systems requires significant conceptual breakthroughs, and that the current type of progress, driven by incremental improvements, is insufficient to reach this goal.

Fundamental conceptual breakthroughs resulting in significant leaps in general-purpose AI capabilities are rare and unpredictable. Even if novel techniques were invented, existing general-purpose AI systems' infrastructure and developer convention could provide barriers to applying them at scale. So, *if* a major conceptual breakthrough is required, it could take many years to achieve.

These opposing views are not necessarily incompatible: although current state-of-the-art deep learning systems broadly have weaker reasoning abilities than most humans, we are seeing progress from one generation of a general-purpose AI model to the next, and many AI researchers are exploring ways to adapt general-purpose AI models to unlock or improve 'system 2' reasoning (analytic, rule-based, and controlled reasoning) and 'autonomous agent' abilities. Progress on these capabilities, if it occurs, could have important implications for AI risk management in the coming years. See 4.4.1 Cross-cutting technical risk factors for a discussion of the potential for progress in and risk from autonomous agent capabilities.

### 2.4.2 Will resources be scaled rapidly?

There is ongoing debate among experts about whether, and for how long, it will be possible to continue rapidly increasing the resources going into AI development.

Technology giants such as Google and Amazon are making substantial capital investments in data centres and GPU acquisitions to support further scaling up general-purpose AI models. If this yields substantial improvements in the near future, the resulting capabilities may inspire market confidence and justify additional rounds of spending. Large technology companies have the cash reserves needed to scale the latest training runs by multiples of 100 to 1,000 (*131**). However, further scaling past this point may be more difficult to finance. In addition, access to capital investment is not the only potential bottleneck. Lack of data, energy, and GPUs are all potential barriers to further rapid scaling of resources that are discussed below. In addition, there are wide global disparities in the quality of digital infrastructure, posing additional barriers to many countries and contributing to a widening global divide in AI capabilities (see 4.3.2 Global AI divide).

## Data bottlenecks may limit rapid scaling, but this has stimulated new approaches like multi-epoch training, synthetic data, and transfer learning across domains

**While data availability may limit general-purpose AI scaling over the medium term, solutions like synthetic data and transfer learning could help to address this bottleneck.** Current state-of-the-art general-purpose AI models use text datasets with trillions of words. The rapid scaling up of training dataset size may soon be limited by the quantity of accessible online high-quality text data (*81, 82*) and by legal disputes about data access.

Data availability bottlenecks for training large general-purpose AI models are a recent challenge, and methods to overcome them are still in the early stages of exploration. A range of approaches for overcoming such challenges are available, such as:

- **Multi-epoch training:** it has been common to train LLMs for a 'single epoch', meaning that it sees the training data just once. However, training on the same data multiple times is standard practice in other domains of machine learning, and it has been shown that training general-purpose AI models for four epochs could yield benefits which are roughly equivalent to 4x data, though with additional epochs the benefits decline (*81*). This could somewhat expand what can be accomplished with existing data stock for language models, depending on the extent to which current state-of-the-art models are already trained for multiple epochs on certain data sources.
- **Synthetic or self-generated data**: General-purpose AI-generated outputs can augment training data or human feedback data, which is useful when real data is limited (*132**). This approach offers control over the generated data, filling gaps in existing datasets (*133*) and improving model robustness, prediction and generalisation (*133, 134*). However, there are concerns that synthetic data could reduce meaningful human oversight and might gradually amplify biases and undesirable behaviours of general-purpose AI models (*135*), and may not improve capabilities as much as real data (*133, 136*).
- **Transfer learning across domains**: Training general-purpose AI models on data from various sources, such as text, images, video, speech, and biological sequences (*35, 137, 138**), could substantially increase the available data. For example, training on code was found to increase model performance on natural language tasks by the equivalent of doubling the language dataset (*81*).

## Energy demands could strain existing electricity infrastructure

The increasing energy demands for training general purpose AI systems could start to strain energy infrastructure. Globally, computation used for AI is projected to require at least 70 TWh of electricity in 2026 (*139*), roughly the amount consumed by smaller European countries such as Austria or Finland. This could divert energy from other purposes and have environmental costs (*140*).

In the US, for example, where many leading AI companies are currently located, electric grids and transmission infrastructures may struggle to accommodate the surge in AI-related electricity demand. Upgrading the electrical grid to transport more power from generation plants to data centres involves a lengthy process of planning, approval, and construction. This slow build-out process poses particular challenges for general-purpose AI training facilities, which will likely require many GPUs to be closely co-located geographically, resulting in an immense local power draw.

Key AI companies are responding by proactively seeking to secure their power supply. For example, a compute provider recently purchased a data centre with a 960MW energy supply (*141**). This is enough to power a training run with around 100x more computational resources than was used to train GPT4 (*2**), but more energy would be required to continue scaling at the current pace for more than a few years.

### AI chip production challenges and slowing GPU improvements may limit AI compute scaling

Over the past few years, the production of data-centre GPUs has been a bottleneck in scaling up compute for general-purpose AI systems, partly due to the limited capacity of semiconductor manufacturing plants and the constraints and priorities in the global semiconductor supply chain (*142, 143*). AI chip manufacturing depends on a complex, hard-to-scale supply chain, featuring sophisticated lithography, advanced packaging, specialised photoresists, and unique chemicals. Constructing new semiconductor fabrication plants ('fabs') is very expensive and typically takes three to five years (*144, 145*) making the industry slow to respond to market demand. These factors make projecting future supply complex, and a range of scenarios for chip availability have been proposed. While state-of-the-art GPU production is picking up substantially, supply chains for AI chips may not be able to adapt to demand. This may slow down the ambitions of frontier AI companies for further rapid growth, though AI companies might continue to scale in the near-term if they can acquire large fractions of the total number of GPUs produced.

Improved GPU performance has also contributed to the recent scaling of compute (*67, 68*). It is possible that over this decade, progress in individual GPU performance may slow due to physical limits on transistor size (*146, 147*) and energy efficiency (*148*). This would have only a limited impact on scaling because the primary driver of compute scaling has not been improved GPU performance, but the increase in the number of GPUs used. GPU price performance and energy efficiency for relevant computation have been improving roughly 30% annually, with an additional factor of 10x - 100x from the addition of tensor cores and the move to lower precision formats (*67, 68*). However, the total compute used in training has increased by approximately 4x per year since 2010 (*17*), outpacing the rate of hardware efficiency improvements. This suggests that increased spending, rather than gains in hardware efficiency, has been the primary driver of the growth in compute budgets for AI training.

## 2.4.3 Will algorithmic progress lead to rapid advancements?

There is continued rapid growth in the efficiency of training algorithms. Techniques and algorithms underpinning general-purpose AI models have consistently and robustly improved over time (see 2.3.1 Recent trends in compute, data, and algorithms). This includes incremental changes to key algorithms such as the architecture of transformers, improvements to how AI algorithms are implemented on hardware, a better understanding of how to scale models, better methods to process representations in a neural network, and other advances. These advances enable researchers and labs to produce ever-more capable general-purpose AI models without increasing their hardware budget, for example in vision (*88**), reinforcement learning (*90*) and language modelling (*89*). The rate of improving algorithmic efficiency in language modelling shows no sign of slowing down (*89*), though there may be diminishing returns to progress in the near future.

**Post-training algorithms can be used to significantly improve general-purpose AI model capabilities at low cost, including via fine-tuning, tool use, and structured reasoning techniques.** Recent years have seen significant advancements in techniques and approaches for enhancing the performance of general-purpose AI models after pre-training. Many post-training algorithms improved model performance on a given benchmark by more than using 5x the training compute, and in some cases more than 20x (*25*). Post-training algorithms can be applied to a given model for various use-cases with the aim of better serving end-users specific needs, at a much lower cost than developing the original model. This low cost means that a wide range of actors, including low-resource actors, could advance frontier general-purpose AI capabilities by developing better post-training algorithms. Governance processes need to take into account post-training algorithms.

Post-training algorithms include fine-tuning models for better performance, equipping them with the ability to leverage external tools, crafting prompts to guide their outputs, structuring their reasoning processes for more coherent and logical responses, and selecting from multiple responses the most relevant and accurate candidate outputs. There is a rapidly growing body of work on post-training enhancements, to improve frontier LLMs performance broadly (*149*), and in specific domains such as code generation (*150*) and mathematics (*151*). Such innovations have the potential to further improve the performance of general-purpose AI systems over the next few years. However, if there is a slowdown in the scaling of resources going to training, due to the bottlenecks discussed above, this might in turn slow progress in finding better post-training algorithms.

**General-purpose AI systems might be deployed to automate and accelerate AI R&D.** Narrow AI systems have already been used to develop and improve algorithms (*152, 153*). Recent LLMs are used in areas related to AI R&D, particularly in programming (*26*), generating and optimising prompts (*154, 155, 156, 157*), replacing human fine-tuning data (*158**), and selecting high-quality training data (*159**). As the capabilities of general-purpose AI systems advance, it becomes harder to predict the effect on algorithmic progress and engineering in AI.

# 3 Methodology to assess and understand general-purpose AI systems

## KEY INFORMATION

- General-purpose AI governance approaches assume that both AI developers and policymakers can understand and measure what general-purpose AI systems are capable of, and their potential impacts.
- Technical methods can help answer these questions but have limitations. Current approaches cannot provide strong assurances against large-scale general-purpose AI-related harms.
- Currently, developers still understand little about how their general-purpose AI models operate. Model explanation and interpretability techniques can improve researchers' and developers' understanding of how general-purpose AI systems operate, but this research is nascent.
- The capabilities of general-purpose AI are mainly assessed through testing the general-purpose AI on various inputs. These spot checks are helpful and necessary but do not provide quantitative guarantees. They often miss hazards, and overestimate or underestimate general-purpose AI capabilities, because test conditions differ from the real world. Many areas of concern are not fully amenable to the type of quantification that current evaluations rely on (for example, bias and misinformation).
- Independent actors can, in principle, audit general-purpose AI models or systems developed by a company. However, companies do not always provide independent auditors with the necessary level of 'white-box' access to models or information about data and methods used, which are needed for rigorous assessment. Several governments are beginning to build capacity for conducting technical evaluations and audits.
- It is difficult to assess the downstream societal impact of a general-purpose AI system because rigorous and comprehensive assessment methodologies have not yet been developed and because general-purpose AI has a wide range of possible real-world uses. Understanding the potential downstream societal impacts of general-purpose AI models and systems requires nuanced and multidisciplinary analysis. Increasing participation and representation of perspectives in the AI development and evaluation process is an ongoing technical and institutional challenge.

Modern general-purpose AI systems can contain multiple large-scale models and hundreds of billions of parameters, often deployed as general-purpose products. Consequently, it is difficult to anticipate how such general-purpose AI products may function in the many possible deployment scenarios, and even harder to appropriately characterise the downstream consequences of their deployment. Scientists are often surprised at the unexpected capabilities and impacts of general-purpose AI systems.

## 3.1 General-purpose AI assessments serve to evaluate model capabilities and impacts.

There are two broad reasons to assess general-purpose AI models and systems:

1. **Determining general capabilities and limitations:** Model evaluations indicate the relationship between model design choices and model outcomes. This performance analysis helps researchers understand how well these systems meet our expectations in both controlled and natural settings. A more informed understanding of model capabilities helps to judge its suitability for use. Every evaluation comes with limitations and uncertainties, which must be documented to properly interpret its results.
2. **Assessing societal impact and downstream risks:** Forecasting and assessing a general-purpose AI system's broader impact can inform debate about questions related to deployment or governance. However, these assessments are a complex interdisciplinary challenge. Societal risk assessment can assess *product safety, security vulnerabilities, and unwanted externalities such as labour and environmental impacts,* among other concerns. This typically involves factors that can contribute to accidents during expected product use, as well as addressing unanticipated and malicious use.

## 3.2 Approaches for model performance analysis

Various stakeholders (i.e. AI developers, users, impacted population members, etc.) have expectations of how a general-purpose AI system should behave in terms of model capabilities and preventing negative downstream social impacts. Researchers have developed a variety of methods to compare model outcomes with these expectations (*160*). This model performance analysis is integral in understanding how a model performs and which limitations, benefits, or risks may arise in deployment.

### 3.2.1 Case studies

In many research papers, the evaluation of model capabilities is qualitative, relying on anecdotal demonstrations of model performance (*161*) and human judgments. For instance, early evaluation of image-generating models often relied on simply demonstrating a small number of examples (*162, 163*). When GPT-4 was newly released (*99**), examples of model outputs on a curated set of tasks, beyond conventional benchmarks, were used to illustrate model performance. Now, several popular benchmarks rely on asking human raters to rate the responses of different models against one another (*164, 165*). Meanwhile, risks are sometimes gauged through 'uplift studies', where the aim is to test how much more competent a human is at accomplishing a potentially harmful task when they have access to a general-purpose AI system versus when they do not (*166**).

### 3.2.2 Benchmarks

Most machine learning evaluations are completed on standardised benchmark measurements (*167*). For example, a relatively small set of image processing benchmarks involving classification (*168, 169, 170, 171*), segmentation (*36*, 172*), and question answering (*173, 174*) have been integral to AI vision research. Similarly, research on language modelling has been shaped by several publicly available benchmarks that are meant to measure general capabilities (*95, 96, 165, 175, 176, 177*, 178, 179*) and trustworthiness (*180, 181, 182*). More recently, benchmarks are being designed to measure the growing general-purpose AI capabilities at combining information from multiple modalities and using software tools like web browsers (*183*) (see also 4.4.1 Cross-cutting technical risk factors).

**Performance on benchmarks can be imperfect measures of downstream task performance.** Benchmarks are inherently proxy measures for desired performance, and good scores on a benchmark do not always translate to desired performance in practice (*184*) because of various validity challenges. *Internal* validity challenges relate to the reliability of the benchmark measurement, i.e. how reliable the reported metric is over repeated executions in comparison to a strong baseline. For instance, *internal* validity issues arise if the benchmark does not contain enough examples to make statistically valid claims about model performance (*185*) or contains false labels (*186, 187*). *External* validity refers to how well benchmark performance translates to real-world settings. Benchmarks themselves can be poorly constructed, inadequate, or incomplete representations of real-world tasks. For instance, limitations on how state-of-the-art models are prompted and combined with other systems can lead to their capabilities being underestimated (*24, 188, 189*). Currently, benchmarks are often not explicit about their scope of applicability. Benchmarks that lay claim to 'general' performance often disguise biases in cultural representation and annotator differences, contested notions of ground truth, and more (*101, 167, 190, 191, 192*). Also, as modern general-purpose AI models have been trained on large amounts of internet data, it can be difficult to disentangle novel capabilities from memorised ones (*193, 194*).

**Interpreting human performance on benchmarks can be difficult.** A critical aspect of intuitively understanding model performance is a comparison with human performance (*195**). Human performance measures are often unreliable – for instance, the baseline for 'human performance' on the ImageNet benchmark consists of the annotations of a single graduate student (*196*). Human annotators of 'ground truth' are notoriously fickle (*197*), often disagreeing in meaningful ways (*198*) due to differences in cultural context, values, or expertise. Furthermore, there is a meaningful difference between human performance and human competence at a task – the latter often involves, for example, judgments of robustness and not just accuracy (*199*). Evaluations which intentionally diversify their annotator populations (*200*), allow for a multiplicity of ground truth labels (*198*), or properly consider the context of human performance claims (*201*) tend to be more trustworthy assessments of human-AI comparisons.

### 3.2.3 Red-teaming and adversarial attacks

Before deploying systems in real-world conditions, evaluators use *adversarial attacks and red-teaming* to identify worst-case behaviours, malicious use opportunities, and the system's potential to fail unexpectedly. In cybersecurity, an adversarial attack refers to a deliberate attempt to make a system fail. For example, attacks against language models can take the form of automatically generated attacks (*202, 203, 204*, 205, 206*) or manually generated attacks (*204*, 207*). These can include 'jailbreaking' attacks which subvert the models' safety restrictions (*208, 209, 210, 211, 212*).

A *red team* refers to a set of people who aim to find vulnerabilities in a system by attacking it. In contrast to benchmarks which are a fixed set of test cases, a key advantage of red-teaming is that it adapts the evaluation to the specific system being tested. Through interaction with a system, red-teamers can design custom tests for models. Researchers approach red teaming through various strategies and tools. Ojewale et al. (*213*) map out an ecosystem of tools leveraged in the AI accountability process including resources for 'harm discovery', such as 'bug bounty' platforms, incident databases (*214*), and more. These tools support the identification of potential vectors of harm and enable broader participation in harm discovery.

**However, evaluators can sometimes fail to represent public interests.** Red-teaming for state-of-the-art general-purpose AI systems is predominantly done by the organisations that develop them (*2*, 22*, 204**). Academia, auditing networks, and dedicated evaluation organisations can also play a key role (*215, 216, 217*). As is the case with AI developers themselves, red team evaluators are not always representative of public interests or demographics and can display bias, or omit important considerations in their identification or assessment of AI-related harms (*215*). Best practices for red-teaming are not established yet (*216*).

**Various faults in general-purpose AI models have remained undetected during red teaming.** Red-teaming and adversarial attacks are useful for developing a better understanding of a model's worst-case performance and assessing performance at capabilities benchmarks don't adequately cover. However, they come with limitations. Previous work on benchmarking red-teaming and adversarial attack techniques has found that bugs often evade detection (*218*). A real-world example is jailbreaks for current state-of-the-art general-purpose AI chat systems (*208, 209, 210, 211, 212*), which seem to have evaded initial detection by developers who designed them (*2*, 22*, 204**). Overall, red-teaming is just one of several assessment tools necessary for meaningful understanding of general-purpose AI capabilities (*219*). Red-teaming can also fail to catch downstream harms, which only arise when the AI system is more widely deployed in society. This is discussed further in 4.3. Systemic risks.

## 3.2.4 Auditing

Design choices throughout the general-purpose AI development process affect how the resulting system works. Auditing provides a mechanism to scrutinise and ensure accountability for those choices. Different groups of general-purpose AI auditors achieve differing degrees of success when it comes to the quality of the evidence gathered and the accountability outcomes achieved (*215, 217*). One survey of a range of approaches to AI auditing (*217*) suggests independently assessing deployed general-purpose AI systems across different dimensions (see Figure 1) to hold stakeholders accountable for the choices they make regarding the development of the general-purpose AI system, and its use.

Analysis of training data can reveal problematic content. The analysis of training data, otherwise known as 'data audits' (*217*), is one concrete approach to the analysis of key model design choices, and can reveal problematic content. In the machine learning development process, data is collected, curated, and annotated (*190*). Examining how these data engineering decisions could lead to indirect harm, and influence its outcomes is useful in understanding the model and its ultimate downstream impacts. For example, the analysis of text and image data from the internet used to train modern systems has identified copyrighted content (*220, 221, 222*), hateful speech and memes (*223, 224, 225*), malign stereotypes (*224, 226, 227*), sexually-explicit content (*223, 226*), and depictions of sexual violence (*226*) including child abuse material (*228*).

Furthermore, investigations into training datasets often reveal issues in terms of the demographic, geographic, and linguistic under-representation of certain populations in mainstream data sources (*229, 230*). Such data audits have provided the evidence necessary for legal challenges on copyright. For example, the New York Times lawsuit against OpenAI (*231*) heavily cites the data audit from Dodge et al. (*232*) and has led to the attempted redaction of some datasets deemed to contain inappropriate material (*233, 234*). However, modern general-purpose AI models are often trained on extremely large amounts of internet datasets, and these datasets are often not made public, so data provenance remains a systemic challenge (*235, 236*). As a result, it is challenging to systematically search for potentially harmful examples in training data at scale.

**The analysis of AI modelling and product choices can reveal trade-offs and indicate downstream risks.** Aside from training data, other methodological choices can contribute to specific problems. Audits that scrutinise how a model is developed are typically referred to as 'process audits'. For example, human feedback-based methods are state-of-the-art for training general-purpose AI models, but can contribute to problems such as sycophancy (*237, 238*) and producing non-diverse outputs (*239, 240, 241*). Similarly, engineering decisions such as 'model pruning' can result in inequitable impacts for certain test demographics (*242*). In the same vein, generative image model architecture choices have been shown to impact the performance of these systems in representing different races (*243*). It is challenging for researchers to analyse developers' methodology because the full development details of proprietary and even of open-source general-purpose AI models are rarely documented and disclosed (*244*).

Meanwhile, 'ecosystem audits' help to assess human-AI interactions. An increasing number of laboratory studies investigate how human users interact with general-purpose AI systems. These often take the form of controlled studies (*245*) where participants interact with models and the impact of modelling and user interaction settings on the decisions and behaviour of participants is measured directly. Such studies have revealed the tendency of users to trust certain presentations of model outputs over others (*246, 247*). However, since participants are recruited, it can be hard to design the study and recruit a wide enough range of participants to meaningfully reflect the full range of user impacts in practice. Several researchers have begun conducting natural and controlled experiments on the impact of AI use in real deployment settings. For instance, there have been studies on the impact of AI risk assessments on judges' bail decisions in Kentucky, USA (*248*), the use of automated hiring tools on hiring manager discretion (*249*), and the use of generative AI on middle manager performance (*250*). Qualitative interviews of stakeholders have proven effective at illustrating more systematic impacts such as some of the social implications of AI implementation (*251*), which can endure after the removal of the AI system (*252*).

Analysing AI systems in the real world, post-deployment, allows researchers to study them as components of larger societal systems. Post-market surveillance is quite common in other industries with audit ecosystems (*253*). Users often discover capabilities and failure modes that developers do not, and monitoring the real-world usage of a system can further scientific understanding. For example, jailbreaks against modern large language chat models were first studied following findings from ordinary users (*254*). The study of deepfakes in the real world has also helped to shape scientific research on studying and mitigating harms (*255, 256*).

## 3.3 Model transparency, explanations, and interpretations

In contrast to studying general-purpose AI model outputs, another common approach to assessing models is to study the inner mechanism through which models produce the outputs. This can help researchers contextualise assessments of model performance and deepens their understanding of model functionality. Studying *how* general-purpose AI models and systems operate internally is a popular topic of research with thousands of academic papers produced. Areas of research that aim to enhance transparency include documentation, third-party access mechanisms, black box analysis, explaining model actions, and interpreting the inner workings of models.

**Documentation templates record decisions made and facilitate transparency at an operational level.** Currently, one of the most practical ways to increase transparency about a general-purpose AI model is through documenting and communicating the engineering decisions that define the model. Several documentation solutions have been proposed to communicate such decisions to a broader range of internal and external stakeholders. Some of these efforts, such as the development of Model Cards (*257*) have been successful. One recent study revealed "a widespread uptake of model cards within the AI community" (*258*). There are documentation templates available for communication on dataset practices (*259, 260, 261*), broader system features (*262, 263*), and broader procedural decision-making (*264*).

**Model explanation and interpretability techniques can improve researchers' understanding of how general-purpose AI systems operate internally.** There are several tools that allow for the external scrutiny of general-purpose AI systems, enabling access for external actors to query general-purpose AI systems directly, or otherwise gain visibility into model details (*213*). One prominent technical approach to this involves studying how a model's output can be explained as a consequence of a given input (*265, 266, 267, 268*). These explanations can play a unique role in supporting accountability, by helping to determine liability, in cases where humans may be wrongfully harmed or discriminated against by automated AI systems (*269, 270, 271*). Another approach used to study the computations in neural networks has involved interpreting the role of parameters (*272*), neurons (*273, 274, 275*), subnetworks (*276, 277*), or layer representations (*278, 279, 280, 281*) inside of AI systems. Interpretations of models have sometimes aided researchers in finding vulnerabilities. Examples have involved red-teaming (*207*), identifying internal representations of spurious features (*282*), brittle feature representations (*283, 284, 285, 286*), and limitations of factual recall in transformers (*287*).

**It is challenging to understand how general-purpose AI systems operate internally and to use this understanding effectively.** A persistent problem is that, in the absence of an objective standard to compare to, it is difficult to ensure that an interpretation of how a general-purpose AI system operates is correct. A number of 'interpretability illusions', in which an interpretability technique suggests a misleading interpretation for how a model works, have been documented (*288*, 289*). In addition, some research has also critically examined how algorithmic transparency tools can be maliciously used to construct false dichotomies, obscure, and mislead (*290*). To rigorously evaluate

interpretability techniques, the interpretations that they generate need to be demonstrably useful for some downstream task (*291, 292, 293, 294*, 295*), yet AI interpretability tools are not consistently competitive with simpler techniques for many tasks yet (*296*). In particular, different techniques to explain model actions often disagree with one another, and fail at sanity checks or downstream uses (*218, 297, 298, 299*). Nonetheless, interpretability has sometimes improved practical diagnostics and understanding, especially with recent progress in the field (*300*). Further progress may enable better applications. However, high-level interpretations of general-purpose AI systems cannot be used to make formal guarantees about their behaviour with current models and methods. The potential for current explainability and interpretability techniques to practically aid in rigorous model assessments is debated.

## 3.4 Challenges with studying general-purpose AI systems

It is extremely difficult to conduct thorough evaluations and generate strong assurances of general-purpose AI capabilities and risks. Modern general-purpose AI systems are the results of complex, decentralised projects involving data collection, training runs, system integration, and deployment applications. Meanwhile, there are a very large number of real-world uses for general-purpose AI systems. This complexity makes it difficult for any single actor to understand the entire process.

**The quality of evaluations depends on levels of access and transparency.** Different techniques for evaluating AI systems require different types of access. For example, evaluating a model's performance on test data typically only requires the ability to query the target model and analyse its outputs. This is typically referred to as 'black-box' access. The ability to query a black-box system is useful but many types of evaluation techniques depend on greater levels of access. Historically, AI researchers have benefitted from open-source methods, models, and data. Today, however, companies are increasingly keeping state-of-the-art general-purpose AI systems private (*244*). Lacking 'white-box' access (access to model parameters) makes it challenging for researchers to perform adversarial attacks, model interpretations, and fine-tuning (*300, 301*). Meanwhile, lacking 'outside-the-box' access to information about how a system was designed, including data, documentation, techniques, implementation details, and organisational details makes it difficult to perform evaluations of the development process (*226, 227, 257, 300, 302, 303*). Meanwhile, the third-party audit ecosystem is nascent but growing. Various AI audit tools, some of which are open-source, allow external users to query and access details of the model (*213*). Several studies have advocated for legal 'safe harbours' (*304*) or government-mediated access regimes (*253*) to enable independent red-teaming and audit efforts. Methods for structured access have been proposed that do not require making the code and weights public (*305*) but make it possible for independent researchers and auditors to perform their analysis with full access to the model in a secured environment designed to avoid leaks.

**Thoroughly assessing downstream societal impacts requires nuanced analysis, interdisciplinarity, and inclusion.** Although understanding the overall societal impact is the ultimate goal of many AI evaluations, many fall short of this goal. Firstly, there are always differences between the settings in which researchers study AI systems and the ever-changing real-world settings in which they will be deployed. Secondly, evaluating AI impacts on society is a complex socio-technical problem (*306, 307*). For example, while it is known that large language models exhibit significant cross-lingual differences in safety (*308, 309*), capabilities (*310, 311**), and tendencies (*312, 313**), it is challenging for researchers to thoroughly evaluate language models across languages. Furthermore, an over-reliance on simplified technical proxies when dealing with ethical concepts like 'fairness' and 'equity', can be misleading or exclude under-represented stakeholders (*314, 315*). Evaluations of the broader impacts of general-purpose AI are also highly multifaceted, requiring interdisciplinarity and representation of multiple stakeholders who may hold very different perspectives (*316, 317, 318**). Modelling the effects of AI deployment in society, in advance of deployment, is inherently complex and difficult to anchor in quantitative analysis. While there has been progress on societal impact

evaluation for general-purpose AI (*319, 320, 321, 322*, 323**), implementation remains challenging due to the need to balance the interests of multiple stakeholders (*324*) and resourcing challenges (*325*) that often make this work difficult to do in practice.

**Increasing participation and representation of perspectives in the AI development and evaluation process is an ongoing technical and institutional challenge.** Specifying relevant goals of evaluation is highly influenced by who is at the table and how the discussion is organised, meaning it is easy to miss or mis-define areas of concern. Broadening the range of those participating in the audit process also broadens the range of experiences incorporated into the process of discovering and characterising current or anticipated harms (*215*). Expanding participation has been a focus of the machine learning community in recent years, highlighting the need to incorporate and engage a broader range of perspectives and stakeholders into the process of AI model design, development, evaluation, and governance (*326, 327, 328*). Multiple strategies have been proposed from facilitating a broader notion of impacts during the implementation of impact assessments (*329*) to enable a more inclusive range of human feedback (*330*). However, seeking out and incorporating more voices is a nuanced endeavour, requiring sensitivity and respect for participating parties to minimise the potential for exploitation (*331*). Crucially, the challenge of increasing participation also involves the necessary negotiation of hard choices between incompatible values or priorities (*332, 333, 334*).

**While transparency is critical for assessing AI systems, it is difficult to achieve in practice.** Transparency efforts are not always meaningful interventions toward accountability (*290*). Bhatt et al. (*335*) surveyed dozens of corporate implementations of explainability techniques and found that many explainability efforts are meaningfully used during the model development process but rarely live up to the policy aspirations of providing end-user transparency or justification. Furthermore, recent surveys of proprietary, open-source explainability and interpretability tooling reveal that such tools are rarely adequately validated to support claims, and are vulnerable themselves to manipulation and robustness challenges (*213, 336, 337*). Similarly, the logistics of operationalising documentation practice in corporate settings can be fraught with internal politics (*338*).

# 4 Risks

The development and deployment of general-purpose AI gives rise to several risks, which are discussed in this section. This report distinguishes between 'risks' and 'cross-cutting risk *factors*'. For the purposes of this report 'risk' means the combination of the probability of an occurrence of harm and the severity of that harm (*339*). 'Cross-cutting risk factor' refers to a condition that contributes to not one but several risks.

## 4.1 Malicious use risks

As general-purpose AI covers a broad set of knowledge areas, it can be repurposed for malicious ends, potentially causing widespread harm. This section discusses some of the major risks of malicious use, but there are others and new risks may continue to emerge. While the risks discussed in this section range widely in terms of how well-evidenced they are, and in some cases, there is evidence suggesting that they may currently not be serious risks at all, we include them to provide a comprehensive overview of the malicious use risks associated with general-purpose AI systems.

### 4.1.1 Harm to individuals through fake content

#### KEY INFORMATION

- General-purpose AI systems can be used to increase the scale and sophistication of scams and fraud, for example through general-purpose AI-enhanced 'phishing' attacks.
- General-purpose AI can be used to generate fake compromising content featuring individuals without their consent, posing threats to individual privacy and reputation.

General-purpose AI can amplify the risk of frauds and scams, increasing both their volume and their sophistication. Their volume can be increased because general-purpose AI facilitates the generation of scam content at greater speeds and scale than previously possible. Their sophistication can be increased because general-purpose AI facilitates the creation of more convincing and personalised scam content at scale (*340, 341*). General-purpose AI language models can be used to design and deploy 'phishing' attacks in which attackers deceive people into sharing passwords or other sensitive information (*342*). This can include spear-phishing, a type of phishing campaign that is personalised to the target, and business email compromise, a type of cybercrime where the malicious user tries to trick someone into sending money or sharing confidential information. Research has found that between January to February 2023, there was a 135% increase in 'novel social engineering attacks' in a sample of email accounts (*343**), which is thought to correspond to the widespread adoption of ChatGPT.

General-purpose AI language models have the potential to dramatically scale the reach of fraudsters by automating conversations with potential victims to find promising targets (*340*). General-purpose AI could also contribute to targeted identity theft, and generating fake identities which could be used for illegal purposes. For example, a few years ago, research had discussed potential risks from AI 'voice cloning' – where an AI system analyses the tone, pitch, and other characteristics of a human voice to create synthetic copies – and how such technology might be used by fraudsters to pretend to be their victim's friends or a trusted authority (*344*). General-purpose AI systems can also help criminals to evade security software by correcting language errors and improving the fluency of messages that might otherwise be caught by spam filters.

Recent data indicates that AI-enabled fraud, particularly using deepfakes, is growing globally (*345, 346*). Detecting the use of general-purpose AI systems to generate content used to commit fraud can be difficult, and institutions may be reluctant to disclose the challenges they are facing with AI-powered fraud (*347*).

General-purpose AI-generated fake content can also be employed to harm individuals by featuring them in that content against their consent, thereby violating their right to privacy and damaging their reputation or dignity. This can happen in the form of fake content featuring any compromising or reputationally damaging activity, but has received particular attention in cases of deepfake pornography, where general-purpose AI is used to create pornographic audiovisual content of individuals without their consent. This may include the creation of child sexual abuse material (CSAM) and other forms of intimate images used to abuse, for example, former domestic partners or for blackmail (*348, 349, 350*).

### 4.1.2 Disinformation and manipulation of public opinion

#### KEY INFORMATION

- General-purpose AI makes it possible to generate and disseminate disinformation at an unprecedented scale and with a high degree of sophistication, which could have serious implications for political processes. However, it is debated how impactful political disinformation campaigns generally are.
- It can be difficult to detect disinformation generated by general-purpose AI because the outputs are increasingly realistic. Technical countermeasures, like watermarking content, are useful but can usually be circumvented by moderately sophisticated actors.

AI, particularly general-purpose AI, can be maliciously used for disinformation (*351*), which for the purpose of this report refers to false information that was generated or spread with the deliberate intent to mislead or deceive. General-purpose AI-generated text can be indistinguishable from genuine human-generated material (*352, 353*), and may already be disseminated at scale on social media (*354*). In addition, general-purpose AI systems can be used to not only generate text but also fully synthetic or misleadingly altered images, audio, and video content (*355*). Humans often find such content indistinguishable from genuine examples, and generating such content is both relatively simple and extremely cheap (*356*). An example of this is images of human faces that were altered, or completely generated, using general-purpose AI or narrower types of AI systems (*357*). Such 'Deepfake' images and videos are thought to have been deployed in several national elections over recent months to defame political opponents, with potentially significant impact, but there is currently not much scientific evidence about the impact of such campaigns.

**General-purpose AI tools might be used to persuade and manipulate people, which could have serious implications for political processes.** General-purpose AI systems can be used to generate highly persuasive content at scale. This could, for example, be used in a commercial setting for advertising, or during an election campaign to influence public opinion (*358*). Recent work has measured the persuasiveness of general-purpose AI-generated political messages and found that it can somewhat sway the opinion of the people reading these messages (*359, 360*). In addition, general-purpose AI can be used to tailor persuasive content to specific individuals or demographics (known as 'microtargeting'), for example, by using information scraped from social media. However, there is currently only limited evidence that micro-targeted messages are more persuasive than generic general-purpose AI-produced content (*361, 362*), which is in line with emerging scepticism about the effectiveness of microtargeting in general (*363*).

An underexplored frontier is the use of conversational general-purpose AI systems to persuade users over multiple turns of dialogue. In one study, human-human or human-general-purpose AI pairs debated over a cycle of opinion-and-rebuttal, and general-purpose AI systems were found to be as persuasive as humans (*364*). Another study in which general-purpose AI systems attempted to dissuade humans of their belief in conspiracy theories over three conversational turns found reductions in reported belief in those theories of 15-20% that endured for up to two months (*365*). These results raise the possibility that in conversational settings, general-purpose AI systems could be used to very powerful persuasive effect. As general-purpose AI systems grow in capability, it might become easier to maliciously use them for deceptive or manipulative means, possibly even with higher effectiveness than skilled humans, to encourage users to take actions that are against their own best interests (*366, 367**). In doing so, they may utilise new manipulation tactics against which humans are not prepared because our defences against manipulation have been developed through the influencing attempts of other humans (*368*).

**The overall impact of disinformation campaigns in general as well as the impact of widespread dissemination of general-purpose AI-generated media are still not well understood.** Despite indications of potentially serious risks to public discourse, and the integrity of the information ecosystem posed by general-purpose AI, there are caveats. First, there is a lack of evidence about the effectiveness of large-scale disinformation campaigns in general (whether using general-purpose AI or not). Second, some experts have argued that the main bottleneck for actors trying to have a large-scale impact with fake content is not generating that content, but distributing it at scale (*369*). Similarly, some research suggests that 'cheapfakes' (less sophisticated methods of manipulating audiovisual content that are not dependent on general-purpose AI use), might be as harmful as more sophisticated deepfakes (*370*). If true, this would support the hypothesis that the quality of fake content is currently less decisive for the success of a disinformation campaign than challenges around distributing that content to many users. Social media platforms like Meta or X employ various techniques such as human content moderation and labelling for reducing the reach of content likely to be disinformation, including general-purpose AI-generated disinformation. On the other hand, research has shown for years that social media algorithms often prioritise engagement and virality over the accuracy or authenticity of content, which could aid the rapid spread of AI-generated disinformation (*371, 372*).

In general, as general-purpose AI capabilities grow and are increasingly used for generating and spreading messages at scale, be they accurate, intentionally false, or unintentionally false, people might come to trust any information less, which could pose serious problems for public deliberation. Malicious actors could exploit such a generalised loss of trust by denying the truth of real, unfavourable evidence, claiming it is AI-generated, a phenomenon coined as the 'liars' dividend' (*373*).

**Potential measures to identify general-purpose AI-generated content, such as watermarking, can be helpful but are easy to circumvent for moderately sophisticated actors.** Researchers have employed various methods attempting to identify potential AI authorship (*374, 375*). *Content analysis* techniques explore statistical properties of text, such as unusual character frequencies or inconsistent sentence length distributions, which may deviate from patterns typically observed in human writing (376, 377, 378). Linguistic analysis techniques examine stylistic elements like sentiment or named entity recognition to uncover inconsistencies or unnatural language patterns indicative of AI generation (379, 380). Readability scores can also be used to examine where general-purpose AI-generated text might score unusually high or low on metrics like the Flesch Reading Ease Score compared to human-written content (381). AI researchers have also proposed other disinformation detection approaches, such as watermarking, in which an invisible signature identifies digital content as generated or altered by AI (382). However, current techniques are relatively easy to circumvent for moderately skilled actors (374), and perfect defences against the removal of watermarks are probably impossible (383). Still, these safeguards may deter relatively unsophisticated threat actors (384). 5. Technical approaches to mitigate risks provides a more in-depth discussion of watermarking techniques.

## 4.1.3 Cyber offence

### KEY INFORMATION

- General-purpose AI systems could uplift the cyber expertise of individuals, making it easier for malicious users to conduct effective cyber-attacks, as well as providing a tool that can be used in cyber defence. General-purpose AI systems can be used to automate and scale some types of cyber operations, such as social engineering attacks.
- There is no substantial evidence yet suggesting that general-purpose AI can automate sophisticated cybersecurity tasks which could tip the balance between cyber attackers and defenders in favour of the attackers.

General-purpose AI systems can exacerbate existing cybersecurity risks in several ways. Firstly, they may lower the barrier to entry of more sophisticated cyber attacks, so the number of people capable of such attacks might increase. Secondly, general-purpose AI systems could be used to scale offensive cyber operations, through increasing levels of automation and efficiency.

Moreover, general-purpose AI can inadvertently leave systems vulnerable to traditional cybersecurity attacks. For example, general-purpose AI systems are used pervasively as coding assistants and can inadvertently introduce software vulnerabilities (*385, 386*, 387*).

**General-purpose AI systems reduce the cost, technical know-how, and expertise needed to conduct cyber-attacks.** Offensive cyber operations include designing and spreading malicious software as well as discovering and exploiting vulnerabilities in critical systems. They can lead to significant security breaches, for example in critical national infrastructure (CNI), and pose a threat to public safety and security. Given the labour-intensive nature of these operations, advanced general-purpose AI that automates certain aspects of the process, reducing the number of experts needed and lowering the required level of expertise, could be useful for attackers.

So far, current general-purpose AI systems have been shown to be capable of autonomously carrying out basic cybersecurity challenges and narrow cyber tasks such as hacking a highly insecure website (*388, 389*, 390, 391*). However, despite consistent research efforts, existing models appear to not be able to carry out multi-step cybersecurity tasks that require longer horizon planning (*367*, 391, 392*).

Given that LLMs can already process and manipulate some cybersecurity concepts (*393*), planning could unlock more sophisticated cyber capabilities such as independently navigating complex digital environments, identifying and exploiting vulnerabilities at scale, and executing long-horizon strategies, without the need for direct human guidance.

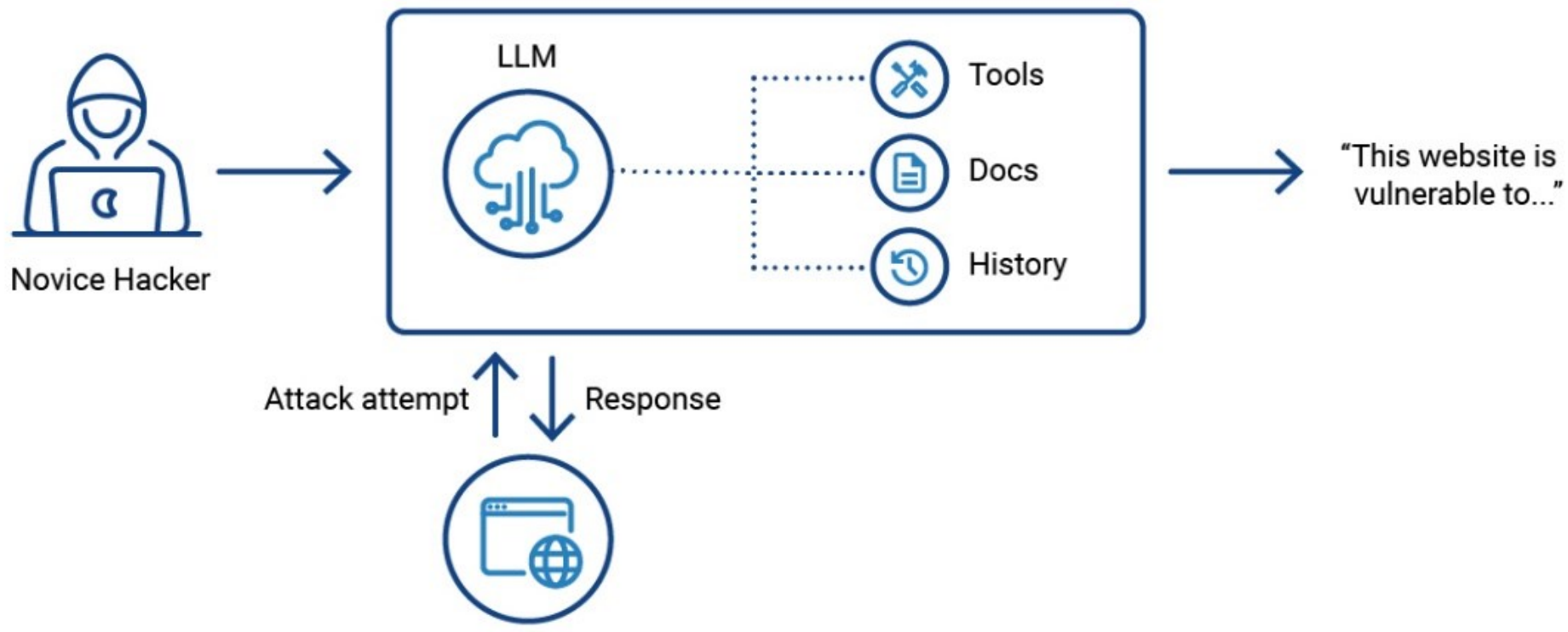

*Figure 6. An AI-enabled pipeline for automating simple attacks against websites (from (391)). A novice hacker supplies the general-purpose AI model with security information sources and textbooks, and prompts it to interact with their target website. Through trial and error, the general-purpose AI model identifies an attack that works on this website and returns the details to the hacker.*

**General-purpose AI systems could improve defensive cyber capabilities if applied across relevant areas.** Despite improved offensive capabilities, adversaries will not inevitably succeed, as advancements in general-purpose AI systems will also enhance defensive capabilities (*394, 395*). For instance, general-purpose AI systems can significantly cut down the time humans spend identifying and fixing vulnerabilities (*396*, 397*). However, similar to their attacking capabilities, the cyber-defence benefits of existing models also have limitations (*367*, 398, 399, 400*, 401*). Given the labour and time required to implement security patches across the digital infrastructure, attackers still have the opportunity to achieve successful breaches in systems that have not yet been fixed.

As general-purpose AI systems improve and become more widely used, the dynamics between attackers and defenders can be influenced by organisational factors such as resource availability and expertise levels. Major cyber defence competitions, such as the (*402**), alongside high-quality datasets and benchmarks (*403*), can drive the development of more sophisticated and resilient cybersecurity measures.

## 4.1.4 Dual use science risks

### KEY INFORMATION:

- General-purpose AI systems could accelerate advances in a range of scientific endeavours, from training new scientists to enabling faster research workflows. While these capabilities could have numerous beneficial applications, some experts have expressed concern that they could be used for malicious purposes, especially if further capabilities are developed soon before appropriate countermeasures are put in place.
- General-purpose AI systems for biological uses do not present a clear current threat, and future threats are hard to assess and rule out. In the biology domain, current general-purpose AI systems demonstrate growing capabilities but the limited studies available do not provide clear evidence that current systems can 'uplift' malicious actors to obtain biological pathogens better than using the internet. There is insufficient publicly available research to assess if near-term advances will provide this uplift, for example through trouble-shooting hands-on laboratory work.
- Due to insufficient scientific work, this interim report does not assess the risks of malicious use leading to chemical, radiological, and nuclear risks.

There are two avenues by which general-purpose AI systems could, speculatively, facilitate malicious use in the life sciences: firstly by providing increased access to information and expertise relevant to malicious use, and secondly by increasing the ceiling of capabilities, which may enable the development of more harmful versions of existing threats or, eventually, lead to novel threats (*404, 405*).

### Current capabilities

**Increased access to information –** Advanced general-purpose AI models can provide scientific knowledge, step-by-step experimental protocols, and guidance for troubleshooting experiments, all of which could potentially be exploited for malicious purposes (*405, 406, 407, 408*). However, information relevant to biological threat creation is already broadly accessible, given its 'dual-use' nature (*409, 410*). The capabilities of current general-purpose AI systems to 'uplift' a malicious actor's ability to maliciously use biology, by increasing their ability to access information relative to existing resources such as the internet, is unclear. Few empirical studies have assessed if current general-purpose AI models provide an uplift and current evidence is mixed (*166*, 411*). Methodologies to quantify uplift are also nascent and face limitations.[4]

**Increased access to hands-on expertise –** Access to information alone is insufficient to enable biological malicious use. Malicious actors must also be able to successfully synthesise, weaponise, and deliver the biological agent. Historically, those processes have required specialised expertise and hands-on skills for practical laboratory tasks (*412, 413*). Experts theorise that bioengineering combined with general-purpose AI advances may have lowered these barriers (*414*). However, existing studies have not evaluated whether general-purpose AI systems uplift malicious actors with practical laboratory research tasks. Existing advanced general-purpose AI models have some ability to design and troubleshoot laboratory experiments (*407*). These capabilities can be enhanced by equipping LLMs with the ability to access tools such as web search and specialised computational tools (*415*). When connected to laboratory robots or cloud labs, LLMs have been experimentally used to directly instruct these platforms to carry out experiments (*189*). However further empirical work is required to determine whether these existing general-purpose AI systems' capabilities 'uplift' actors with practical laboratory research tasks, relative to the internet's capabilities.

**Increasing the ceiling of capabilities –** While this report focuses on general-purpose AI systems (as defined in 1. Introduction), dual-use science risks are also impacted by the existence of narrow AI tools and the ability for general-purpose AI systems to interact with those narrow AI tools. Narrow AI biological tools can already redesign existing proteins to enhance existing functionality (*416*) and confer new biological functions (*417*), as well as generate novel proteins (*418, 419*). Analogous design capabilities have been demonstrated in other scientific domains, including chemistry (*420*). The capabilities of narrow tools themselves have dual-use implications: for instance, their use to predict viral mutations likely to demonstrate immune evasion (*421*), or generate plausible novel toxic molecules (*422*). Narrow AI tools are also often widely available, which makes implementing meaningful safeguards challenging (*43, 419, 423**). In addition, general-purpose AI systems are able to direct laboratory robots using language instructions, and make use of specialised chemical computational tools (*189, 415, 424*).

## Projected future capabilities

The malicious use potential of future general-purpose AI systems could be influenced by a range of projected advances. These include: advances in model capabilities, the integration of general-purpose AI systems with narrow AI tools, and the integration of general-purpose AI with automated laboratory equipment. Though there is some evidence of these projected advances being realised, it remains highly uncertain whether these capabilities will uplift users' abilities relative to existing resources.

**Advances in general-purpose AI model capabilities –** Future general-purpose AI systems will, plausibly, feature even greater domain-specific knowledge, reasoning abilities, and the capacity to formulate complex plans. Experts disagree on how much general-purpose AI systems could enable troubleshooting practical laboratory tasks, compared to using internet search. There is agreement that more effective, real-time troubleshooting for practical laboratory tasks would be enabled by 'multimodal' general-purpose AI systems that can incorporate information from images and videos (*405, 407, 425*). However, this possibility has not been tested yet. Additionally, some researchers argue that general-purpose AI systems trained on biological data may eventually outperform narrower tools designed for specific biological applications (*426*), but general-purpose AI systems are yet to demonstrate such capabilities, and the future performance and scalability of general-purpose AI systems in the biological domain remain unclear.

---

[4] For instance, there is uncertainty about the use of statistical significance as an effective measure of measuring model risk, and it is challenging to operationalise the threshold at which increased information access would actually be dangerous.

**Integration with narrow tools –** Though general-purpose AI systems can make use of narrow AI tools, this type of integration has so far been limited (*189, 415, 427**). It is uncertain how increasing the integration of general-purpose AI systems with narrow AI tools will affect the risk of malicious use. Today, narrow AI tools require specialist expertise to use effectively for scientific research (*428*). Though the ability to direct these tools using natural language will make specialised tasks more accessible, leveraging the outputs will likely still require technical expertise until substantial advances are made with general-purpose AI. Existing tools also have limited outputs and require extensive laboratory validation, and the extent to which these limitations will be overcome is unclear.

**Autonomous science capabilities –** While advanced general-purpose AI systems are already enabling some autonomous science capabilities, it is not clear what the near-term implications are for potential biological malicious use. While aspects of chemistry workflows such as chemical synthesis are already automatable (*189, 415*), experts are uncertain if these advances will transfer well to biology workflows, due to challenges automating work involving living systems (*407*). Furthermore, the high costs of automated laboratories are likely to make large-scale automation largely inaccessible to all but the most advanced malicious actors, although commercial cloud labs could partially offset this.

**In summary, the degree to which current state-of-the-art general-purpose AI systems enhance the capabilities of malicious actors to use the life sciences over existing resources, such as the internet, remains unclear.** Though some empirical work has assessed this uplift with respect to information access and biological threats (*166*, 429*), additional studies evaluating a broader range of tasks and scientific domains are needed to provide greater insight into this question. It has not yet been investigated whether general-purpose AI systems could also strengthen defences against dual-use science hazards.

## 4.2 Risks from malfunctions

### 4.2.1 Risks from product functionality issues

#### KEY INFORMATION

- Product functionality issues occur when there is confusion or misinformation about what a general-purpose AI model or system is capable of. This can lead to unrealistic expectations and overreliance on general-purpose AI systems, potentially causing harm if a system fails to deliver on expected capabilities.
- These functionality misconceptions may arise from technical difficulties in assessing an AI model's true capabilities on its own, or predicting its performance when part of a larger system. Misleading claims in advertising and communications can also contribute to these misconceptions.

Risks may arise where general-purpose AI models and systems fail to comply with general tenets of product safety and product functionality. As with many products, risks from general-purpose AI - based products occur because of misunderstandings of functionality and inadequate guidance for appropriate and safe use. In that respect, general-purpose AI-based products may be no different (*430*).

Product functionality issues, and the risks they pose may be clustered by potential failure modes (see Table 1).

*Impossible tasks* arise from instances of an attempt to accomplish goals with a general-purpose AI system that goes beyond the general-purpose AI system's capability. It can be hard to say definitively what constitutes an impossible task in a modern setting. Historically, large language models have not been able to consider events or developments that occurred after the end of their training. However, enabling AI products to retrieve information from databases has improved their ability to consider what happened after their training - although models still perform worse on tests that require novel information (*431*). Another potentially impossible task may be tasks requiring data that is inherently inaccessible -- such as information that does not exist in the format of computable media, or data not available for training due to legal or security reasons.

Impossible tasks pose risks because often, salient types of failure-- including many of the engineering failures, post-deployment failures and communication failures (see Table 1) -- might be the by-product of mismeasurements, misapprehensions or miscommunication around what a model can do, and the misinformed deployments that result.

For instance, the GPT-4 model achieved results of “passing a simulated bar exam with a score around the top 10% of test takers” and being in the 88th percentile of LSAT test takers (*2**). Confidence in this result even led some lawyers to adopt the technology for their professional use (*432*). Under different circumstances, such as changes to test-taking settings or when comparing to first-time bar examinees who passed the exam, the model achieved substantially lower percentile results (*433*). Those who were attempting to make use of the model in actual legal practice encountered these inadequacies, facing severe professional consequences for the errors produced by these models (i.e. inaccurate legal citations, inappropriate format and phrasing, etc.) (*434*). Similar misapprehensions regarding model performance are thought to apply in the medical context (*435*), where real world use and re-evaluations reveal complexity to the claims of these models containing reliable clinical knowledge (*436*) or passing medical tests such as the MCAT (*2**) or USMLE (*437*). More generally, some deployed large language models struggle under some linguistic circumstances: They might, for instance, have trouble navigating negations and consequently fail to distinguish between advising for and against a course of action – though some research suggests these issues are addressed by general capability gains (*438, 439*).

Some shortcomings are only revealed after deployment. Although many thorough evaluations have examined large language model use for code generation (*440**), including in relevant real-world tasks (*441*), instances of real-world deployment of large language models for coding suggest that the use of these models could lead to the potential introduction of critical overlooked bugs (*442*), as well as confusing or misleading edits (*443*) that could be especially impactful when guiding engineering programmers, particularly in applications that automate parts of the workflow (*444*).

| **Failure Modes** | **Categories** |
|---|---|
| Impossible Tasks | Conceptually Possible<br>Practically Impossible |
| Engineering Failures | Design Failures<br>Implementation Failures<br>Missing Safety Features |
| Post-Deployment Failures | Robustness Issues<br>Failure under Adversarial Attacks<br>Unanticipated Interactions |
| Communication Failures | Falsified or Overstated Capabilities<br>Misrepresented Capabilities |

*Table 1: Taxonomy of AI functionality issues, reproduced with permission (430).*

Functionality misapprehensions arise from different underlying issues. Firstly, as noted in 3. Methodology to assess and understand general-purpose AI systems, there are technical difficulties in designing and representative evaluations of performance for general-purpose AI systems, making definite statements about functionality difficult. Secondly, functionality issues might only manifest or manifest differently in a real-world realistic context, rendering even informative model evaluation insufficient for robust statements about general-purpose AI system and product functionality. Thirdly, failures can result not just from inadequate evaluation, but a lack of appropriate communication to product users around the product's limitations and the potential consequences. Misleading advertising, as it occurs in many markets, could become a substantial source of risks from functionality in general-purpose AI (*445*).

In general, for many machine learning based products, it can be unclear exactly which context of deployment is well represented in the data and suitable for the model. However, more general purpose AI tools specifically are more difficult to vet for deployment readiness than lower-capability or narrower AI systems: With General Purpose AI, it can be difficult to clearly define and restrict potential use cases that may not be suitable or may be premature, although substantial progress on restricting use cases is feasible.

### 4.2.2 Risks from bias and underrepresentation

#### KEY INFORMATION

- The outputs and impacts of general-purpose AI systems can be biased with respect to various aspects of human identity, including race, gender, culture, age, and disability. This creates risks in high-stakes domains such as healthcare, job recruitment, and financial lending.
- General-purpose AI systems are primarily trained on language and image datasets that disproportionately represent English-speaking and Western cultures, increasing the potential for harm to individuals not represented well by this data.

Harmful bias and underrepresentation in AI systems have been challenges since well before the increased attention to general-purpose AI. They remain an issue with general-purpose AI, and will likely be a major challenge with general-purpose AI systems for the foreseeable future. Decisions by an AI might be biased if their decision-making is skewed based on protected characteristics, such as gender, race, etc. They might hence be discriminatory when this bias informs decisions to the disadvantage of members of these protected groups; thereby creating harm to fairness. This section discusses present and future risks resulting from bias and underrepresentation risks in AI. Because of the rich history of research in this space, this section explores research both on narrow AI and general-purpose AI.

AI systems can demonstrate bias as a result of skewed training data, choices made during model development, or the premature deployment of flawed systems. Despite extensive research, reliable methods to fully mitigate any discrimination remain elusive. There are particular concerns over the tendency of advanced general-purpose AI systems to replicate and amplify bias present within their training data (*446*). This poses a significant risk of discrimination in high-impact applications such as job recruitment, financial lending, and healthcare (*447*). In these areas biased decisions resulting from general-purpose AI systems outputs can have profoundly negative consequences for individuals, potentially limiting employment prospects (*448, 449*), hindering upward financial mobility, and restricting access to essential healthcare services (*450, 451*).

**There are several well-documented cases of AI systems displaying discriminatory behaviour based on race, gender, age, and disability status, causing substantial harm.** Given increasingly widespread adoption of AI systems across various sectors, such behaviour can perpetuate various types of bias, including race, gender, age, and disability. This can cause serious harm if these systems are entrusted with increasingly high-stakes decisions which can have severe consequences for individuals. Racial bias in AI systems has been shown to be present in commercially available facial recognition algorithms (*452*) and has caused ineffectiveness in predicting recidivism outcomes for

defendants of colour, underestimation of the needs of patients from marginalised racial and ethnic backgrounds (*453, 454*), and the perpetuation of inappropriate race-based medicine in responses from text-generation models (*435, 450*).

Gender biases in the outputs of AI systems are another key concern. Research has uncovered sexist, misogynistic, and gender-stereotyping content being produced from general-purpose AI (*455, 456*) and male-dominated results from gender-neutral internet searches using narrow AI algorithms (*457*). Age bias is also a key issue: some AI systems have exhibited bias against older job seekers (*458*), and age bias appears in some outputs from sentiment analysis models (*459*). One reason for this could be biases in training data. For example, LLM-powered HR screening tools may be trained on resumes skewed towards younger workers that may inadvertently discount the experiences and skill sets of older applicants. Similarly, healthcare allocation algorithms developed by health insurance companies may disadvantage older individuals based on age-related health risks, even if these individuals are healthy. Lending algorithms may not appropriately handle older adults' financial circumstances, particularly regarding social security income, which could impact approval outcomes (*460*).

Research has also shown that AI systems and tools can discriminate against users with disabilities, for example by disproportionately denying coverage claims from disabled individuals with complex medical needs (*461*), reproducing societal stereotypes about disabilities (*462*), and inaccurately classifying sentiments about people with disabilities (*463*). Despite growing research on sign language recognition (*464*), AI systems have limited automated transcription abilities for sign language speakers (*143*), and limited diversity in sign language datasets may also exacerbate disability bias from advanced general-purpose AI systems, as the majority of sign language datasets represent American Sign Language. Recent work that, for instance, has developed datasets for six African sign languages (*465*) is a step, albeit a modest one, toward achieving more equitable inclusion of sign language dialects.

AI systems exhibit a tendency towards intersectional biases. Intersectionality describes how individuals who share more than one marginalised characteristic may experience compounded bias or discrimination (for example, a low-income woman of colour). Intersectional bias exacerbates existing social inequalities and can limit individuals' access to vital resources and opportunities. While there is an emerging area of research focused on developing methods for detecting intersectional bias within AI models (*466, 467, 468*), there has been much less progress on mitigating potential impacts (*469*).

**Bias in general-purpose AI system outputs may result from a lack of representation in training datasets, leading to biased outputs.** Different groups and different cultures are unequally represented at various stages of the AI lifecycle – from the input data to the outputs of language and image generation models. Many of the issues associated with bias in AI demonstrate the harms of limited representation in training datasets, which are overwhelmingly likely to be in English (*236*). This has led to major disparities in the safety and reliability of modern general-purpose AI systems in different societies (*309, 310, 313**). Datasets used to train AI models have also been shown to under-represent various demographic markers such as age, race, gender, and disability status (*470, 471*). AI language models predominantly rely on digitised books and online text in their training, which fails to reflect oral traditions and non-digitised cultures. The inherent historical biases embedded in these sources, coupled with potential inequalities in the data collection process, can perpetuate systemic injustices (*472**) and lead AI systems to reflect dominant cultures, languages, and worldviews, at the detriment of marginalised groups such as indigenous communities (*196, 230, 239, 473, 474*).

**Issues of bias and representation remain an unsolved problem.** Despite considerable attention in the literature and economic incentives for companies to avoid reputational damage from biased AI products, mitigating bias remains an unsolved challenge. While developers may attempt to explicitly address bias during model fine-tuning, AI models can still pick up on implicit associations between features (*475*) or perpetuate significant biases and stereotypes even when prompts do not contain explicit demographic identifiers (*476*). While methods such as Reinforcement Learning from Human Feedback (RLHF) aim to align model decision-making with human preferences, these methods could inadvertently introduce biases based on the diversity and representativeness of the humans providing feedback (*303*). RLHF has been shown to lead to more politically biased models (*239, 477*) and incorporate user beliefs over factual information (*238*). In addition, rater feedback is often inconsistent (*478, 479*). Understanding issues of representation within datasets, models, and evaluation methods such as RLHF is crucial, as skewed representation can lead to biased outputs in AI models. More research is needed to address these issues.

## 4.2.3 Loss of control

### KEY INFORMATION

- Ongoing AI research is seeking to develop more capable 'general-purpose AI agents', that is, general-purpose AI systems that can autonomously interact with the world, plan ahead, and pursue goals.
- 'Loss of control' scenarios are potential future scenarios in which society can no longer meaningfully constrain some advanced general-purpose AI agents, even if it becomes clear they are causing harm. These scenarios are hypothesised to arise through a combination of social and technical factors, such as pressures to delegate decisions to general-purpose AI systems, and limitations of existing techniques used to influence the behaviours of general-purpose AI systems.
- There is broad agreement among AI experts that currently known general-purpose AI systems pose no significant loss of control risk, due to their limited capabilities.
- Some experts believe that loss of control scenarios are implausible, while others believe they are likely, and some consider them as low-likelihood risks that deserve consideration due to their high severity.
- This expert disagreement is difficult to resolve, since there is not yet an agreed-upon methodology for assessing the likelihood of loss of control, or when the relevant AI capabilities might be developed.
- If the risk of loss of control will in fact be large, then resolving this risk could require making fundamental progress on certain technical problems in AI safety. It is unclear if this progress would require many years of preparatory work.

AI companies and researchers are increasingly interested in developing general-purpose AI 'agents' (sometimes also referred to as 'autonomous general-purpose AI systems'). General-purpose AI agents are systems that can autonomously interact with the world, plan ahead, and pursue goals. Although general-purpose AI agents are beginning to be developed, they still demonstrate only very limited capabilities (*26, 178, 480**). Various researchers and AI labs ultimately hope to create general-purpose AI agents that can operate and accomplish long-term tasks with little or no human oversight or intervention.

Autonomous general-purpose AI systems, if fully realised, could be useful in many sectors. However, some researchers worry about risks from their malicious use, or from accidents and unintended consequences of their deployment (*481*, 482*).

Some researchers have also expressed concern about society's ability to exercise reliable oversight and control over autonomous general-purpose AI systems. For decades, concerns about a potential loss of control have been raised by computer scientists looking ahead toward these kinds of AI systems, including AI pioneers such as Alan Turing (*483*), I. J. Good (*484*), and Norbert Wiener (*485*). These concerns have gained more prominence recently (*486*), partly because a subset of researchers now believe that sufficiently advanced general-purpose AI agents could be developed sooner than previously thought (*127, 487, 488*).

The plausibility of such risks remains highly contentious. This section seeks to clarify the nature of the proposed risk, outline the main arguments and evidence that currently inform researchers' views about its likelihood, and summarise current expert opinion.

## Risk factors

An AI system is considered 'controllable' when its behaviours can be meaningfully determined or constrained by humans. While a lack of control is not intrinsically harmful, it significantly increases the risks of various harms. Current general-purpose AI systems are generally considered to be controllable, but, if autonomous general-purpose AI systems are fully developed, then risk of the loss of control may grow considerably.

In 4.4. Cross-cutting risk factors, this report discusses technical and societal risk factors for dangerous AI systems. In this section, we expand on how these factors may affect loss of control risk. Overall, the risk of losing control over currently known general-purpose AI systems appears negligible. The degree to which future, potentially much more capable, autonomous general-purpose AI agents can be controlled remains unclear.

**It is not yet known whether it will be easier or harder in future to ensure that highly capable AI systems pursue the objectives their developers intend.** One reason for this is that AI systems can 'game' their objectives by achieving them in unintended and potentially harmful ways. For example, widely deployed general-purpose AI language models adjust their stated view to better match their user's view, regardless of truth, and consequently gain more positive feedback (*237, 238*). There are observations that objective-gaming can become more prevalent as system capability increases, because more capable systems can find more ways to achieve a given objective (*489, 490*). Furthermore, more capable systems have more ways to coherently 'generalise' to behave differently in situations that differ from their training setting including potentially harmful ways – although this is only a hypothesised concern to date (*489, 490*). However, some recent more capable general-purpose AI systems have been more controllable due to better tools for training and oversight (*19, 20, 491*), and are less likely to generalise *incoherently* beyond their training data (*492*). It is therefore unclear if general-purpose AI systems will become more or less controllable as more advanced systems are created. For a discussion of how controllable present general-purpose AI systems are, see 5.2 Training more trustworthy models and 4.4.1 Cross-cutting technical risk factors.

**Some mathematical findings suggest that future general-purpose AI agents may use strategies that hinder human control, but as yet it is unclear how well these findings will apply to real-world general-purpose AI systems.** Some mathematical models of idealised goal-directed AI agents have found that, with sufficiently advanced planning capabilities, many such AI agents would hinder human attempts to interfere with their goal pursuit (*493, 494, 495*, 496*). Similar mathematical findings suggest that many such AI agents could have a tendency to 'seek power' by accumulating resources, interfering with oversight processes, and avoiding being deactivated, because these actions help them achieve their given goals (*493, 494, 495*, 497*, 498, 499*).

However, drawing real-world implications from this mathematical research is not straightforward. For instance, most research assumes that an AI system is trained using a randomly chosen goal (*493, 494, 495*, 497*, 498, 499*), but in practice, AI developers can substantially influence which goals are potentially encoded in general-purpose AI models (see below and 5.2 Training more trustworthy models). Further, philosophical research suggests that not all of the above behaviours are implied by pursuing a random goal (*500*). Currently, one mathematical finding and early-stage empirical findings support, although weakly as of this writing, that such behaviours may be found under more realistic circumstances (*181, 237, 490, 497**). There have also been case studies where, without being prompted to do so, general-purpose AI systems and other AI systems learned to systematically induce false beliefs in others because this was useful for achieving a goal (*366, 501*). If observed more widely, such behaviour is also relevant to near-term applications of general-purpose AI chatbots and agents.

**If people entrust general-purpose AI systems with increasingly critical responsibilities, then this could increase the risk of loss of control.** A range of social and economic forces would influence the

interaction between human and autonomous agents in such scenarios. For example, economic pressures may favour general-purpose AI-enabled automation in the absence of intervention, despite potentially negative consequences (*502*), and human over-reliance on general-purpose AI agents would make it harder to exercise oversight (*481**). Using general-purpose AI agents to automate decision-making in government, military, or judicial applications might elevate concerns over AI's influence on important societal decisions (*503, 504, 505, 506*). As a more extreme case, some actors have stated an interest in purposefully developing uncontrolled AI agents (*507*).

**Certain specific capabilities could disproportionately increase the risk of loss of control.** These capabilities – which are currently limited – include identifying and exploiting software vulnerabilities, persuasion, automating AI research and development, and capabilities needed to autonomously replicate and adapt (*367*, 508*, 509*). The relevant sections in this report discuss how capable current general-purpose AI systems are in some of these areas (4.1.3 Cyber offence, 4.1.2 Disinformation and manipulation of public opinion, 4.1.4 Dual use science risks). Particularly relevant are agent capabilities, which increase the ability for general-purpose AI systems to operate autonomously, such as planning and using memory. These are discussed in 4.4.1 Cross-cutting technical risk factors.

## Consequences of loss of control

An irreversible loss of control of some general-purpose AI systems is not necessarily catastrophic. As an analogy, computer viruses have long been able to proliferate near-irreversibly and in large numbers (*510*) without causing the internet to collapse. Some researchers have explored hypothetical scenarios where highly advanced future general-purpose AI agents acting autonomously might cause catastrophic harm to humans, especially if humans present obstacles to achieving the agents' given goals (*127, 511*). The mechanisms of harm are often imagined to stem from capabilities such as the ones listed in the previous paragraph and in 4.1.4 Dual use science risks. However, these scenarios remain hypothetical as they are not exhibited by current general-purpose AI systems.

## AI researchers have differing views on loss of control risks

AI experts debate the likelihood of losing control over future general-purpose AI systems. Key questions include whether sufficiently capable general-purpose AI agents will be developed in the medium-term, and whether technical safety and governance solutions can be developed in time to keep them adequately controlled. As there is limited research that assesses the risk of loss of control, expert opinion can provide alternative guidance, but it cannot replace research.

**A subset of researchers suggest that risks of loss of control deserves consideration. However, the overall likelihood of extreme control failures remains highly contentious.** Some researchers have argued that there has been little progress in developing the types of AI systems that others fear could pose risks of loss of control (*129, 512*). Nevertheless, broad hypothetical scenarios for how society might lose control over AI systems have been proposed (*507, 511*) and these scenarios have been emphasised by some leading researchers (*127*). For example, several hundred AI researchers have signed a statement declaring that "Mitigating the risk of extinction from AI should be a global priority" (*486*), though without explicitly referring to loss of control. The actual risk remains highly contentious as there is only limited research assessing it, and the opinions of scientists cannot replace this research.

# 4.3 Systemic risks

## 4.3.1 Labour market risks

### KEY INFORMATION

- Unlike previous waves of automation, general-purpose AI has the potential to automate a very broad range of tasks, which could have a significant effect on the labour market.
- This could mean many people could lose their current jobs. However, many economists expect that potential job losses as a result of automation could be offset, partly or completely, by the creation of new jobs and by increased demand in non-automated sectors.
- Labour market frictions, such as the time needed for workers to learn new skills or relocate for new jobs, could cause unemployment in the short run even if overall labour demand remained unchanged.
- The expected impact of general-purpose AI on wages is ambiguous. It is likely to simultaneously increase wages in some sectors by augmenting productivity and creating new opportunities, and decrease wages in other sectors where automation reduces labour demand faster than new tasks are created.

### AI is likely to transform a range of jobs and could displace workers

Economists expect general-purpose AI to impact the workforce by automating tasks, augmenting worker productivity and earnings, changing the skills needed for various occupations, and displacing workers from certain occupations (*513, 514, 515*). Economists hold a wide range of views about the magnitude and timing of these effects, with some expecting widespread economic transformation in the next ten years, while others do not think a step-change in AI-related automation and productivity growth is imminent (*516*). The uncertainty about future general-purpose AI progress contributes to this uncertainty about the labour market effects of general-purpose AI.

Previous waves of automation by computing have primarily affected 'routine' tasks which could be easily codified and programmed into computers (*517*). General-purpose AI, in contrast, has the potential to perform a wide array of tasks that are typically carried out by humans, including complex problem-solving and decision-making. An extensive body of literature has studied the exposure of jobs to automation and AI generally, without a particular focus on general-purpose AI (*518*). In comparison to this research, the exploration of likely labour market impacts of general-purpose AI is at a very early stage. In advanced economies, it is estimated that, due to the prevalence of cognitive-task-oriented jobs, 60% of current jobs could be affected by the introduction of general-purpose AI systems such as today's LLMs (*519*). This means that general-purpose AI has the potential to either automate substantial portions of the work or to complement and significantly alter how the work is done. In emerging economies, the share of jobs thought to be potentially affected by general-purpose AI systems is lower, but still substantial, at 40% (*519*).

Recent empirical studies have begun to demonstrate the impact of current-generation general-purpose AI systems on various industries, notably in knowledge work:

- General-purpose AI systems have been shown to increase productivity, quality, and speed in strategy consulting (*520*), improve performance in customer support (*250*) and improve computer programming (*521**).
- Machine translation and language models have been shown to substitute for human workers in tasks like simple large-scale language translation (*522*) and some writing/coding-related occupations (*523, 524*), leading to reduced demand for their services.

- The introduction and mass adoption of ChatGPT has resulted in a decrease in employment and earnings for freelancers offering writing, editing, and proofreading services (*523*).

A key question is whether there will be significant job losses as general-purpose AI systems become more advanced and more widespread. Some economists consider it likely that job losses will be offset by increased demand for labour in existing occupations and the creation of new kinds of jobs (*516, 517, 525*). This would be in line with the impact of previous waves of automation: For instance, more than 60% of employment in 2018 was in jobs with titles that did not exist in 1940 (*526*). Other economists emphasise that the future effects of general-purpose AI systems on job markets is very difficult to predict (*519, 527*) general-purpose AI could lead to a substantial reduction of the value of human labour compared with capital (*528*).

Even if overall demand for labour in the economy is not reduced, the process of displacement can create unemployment if the labour market fails to match workers with new employment opportunities quickly enough. This can happen because of various labour market frictions (*529*), such as:

- When workers have to learn new skills to switch occupations, it takes time for them to complete the necessary training or education
- The imperfect mobility of labour, which limits workers' ability to relocate for new job opportunities
- Skill mismatches between the requirements of new jobs and the existing skills of displaced workers

These factors influence how quickly and whether workers displaced by automation can move to new roles. Some workers may therefore be temporarily unemployed even though there may be job vacancies in the economy as a whole. General-purpose AI could lead to concentrated job losses, while productivity gains are likely to be more spread out throughout the economy, potentially leading to a difficult transition for some workers unless support is made available.

Some economists consider it plausible or even likely that the rate of job displacement due to general-purpose AI systems enabling automation could outpace the creation of new job opportunities, especially if the development of general-purpose AI systems is focused on substituting human labour rather than augmenting it (*502, 530*). However, there are few precise quantitative predictions, and the existing research is consistent with a wide range of possible outcomes.

AI researchers disagree on the pace of future general-purpose AI advancements, but there is some support for the possibility of extremely rapid advancements, including the possibility of general-purpose AI systems matching or surpassing the abilities of human experts in almost any cognitive task. (see 2.4.3. Will algorithmic progress lead to rapid advancements?). This latter scenario has only been considered in little economic research. That research assumes that AI systems can perform virtually all knowledge tasks more cost effectively than humans and indicates that such a scenario could lead to a collapse in wages across many sectors, severe unemployment and a dramatic reduction in labour force participation (*531, 532*). However, even in such a scenario some demand for human labour may persist due to consumer preferences and ethical or control-related reasons (*531, 532*). Among economists, extreme scenarios involving a dramatic decline in labour demand due to mass general-purpose AI automation are currently considered relatively fringe.

It is very difficult to forecast how and when general-purpose AI systems might affect various labour markets. Firstly, there is considerable uncertainty about how quickly general-purpose AI technology will advance and what its future capabilities may be. Secondly, even for a given level of technological sophistication, the extent to which general-purpose AI systems enable automation and how this kind of automation could affect labour markets is unclear. One reason for this is that in many sectors there can be considerable lags in deploying and adopting general-purpose AI systems at scale. For instance, if employees lack the necessary skills to effectively employ general-purpose AI assistance, this can slow down adoption. Overall, many economists expect modest macroeconomic impacts of general-purpose AI systems over the next decade (*533*), while some economists expect substantial labour market and macroeconomic effects in the next five to ten years (*534*). These forecasts often indicate high uncertainty due to the unknown pace of future advancements in general-purpose AI capabilities (see 2.4 Capability progress in coming years).

**The impact on wages of general-purpose AI systems leading to automation is uncertain and the evidence is mixed.** General-purpose AI automation could increase wages in some sectors by:

- Directly augmenting human productivity with general-purpose AI tools (*502, 535*)
- Leading to an overall growth of the economy, which lead to more investment that enhances the productivity of workers in non-automated sectors (*536*)
- Increasing output and boosting the demand for labour in tasks that are not yet automated, or creating new tasks and occupations (*517, 537*)
- Improving technology broadly through accelerating R&D (*538*)

If general-purpose AI systems complement human labour, then as the capabilities of general-purpose AI systems advance, wage growth could accelerate alongside economic growth, potentially much faster than historically (*539*). Recent large-scale surveys of the use of current general-purpose AI systems support this view. For example, a recent survey of a total of 5,334 workers and 2,053 firms in the manufacturing and financial sectors across seven OECD countries found that around 80% of workers who use AI said that AI had improved their performance at work (*540*).

However, if in the future automation reduces labour demand faster than new jobs are created, then it is possible that wages and the share of income going to human workers may substantially decline (*541*). It is also possible that general-purpose AI systems might have different economic effects at different points in time: the use of general-purpose AI systems might initially boost wages, but as more and more tasks are automated, increasing competition for the remaining jobs could push wages down (*532*). There is currently no clear consensus among experts regarding the net effects of general-purpose AI automation on average wages: the effects are likely to differ over time and across occupations, depending on many factors including social acceptance of the technology and organisational decision-making, as well as government policies.

## General-purpose AI could increase income inequality

General-purpose AI could increase income inequality both within and between countries. Historically, the automation of routine jobs has likely increased wage inequality within countries by displacing workers from the types of jobs in which they held a comparative advantage (*517*). Similarly, general-purpose AI could systematically compete with some tasks that human knowledge workers currently hold a comparative advantage in, potentially depressing wages if they cannot easily find work elsewhere (*542, 543*). At the same time, general-purpose AI could improve the productivity of high-income occupations, and so high-wage earners could see disproportionately larger increases in labour income, thereby amplifying labour income inequality. One simulation suggests that widespread adoption of AI could increase wage inequality between high and low-income occupations by 10% within a decade of adoption in advanced economies (*519*).

Automation could also exacerbate inequality by reducing labour's share of income, which would boost the relative incomes of wealthier capital owners (*528*). This would be part of an ongoing trend: globally, the share of income from labour has fallen by roughly six percentage points between 1980 to 2022, with similar patterns seen in the United States, Asia and Europe (*544*). Further automation could result in a continuation of this trend. Typically, 10% of earners earn the majority of capital income (*545, 546*). Hence, the greater productivity of capital would therefore act as a boon for high-earners. This dynamic could be particularly pronounced if general-purpose AI aids the creation of 'superstar' firms with strong market power, as they would capture an outsized share of economic profits (*519*).

Finally, general-purpose AI technology could exacerbate global inequality if it is primarily adopted by advanced economies (also see 4.3.2. Global AI divide). These countries have a higher share of cognitive task-oriented jobs exposed to the potential impacts of general-purpose AI, stronger digital infrastructure, skilled workforces, and more developed innovation ecosystems. This positions them to capture AI productivity gains more rapidly than emerging markets and developing economies, potentially leading to divergent income growth trajectories and a widening gap between high- and low-income countries (*519, 547*).

### 4.3.2 Global AI divide

#### KEY INFORMATION

- General-purpose AI research and development is currently concentrated in a few Western countries and China. This 'AI Divide' is multicausal, but in part related to limited access to computing power in low-income countries.
- Access to large and expensive quantities of computing power has become a prerequisite for developing advanced general-purpose AI. This has led to a growing dominance of large technology companies in general-purpose AI development.
- The AI R&D divide often overlaps with existing global socioeconomic disparities, potentially exacerbating them.

There is a well-documented concentration of AI research and development, including research on potential societal impacts of AI, in Western countries and China (*316, 548, 549*). This global 'AI Divide' could become even larger for general-purpose AI specifically because of the high costs associated with general-purpose AI development. Some countries face substantial barriers to benefiting from general-purpose AI development and deployment, including lower digital skills literacy, limited access to computing resources, infrastructure challenges, and economic dependence on entities in higher-income countries (*519, 550*). Because general-purpose AI system development is so dominated by a few companies, particularly those based in the US, there are concerns that prominent general-purpose AI systems which are used worldwide primarily reflect the values, cultures and goals of large Western corporations. In addition, the recent trend towards aiming to develop ever-larger, more powerful general-purpose AI models could also exacerbate global supply chain inequalities (*551*), place demands on energy usage, and lead to harmful climate effects which also worsen global inequalities (*552, 553*). The global general-purpose AI divide could also be harmful if biased or inequitable general-purpose AI systems are deployed globally.

**Disparities in the concentration of skilled talent and the steep financial costs of developing and sustaining general-purpose AI systems could align the AI divide with existing global socioeconomic disparities.** The United States has the largest percentage of elite AI researchers, contains a majority of the institutions who conduct top-tier research, and is the top destination for AI talent globally (*554*). However, countries leading in AI development also experience issues with the distribution of skilled AI talent, which is rapidly shifting towards industry. For example, 70 percent of graduates of North American universities with AI PhDs end up getting a job in private industry compared with 21% of graduates two decades ago (*555*).

In April 2023, OpenAI's AI systems were reportedly estimated to incur $700k/day in inference costs (*77*), a cost that is widely inaccessible for the vast majority of academic institutions and companies and even more so for those based in the Global South (*556, 557*). Low-resource regions also experience challenges with access to data given the high costs of collection, labelling, and storage. The lower availability of skilled talent to leverage these datasets for model development purposes could further contribute to the AI divide. Infrastructure concerns are a major factor that prohibit equitable access to the resources needed to train and implement general-purpose AI due to issues such as inadequate access to broadband internet (*558, 559*), power blackouts and insufficient access to electricity (*560, 561*).

**Academic institutions in China and the United States lead in general-purpose AI research production, but industry is increasingly influential.** China currently publishes the most research on AI, as measured by the total volume of articles in journals, conferences, and online repositories. Geographically, the development of significant machine learning models is concentrated in nations such as the US, Canada, the UK, and China, with at least one coming from Africa. US industry currently

dominates the development of advanced general-purpose AI systems. American institutions produced the majority (54%) of large language and multimodal models in 2022 (*562*). Industry now surpasses academia in producing significant machine learning models (32 vs three in 2022) and industry co-authorship on papers presented at the top ten leading AI conferences rose from 22% in 2000 to 38% in 2020 (*563*).

**The rising 'compute divide' is contributing to disparities in the distribution of computing resources, and unequal participation in general-purpose AI development.** The term 'compute divide' describes the different extent to which large industrial AI labs and typical academic AI labs have access to computing resources (*556*). In recent years, this divide has widened (*556, 557*). Estimates show that US technology companies are the major buyers of NVIDIA H100 GPUs, one of the most powerful GPU chip types on the market explicitly designed for AI (*564**). Amazon, Meta, Google, and Microsoft have all recently announced custom AI chips to reduce their dependence on the AI chip supply chain, potentially paving the way for more widespread access to GPUs. However, the exceptionally high cost of GPUs ($15,000 for top-tier models such as the H100 at the time of writing) could hinder academic institutions and less wealthy countries from affording this level of AI infrastructure.

**The delegation of lower-level AI work to workers in low-income countries has led to a 'ghost work' industry.** From content moderation to proofreading to data labelling, a lot of human labour that the typical consumer is usually not aware of – sometimes referred to as 'ghost work' – is necessary for many products of large technology companies (*565*). The increasing demand for data to train general-purpose AI systems, including human feedback to aid in training, has further increased the reliance on ghost work including the creation of firms helping big technology companies to outsource various aspects of data production, including data collection, cleaning, and annotation. This trend has played a significant role in the development of notable machine learning benchmark datasets such as ImageNet (*566*). Data production, a crucial aspect of general-purpose AI advancement, often relies on workers in countries with lower average wages. These workers may face exposure to graphic content, erratic schedules, heavy workloads, and have limited social and economic mobility (*567, 568, 569, 570*). This can lead to harms against marginalised workers, widening the AI divide, and increasing the disparity of who reaps the benefits from advanced general-purpose AI development.

## 4.3.3 Market concentration risks and single points of failure

### KEY INFORMATION

- Developing state-of-the-art, general-purpose AI models requires substantial up-front investment. These very high costs create barriers to entry, disproportionately benefiting large technology companies.
- Market power is concentrated among a few companies that are the only ones able to build the leading general-purpose AI models.
- Widespread adoption of a few general-purpose AI models and systems by critical sectors including finance, cybersecurity, and defence creates systemic risk because any flaws, vulnerabilities, bugs, or inherent biases in the dominant general-purpose AI models and systems could cause simultaneous failures and disruptions on a broad scale across these interdependent sectors.

The development of state-of-the-art AI systems currently costs hundreds of millions, or even billions of US dollars. The biggest upfront investments are specialised computational resources, AI expertise and access to large, often proprietary datasets (see 2.3.1 Recent trends in compute, data, and algorithms). The significant costs associated with these inputs are a barrier to entry for new firms (*571,*

*572, 573, 574*). Large technology companies are well-positioned thanks to their existing access to the necessary resources and ability to make substantial financial investments.

In addition, general-purpose AI systems benefit from scale. More compute-intensive large-scale models tend to outperform smaller ones (*110**), giving rise to economies of scale: large-scale general-purpose AI systems are in higher demand due to their superior performance, driving down their costs per customer. High user numbers also have network effects: as more users interact with these models, they generate large amounts of additional training data that can be used to improve the models' performance (*575*).

These tendencies towards market concentration in the general-purpose AI industry are particularly concerning because of general-purpose AI's potential to enable greater centralisation of decision-making in a few companies than ever before. Since society at large could benefit as well as suffer from these decisions, this raises questions about the appropriate governance of these few large-scale systems. A single general-purpose AI model could potentially influence decision-making across many organisations and sectors (*571*) in ways which might be benign, subtle, inadvertent, or deliberately exploited. There is the potential for the malicious use of general-purpose AI as a powerful tool for manipulation, persuasion and control by a few companies or governments. Potentially harmful biases such as demographic, personality traits, and geographical bias, which might be present in any dominant general-purpose AI model that become embedded in multiple sectors, could propagate widely. For example, popular text-to-image models like DALL-E 2 and Stable Diffusion exhibit various demographic biases across occupations, personality traits, and geographical contexts (*576*).

The increasing dependence on a few AI systems across critical sectors introduces systemic risks. Errors, bugs, or cyberattacks targeting these systems could cause widespread disruption. Different scenarios have been proposed that illustrate potential disruptions. For example, a denial-of-service attack on a widely used AI API could disrupt critical public infrastructure which relies on that technology. In finance, the adoption of homogeneous AI systems by multiple institutions could destabilise markets by synchronising participants' decisions (*577*): If several banks rely on one model, they may inadvertently make similar choices, creating systemic vulnerabilities (*2**). Comparable risks could potentially arise in domains, like defence or cybersecurity, if AI systems with similar functionality are widely deployed (see also 4.4. Cross-cutting risk factors).

## 4.3.4 Risks to the environment

### KEY INFORMATION

- Growing compute use in general-purpose AI development and deployment has rapidly increased energy usage associated with general-purpose AI.
- This trend might continue, potentially leading to strongly increasing CO2 emissions.

The recent rapid growth in demand for computing power ('compute') used for AI, and particularly general-purpose AI, development and deployment could make AI a major, and potentially the largest, contributor to data centre electricity consumption in the near future. This is because compute demand is expected to far outpace hardware efficiency improvements.

Today, data centres, servers and data transmission networks account for between 1% to 1.5% of global electricity demand (578); roughly 2% in the EU, 4% in the US, and close to 3% in China (69, 579, 580). AI likely accounts for well under half of data centre electricity consumption currently, but if the rapid growth of AI's computational requirements continues, AI could become the primary consumer of data centre electricity over the coming years and increase its share of global electricity demand. In 2023, the largest general-purpose AI training runs used around 5e25 FLOP (65). Using H100 GPUs running at 1400 watts per GPU for operation and cooling, this consumes around 40 GWh. If this figure were to grow at a rate of 3x/year, then at the end of the decade, the largest training run would consume 90 TWh, over half of total US data centre electricity consumption in 2022.

There are several potential mitigations for the increasing energy use of widespread general-purpose AI systems. Specialised AI hardware and other hardware efficiency improvements can enhance the performance-per-watt of machine learning workloads over time (*78*). Moreover, new machine learning techniques and architectures can help reduce energy consumption (*78*). The energy efficiency of computation generally improves by an estimated 26% annually (*68*). However, even with additional optimisation for AI, the growing demand for computing power used for AI training, which has been increasing by a factor of approximately 4x each year, is so far significantly to outpacing energy efficiency improvements (*17*).

The CO2 emissions resulting from AI development and deployment depend on the extent and sources of its energy consumption as well as several factors. The carbon intensity of the energy source is a key variable, with renewable sources like solar power contributing substantially less CO2 emissions throughout their life cycle compared to fossil fuels (*581**). AI firms often rely on renewable energy (*76, 78*), a significant portion of AI training globally still relies on high-carbon sources such as coal or natural gas (*581**). Other important factors affecting CO2 emissions include the geographic location of data centres, their efficiency, and the efficiency of the hardware used. As a result, the actual CO2 emissions for a given amount of energy consumed in AI can vary considerably.

The 'embodied carbon footprint' of AI hardware, which includes emissions from manufacturing, transportation, the physical building infrastructure, and disposal (as opposed to not running the hardware), contributes a substantial portion to emissions - depending on the location this could be as high as 50% (*76*). As hardware efficiency improves, the embodied carbon footprint could become a larger proportion of the total carbon footprint (*76, 78*).

Water consumption might be another noteworthy area of environmental risk from AI. Given the increases in compute used for training and deploying models, cooling demands increase, too, leading to higher water consumption. Water consumption by current models and the methodology to assess it are still subject to scientific debate, but some researchers predict that water consumption by AI could ramp up to billions of cubic metres by 2027 (*76, 582*). In the context of concerns around global freshwater scarcity, and assuming no obvious short-term cooling alternatives exist, AI water footprint might be a substantial cause of environmental concern.

## 4.3.5 Risks to privacy

### KEY INFORMATION

- General-purpose AI models or systems can 'leak' information about individuals whose data was used in training. For future models trained on sensitive personal data like health or financial data, this may lead to particularly serious privacy leaks.
- General-purpose AI models could enhance privacy abuse. For instance, Large Language Models might facilitate more efficient and effective search for sensitive data (for example, on internet text or in breached data leaks), and also enable users to infer sensitive information about individuals.

General-purpose AI systems rely on and process vast amounts of personal data, and this could pose significant and potentially wide-reaching privacy risks. Such risks include loss of data confidentiality for people whose data was used to train these systems, loss of transparency and control over how data-driven decisions are made, and new forms of abuse that these systems could enable.

Privacy, broadly speaking, refers to a person's right to control others' access to their sensitive or personal information. In the context of AI, privacy is a complex and multi-faceted concept, which encompasses issues of confidentiality, transparency, and control. Privacy is a challenging concept to define (*583*). In the context of AI it encompasses:

- Data confidentiality and protection of personal data collected or used for training purposes, or during inference (*584*)
- Transparency, and controls over how personal information is used in AI systems (*585*), for example the ability for individuals to opt-out from personal data being collected for training, or the post-hoc ability to make a general-purpose AI system 'unlearn' specific information about an individual (*586*);
- Individual and collective harms that may occur as a result of data use or malicious use, for example the creation of deepfakes (*587*).

General-purpose AI systems may expose their training data. The training of general-purpose AI models generally requires large amounts of training data. Academic studies have shown that some of this training data may be memorised by the general-purpose AI model, or may be extractable using adversarial inputs, enabling users to infer information about individuals whose data was collected (*588, 589, 590*) or even reconstruct entire training examples (*591, 592, 593, 594*). However, definitions of memorisation vary, so it is challenging to make any concrete claims about the harms that might arise from memorisation (*595*).

Many systems are trained on publicly available data containing personal information without the knowledge or consent of the individuals it pertains to. This information could then be outputted by a general-purpose AI system in undesired contexts. There is a risk that training models on sensitive data containing personal information (such as medical or financial data) could result in serious privacy leaks. It is difficult to assess the likelihood or potential impact of these risks: for example, existing medical general-purpose AI systems such as Google's Gemini-Med (*596**) are only trained on anonymised public patient data, and the rate at which such models regurgitate training data has not yet been studied. General-purpose AI systems that continuously learn from interactions with users (e.g. chatbots such as ChatGPT) might also leak such interactions to other users, although at the time of writing, there are no well-documented cases of this occurring.

General-purpose AI systems could enable privacy abuse. Some studies have found that general-purpose AI systems have privacy-relevant capabilities that may be exploited by malicious users of these systems. For example, fine-grained internet-wide search capabilities, such as powerful reverse image search or forms of writing style detection, which allow individuals to be identified and tracked across online platforms, or sensitive personal characteristics to be inferred, further eroding individual privacy (*597, 598*). Large language models could also enable more efficient and effective search for sensitive information on the internet, or in breached datasets. General-purpose AI-generated content, such as non-consensual deepfakes, could be used to manipulate or harm individuals, raising concerns about the harm caused by the malicious use of personal data and the erosion of trust in online content (*255, 256, 373, 599*).

## 4.3.6 Copyright infringement

### KEY INFORMATION

- The use of large amounts of copyrighted data for training general-purpose AI models poses a challenge to traditional intellectual property laws, and to systems of consent, compensation, and control over data.
- The use of copyrighted data at scale by organisations developing general-purpose AI is likely to alter incentives around creative expression.
- An unclear copyright regime disincentivizes general-purpose AI developers from following best practices for data transparency.

- There is very limited infrastructure for sourcing and filtering legally and ethically permissible data from the internet for training general-purpose AI models.

**General-purpose AI models are usually trained on large data sets sourced online, giving rise to concerns over breaches of copyright, lack of creator compensation, and the potential for economic disruption.** Copyright laws aim to protect intellectual property and encourage written and creative expression (*600, 601*). They grant the creators of original works the exclusive right to copy, distribute, adapt, and perform their own work. However, the third-party use of copyrighted data as training data may be legally permissible in certain circumstances, for instance on the basis of the 'fair use' exception in the US (*602*), by the 'text and data mining' exception in the EU (*603*), by the amended Copyright Act in Japan (*604*), under Israeli copyright law (*605*), and by the Copyright Act 2021 in Singapore (*606*). Beyond copyright, artists and other individuals sometimes feel their style, voice, and likeness are not sufficiently protected, which may implicate other forms of intellectual property such as trademarks and brands.

Recent advances in general-purpose AI capabilities have largely resulted from large-scale web scraping and aggregation of data to train general-purpose AI models (*607, 608*), often containing copyrighted works, or used without consent from the data's creators. This applies to creative works including text, images, videos, and speech, and other modalities that are increasingly used to develop general-purpose AI models. The extent to which this is legally permissible is complex and can vary by country. In the US, the fair use exception has been argued for in the case of training general-purpose AI models (*222, 609, 610, 611*), as well as legally challenged (*612*). Many issues related to dataset creation and use across its lifecycle make copyright concerns for training AI models very complicated (*613*). These issues include the questions of whether datasets are assembled specifically for machine learning or originally for other purposes (*614*), whether the infringement analysis applies to model inputs or model outputs (*615*), and issues of jurisdiction, among others (*236, 616, 617*). This also presents questions on who is liable for infringement or harmful model outputs (*618*). While there are technical strategies for mitigating the risks of copyright infringement from model outputs, these risks are difficult to eliminate entirely (*619, 620*).

As general-purpose AI systems become more capable, they increasingly have the potential to disrupt labour markets, and in particular creative industries (*250, 621*), (also see 4.3.1 Labour market risks). The legal determinations regarding copyright infringement in the AI training phase will affect the ability for general-purpose AI developers to build powerful and performant models. They may also impact data creators' ability to exert control over their data, which may disincentivize creative expression.

**An unclear copyright regime disincentivizes general-purpose AI developers from improving data transparency.** Transparency about general-purpose AI model training data is useful for understanding various potential risks and harms of a general-purpose AI system (*259*). However, this type of transparency is often lacking for major general-purpose AI developers (*244*). Fears of legal risk, especially over copyright infringements, may disincentivise these developers from disclosing their training data (*622*).

**The infrastructure to source and filter for legally permissible data is under-developed, making it hard for developers to comply with copyright law.** The permissibility of using copyrighted works as part of training data without a licence is an active area of litigation. Tools to source and identify available data without copyright concerns are limited. For instance, recent work shows that ~60% of popular datasets in the most widely used openly accessible dataset repositories have incorrect or missing licence information (*236*). Similarly, there are limitations to the current tools for discerning copyright-free data in web scrapes (*607, 623*). However, practitioners are developing new standards for data documentation and new protocols for data creators to signal their consent for use in training AI models (*235, 624*).

# 4.4 Cross-cutting risk factors

## 4.4.1 Cross-cutting technical risk factors

### KEY INFORMATION

- This section covers seven cross-cutting *technical risk factors* – technical factors that each contribute to many general-purpose AI risks.
    a. General-purpose AI systems can be applied in many ways and contexts, making it hard to test and assure their trustworthiness across all realistic use-cases.
    b. General-purpose AI developers have a highly limited understanding of how general-purpose AI models and systems function internally to achieve the capabilities they output.
    c. General-purpose AI systems can act in accordance with unintended goals, leading to potentially harmful outputs, despite testing and mitigation efforts by AI developers.
    d. A general-purpose AI system can be rapidly deployed to very large numbers of users, so if a faulty system is deployed at scale, resulting harm could be rapid and global.
    e. Currently, risk assessment and evaluation methods for general-purpose AI systems are immature and can require significant effort, time, resources, and expertise.
    f. Despite attempting to debug and diagnose, developers are not able to prevent overtly harmful behaviours across all circumstances in which general-purpose AI systems are used.
    g. Some developers are working to create general-purpose AI systems that can act with increasing autonomy, which could increase the risks by enabling more widespread applications of general-purpose AI systems with less human oversight.

Risk factors are distinct from risks. They are conditions that increase the likelihood and/or impact of risks occurring. This section covers seven cross-cutting *technical risk factors* i.e.– factors that each contribute to multiple general-purpose AI risks.

**General-purpose AI systems can be applied in many ways and contexts, making it hard to test and assure their trustworthiness across all possible use cases.** The relative safety of general-purpose AI systems depends on the context in which they are used. General-purpose AI systems' outputs are often open-ended, such as free-form dialogue or code generation. This makes it difficult to design safe systems because it is not tractable to exhaustively evaluate all possible downstream use cases. Users can also 'jailbreak' general-purpose AI models to make them comply with potentially harmful requests (see 5.2.3 Improving robustness to failures). At present, computer scientists are unable to give guarantees of the form "System X will not do Y" about general-purpose AI systems (*625*). As discussed in 3. Methodology to assess and understand general-purpose AI systems, assessing the risks of general-purpose AI systems in real-world applications and making strong assurances against general-purpose AI-related harms is extremely difficult with current methods.

**General-purpose AI developers have a highly limited understanding of how general-purpose AI models and systems function internally.** A key feature of general-purpose AI systems is that their capabilities are mainly achieved through learning rather than from top-down design. As a result, unlike most human-engineered systems, state-of-the-art general-purpose AI models do not come with blueprints, and their structure does not conform to common design principles. This gives rise to concerns around understanding or explaining general-purpose AI, which are used interchangeably in this report to refer to the ability to provide human-understandable accounts of how general-purpose AI arrives at outputs and decisions from inputs and objectives. There are different views around what

constitutes a human-understandable account. A thorough discussion of these views lies outside the scope of this report, but key questions include:

- What forms it may take, e.g. whether it has to be easily comprehensible language or can be complex mathematical information, and whether the size of the description should be bounded to allow humans to make sense of the explanation.
- How comprehensive it must be, e.g. whether it should include retrospective analysis of how an explained decision process was established in training;.
- Whether it must be counterfactual, i.e. whether it must allow hypothetical statements about what outputs would have resulted from different inputs, different objectives or different training.

Currently, scientists' understanding of general-purpose AI systems is more analogous to that of brains or cells than aeroplanes or power plants. Some researchers believe that it may be possible to develop general-purpose AI systems that can be proven safe, or are 'safe by design' (*626*) by focusing the validation on interpretable outputs of the neural networks rather than their internal states, which humans currently cannot understand. However, it has not yet been possible for scientists to achieve such quantitative safety guarantees for state-of-the-art general-purpose AI models (*58*). Research on thoroughly understanding how AI systems operate has been limited to 'toy' systems that are much smaller and less capable than general-purpose AI models (*627, 628, 629, 630*), or are unreliable and require major simplifying assumptions (*631, 632**). In practice, techniques for interpreting the inner workings of neural networks can be misleading (*213, 288*, 289, 290, 336*), and can fail sanity checks or prove unhelpful in downstream uses (*218, 297, 298, 299*). As discussed in 3. Methodology to assess and understand general-purpose AI systems, these research methods are being developed, and new improvements might yield further insights, especially with sufficient investment in further research. However, it is unclear yet whether interpreting the inner structures of neural networks will offer sufficient safety assurances.

**Ensuring that general-purpose AI systems pursue the goals intended by their developers and users is difficult.** Although general-purpose AI systems can appear to excel at learning what they are 'told' to do, their behaviour may not necessarily be what their designers intended (*489, 633, 634, 635*). Even subtle differences between a designer's goals and the incentives given to a system can lead to unexpected failures. For example, general-purpose AI chatbots are often trained to produce text that will be rated positively by human evaluators, but user approval is an imperfect proxy for user benefit: widely-used chatbots can learn to pander to users' biases instead of prioritising truth (*237, 238*). Even when a general-purpose AI system receives correct feedback during training, it may still develop a solution that does not generalise well when applied to novel situations during new situations once deployed (*636, 637, 638*) because the training data may not adequately represent real-world scenarios. For example, some researchers have found that chatbots are more likely to comply with harmful requests in languages that are under-represented in their training data (*309*). See 4.2.3. Loss of control and 5.2. Training more trustworthy models for further discussion of these challenges.

**Because general-purpose AI systems can proliferate rapidly, like other software, a new fault or harmful capability can rapidly have a global and sometimes irreversible impact.** A small number of proprietary and freely available (open-source) general-purpose AI models reach many millions of users (4.3.3.Market concentration risks and single points of failure). Both proprietary and open-source models can therefore have rapid and global impacts when they are released, although in different ways. Once a model is made available open-source, there is no practical way to erase the model from the market in case it has faults or capabilities that enable malicious use (*639*) (see 4.1. Malicious use risks). For model faults, however, open-sourcing a model allows a much greater and more diverse number of practitioners to discover them which can improve the understanding of risks and possible mitigations (*640*) (see 3. Methodology to assess and understand general-purpose AI systems). Developers or others can then repair faults and encourage users to update to a new model version. Furthermore, open-sourcing models allows more actors to customise these models. If one version of the model has flaws, other customised versions may not share the same issues (*640*). However, neither repairing model faults nor customising can prevent deliberate malicious use (*639*). The risk of deliberate malicious use depends on a model's *marginal risk* compared to available closed source

models and other technologies such as internet search (*640*) (see 4.1. Malicious use risks). The above factors are relevant to the specific possibility of rapid, widespread, and irreversible impacts of general-purpose AI models, but this report does not provide an assessment of the overall impacts of open-source models. Even when a system is not open-sourced, its capabilities can still be accessed by a wide user base. Within two months of launch, ChatGPT had over 100 million users and set a record for the fastest-growing user base of any consumer application (*641*). Each time a general-purpose AI system is updated, a new version of it rapidly reaches a large user base, so any vulnerabilities or harmful tendencies can potentially have a global impact quickly (see also 4.3.3. Market concentration and single points of failure.)

**Despite attempting to debug and diagnose, developers are not able to prevent even overtly harmful behaviours across all circumstances in which general-purpose AI systems are used.** Empirically, harmful behaviours have included revealing private or copyrighted information (*221, 642*, 643*); generating hate speech (*225, 644*); spreading social and political biases (*239, 455, 645*); pandering to user biases (*238*); hallucinating inaccurate content (*46, 47*, 49*); exhibiting vulnerabilities to a variety of attacks on their safety protections (*108, 208, 209, 210, 211, 309, 646, 647, 648**); and assisting in overtly harmful tasks (*368, 649, 650*). Users can circumvent general-purpose AI model safeguards with relative ease (*650, 651*), for example through 'jailbreaking' techniques (see 3. Methodology to assess and understand general-purpose AI systems). While some have called for safety measures that rule out *all* overtly harmful behaviours across *all* situations (*626*), current general-purpose AI development fails to meet a lower standard than this: ruling out any specific overtly harmful behaviour across foreseeable situations (such as situations where users try to jailbreak a model.)

## AI developers are working to create 'agent' general-purpose AI systems that can accomplish tasks with little to no human involvement, which could increase risks of accident and malicious use

Today, general-purpose AI systems are primarily used directly as tools by humans. For example, a human may ask a chatbot to write computer code to help them accomplish a task, which naturally requires a 'human in the loop'. However, developers are increasingly designing systems that allow general-purpose AI systems to act autonomously, by controlling software tools such as a web browser or controlling the execution of code rather than only writing code. This enables some forms of reasoning about problems, constructing plans, and executing plans step-by-step (*13, 26, 188, 189, 652, 653, 654, 655*, 656*). Such systems have included web-browsing virtual agents (*657**), research assistants (*415*), and writing, fixing, and running code autonomously (*441*).

The main purpose of general-purpose AI 'agents' is to reduce the need for human involvement and oversight, allowing for faster and cheaper applications of general-purpose AI. This property also reduces human oversight, potentially increasing the risk of accidents (see 4.1 Malicious use risks), and allowing the automation workflows for malicious uses (see 4.2.1 Risks from product functionality issues use) while also being relevant to the risk of loss of control (see 4.2.3 Loss of control) (*481*, 658*).

**General-purpose AI agents have early forms of many autonomous capabilities, but lack reliability at performing complex tasks autonomously.** Current state-of-the-art general-purpose AI systems are capable of autonomously executing many simple tasks, but some evaluations have shown that they struggle with more complex ones (*183, 509*). They are particularly unreliable at performing tasks that involve many steps. As of 2023, general-purpose AI agents scored low in benchmarks designed to measure their performance on complex and economically useful tasks (*183, 441*). However, given current efforts and growing investment in developing general-purpose AI systems with greater autonomy, the capabilities of general-purpose AI agents could continue to increase.

**Although current general-purpose AI agents are unreliable, advancement has been rapid.** For example, for two challenging benchmarks for general-purpose AI agents that measure general problem-solving (*183*) and autonomous software engineering (*441*) accuracy has improved over a period of a few months by factors of 2.2x and 7.1x respectively compared to strong baselines using GPT-4. The performance of the general-purpose AI models (such as LLMs) which underlie these general-purpose AI agents is key to how reliable they are, so newer generations of general-purpose AI models typically increase agent capabilities. One potential way to increase agent capabilities could be through combining LLMs with search and planning methods. In board games like Go and Stratego, combining deep learning with methods like Monte Carlo Tree Search (MCTS) and self-play has led to above-human level performance (*659, 660*). Early work on combining LLMs with search has yielded improvements in simple settings (*661*). However, careful monitoring of general-purpose AI agent capabilities is needed to assess many of the risks discussed in 4. Risks.

## 4.4.2 Cross-cutting societal risk factors

### KEY INFORMATION

- This section covers four cross-cutting societal risk factors – non-technical aspects of general-purpose AI development and deployment that each contribute to many risks from general-purpose AI:
    a. AI developers competing for market share may have limited incentives to invest in mitigating risks.
    b. As general-purpose AI advances rapidly, regulatory or enforcement efforts can struggle to keep pace.
    c. Lack of transparency makes liability harder to determine, potentially hindering governance and enforcement.
    d. It is very difficult to track how general-purpose AI models and systems are trained, deployed and used.

Risk factors are distinct from risks - they are conditions that increase the likelihood and/or impact of risks occurring. There are societal aspects of general-purpose AI development and deployment that increase not one but several general-purpose AI risks. This section discusses these 'cross-cutting societal risk factors'.

**General-purpose AI developers, who are competing for market share in a dynamic market where getting a product out quickly is vital, may have limited incentives to invest in mitigating risks.** The one-time cost of developing a state-of-the-art general-purpose AI model is very high, while the marginal costs of distributing such a model to (additional) users are relatively low. This can lead to 'winner takes all' dynamics, creating strong incentives for developers to build the most capable model at any cost, because doing so might allow immediately capturing a large market share. In recent years, there has been intense competition between general-purpose AI developers to rapidly build and deploy models. This has raised concern about potential 'race to the bottom' scenarios, where actors compete to develop general-purpose AI models as quickly as possible while under-investing in measures to ensure safety and ethics (*662, 663*). This could contribute to situations in which it is challenging for general-purpose AI developers to commit unilaterally to stringent safety standards, as doing so might put them at a competitive disadvantage (*664*). Similar dynamics can also occur at the international level regarding regulation efforts. Without global coordination regarding the regulation of general-purpose AI, a regulatory 'race to the bottom' could see countries attempt to attract AI companies through lax regulation that might be insufficient for ensuring safety domestically and abroad, a dynamic that has been described for several types of regulation like labour law (*665*).

**As general-purpose AI markets advance rapidly, regulatory or enforcement efforts can struggle to keep pace.** A recurring theme in the discourse on general-purpose AI risk is the mismatch between the pace of technological innovation and the development of governance structures (*666*). While existing legal and governance frameworks apply to some uses of general-purpose AI systems and several jurisdictions (like the European Union, China, the USA or Canada) have initiated or completed efforts to regulate AI and general-purpose AI specifically, there often remain regulatory gaps. In a market that is as fast-moving as the general-purpose AI market currently is, it is very difficult to fill such gaps ex-post, because by the time a regulatory fix is implemented it might already be outdated. Policymakers therefore face the challenge of creating a flexible regulatory environment that ensures the pace of general-purpose AI development and deployment remains manageable from a public safety perspective.

**General-purpose AI systems' inherent lack of transparency makes legal liability hard to determine, potentially hindering governance and enforcement.** How current legal frameworks apply in cases where it is suspected that a general-purpose AI system caused harm is often unclear. This raises many issues for accountability, liability and justice. In principle, people and corporate entities are held accountable, not the technology, which is why many critical areas maintain a 'human in the loop' policy. However, tracing harm back to the responsible individuals who developed or deployed the AI is very challenging (*667, 668, 669*), as is gathering evidence of error. This accountability issue is compounded by the opaque nature of proprietary general-purpose AI models, where commercially sensitive training data, methodologies, and decision-making processes are usually not open to public scrutiny and there is a lack of widely shared standard operating procedures for using and interpreting AI systems (*291, 292, 293, 294*, 295, 670, 671*). Some also argue that general-purpose AI systems can exhibit 'emergent' behaviours that were not explicitly programmed or intended by their developers, raising questions about who should be held liable for resulting harm. The distributed nature of general-purpose AI development, involving multiple actors such as data providers, model trainers, and deployers, also makes it challenging to assign liability to a single entity (*669*).

**It is very difficult to track how general-purpose AI models and systems are trained, deployed and used.** Tracking the use of general-purpose AI models and systems is not only important for establishing liability for potential harms caused by the use of general-purpose AI models and systems, but also for monitoring and evidencing malicious use, and noticing malfunctions (*658, 672, 673*). Comprehensive safety governance is common in safety-critical fields such as automotive, pharmaceuticals, and energy (*674, 675, 676, 677*), but it often relies on broadly accepted standards that are currently missing in general-purpose AI governance.

# 5 Technical approaches to mitigate risks

This section of the report discusses technical approaches to increase general-purpose AI safety through mitigating general-purpose AI-related risks: to reduce the scale of harms or the likelihood of their occurrence. This report ultimately finds that, while there are many technical approaches that reduce risk, existing methods are insufficient to prove that systems are safe.

The scope of this report does not include nontechnical (political, legal, or social) interventions, but these are equally important for addressing risks. Moreover, technical and nontechnical aspects of managing risks from general-purpose AI are highly intertwined; no technical solutions are implemented in a vacuum. The last months and years have seen increased interest in AI regulation from policymakers, and several jurisdictions (like the European Union, China, the USA or Canada) have initiated or completed efforts to regulate AI and general-purpose AI specifically. Effective approaches will require the resources and political will to implement technical solutions as well as an interplay between multiple technical and nontechnical safeguards against harms from general-purpose AI.

## 5.1 Risk management and safety engineering

### KEY INFORMATION

- Developing and incentivising systematic risk management practices for general-purpose AI is difficult. This is because current general-purpose AI is progressing rapidly, is not well-understood, and has a wide range of applications. Methodologies for assessing general-purpose AI risk are too nascent for good quantitative analysis of risk to be available.
- While many other fields offer lessons for how such approaches could be developed, there are currently no well-established risk management and safety engineering practices for general-purpose AI systems.
- Since no single existing method can provide full or partial guarantees of safety, a practical strategy is *defence in depth* – layering multiple risk mitigation measures. This is a common way to manage technological risks.
- An important consideration for effective risk management of general-purpose AI is who to involve in the process in order to identify and assess high-priority risks. This can include experts from multiple domains but also representatives of impacted communities.

**Risk** is the combination of the probability of an occurrence of harm and the severity of that harm if it occurs (*339*). This technically includes both positive and negative outcomes, but in common usage, the focus is on the negative outcomes. Some (but not all) of the common risks associated with general-purpose AI are described in 4. Risks.

Risk increases with the severity of the potential harm and the probability of the harm materialising Whether and to what extent the outcomes of a system are considered undesirable has to be conceptualised with respect to contextual human values. Various stakeholders may disagree how undesirable any particular outcome is.

**Risk surface/exposure:** The risk surface of a technology consists of all the ways it can cause harm through accidents or malicious use. The more general-purpose a technology is, the more extensive its risk exposure is expected to be. General-purpose AI models can be fine-tuned and applied in numerous application domains and used by a wide variety of users (4.4.1. Cross-cutting technical risk factors), leading to extremely broad risk surfaces and exposure, challenging effective risk management.

**Risk management** consists of the *identification*, *assessment*, and *prioritisation* of risks and utilising resources to minimise, monitor, and control high-priority risks.

**System safety engineering** is defined very similarly but with an emphasis on the importance of the interactions of multiple parts of a larger system (*678*). In the case of AI, this approach entails taking into account all the constituent parts of a general-purpose AI system, as well as the broader context in which it operates. Industries such as finance, insurance, health, and cybersecurity have well-established risk management practices (*679*). The AI risk management framework by NIST (*680*) is among the few prominent recent efforts to come up with a framework of risk management specific to AI systems.

### 5.1.1 Risk assessment

When the scope of applicability and use of an AI system is narrow (e.g., consider spam filtering as an example), salient types of risk (e.g., the likelihood of false positives) can be measured with relatively high confidence. In contrast, assessing general-purpose AI models' risks, such as the generation of toxic language, is much more challenging, in part due to a lack of consensus on what should be considered toxic and the interplay between toxicity and contextual factors (including the prompt and the intention of the user).

There is a broad range of risk assessment techniques that could and are already being used with regard to general-purpose AI (*216, 679*), including evaluations, red-teaming, auditing, and qualitative evaluation of general-purpose AI systems (see 3. Methodology to assess and understand general-purpose AI systems). Other risk methods for assessment, some of which draw on established practices in other fields, include:

- Uplift studies, where the aim is to test how much more competent a human is at accomplishing a potentially harmful task when they have access to a general-purpose AI system versus when they do not (*166**).
- Forecasting to inform high-stakes decisions (*323**)
- Delphi studies aggregating forecasts from groups of relevant experts (*681*).
- Field testing aims to detect and document risks arising in the environment in which the model is to be used and the specific dangers posed in that setting.
- Benchmarking tasks and datasets to assess the prevalence and severity of specific types of risks (e.g., toxicity, certain forms of bias, grasp of scientific domains, dangerous capabilities).

Current risk assessment methodologies often fail to produce reliable assessments of the risk posed by general-purpose AI systems. Some of the key challenges of utilising such risk assessment methodologies for highly capable models are:

- Specifying the relevant/high-priority flaws and vulnerabilities is highly influenced by who is at the table and how the discussion is organised, meaning it is easy to miss or mis-define areas of concern.
- Guarding against potential malicious uses requires understanding likely and viable threats, including the estimation of resources available to malicious actors (e.g., compute, access, and expertise) and their incentives.
- The general-purpose nature of these technologies adds uncertainty with regard to their use in deployment. An example illustrating this point is applications involving open-ended interactions (e.g., chatbots), which can generate a vast array of potential outputs.
- The rapid pace of technological advancement (2.3.1 Recent trends in capabilities and 4.4.1 Cross-cutting technical risk factors) worsens the above challenges.

Red teaming, for example, only assesses *whether* a model can produce some output, not the extent to which it will do so in real-world contexts nor how harmful doing so would be. Instead, they tend to provide qualitative information that informs judgments on what risk the system poses.

## 5.1.2 Risk management

There are a range of risk assessment techniques that could and are already being used with regard to general-purpose AI (see 3. Methodology to assess and understand general-purpose AI systems). To address the limitations of these existing tools, researchers can look to established practice in risk management in other domains. Some of the common risk management tools in other safety-critical industries are:

- Planned audits and inspection.
- Ensuring traceability using standardised documentation.
- Redundant defence mechanisms against critical risks and failures.
- Risk management guidelines prescribing processes, evaluations, and deliverables at all stages of a safety-critical system's life cycle.

Similar ideas have been proposed for managing the risks associated with general-purpose AI systems. Many of these existing approaches are not directly applicable to highly capable general-purpose AI models, or their efficacy is not well-studied, but efforts are underway to extend existing guidelines to Generative AI.

**Safety and reliability engineering:** The practice of safety engineering has a long history in various safety-critical engineering systems, such as the construction of bridges, skyscrapers, aircraft, and nuclear power plants. At a high level, safety engineering assures that a *life-critical* system acts as intended and with minimal harm, even when certain components of the system fail. Reliability engineering is broader in scope and address non-critical failures as well. These approaches offer several techniques that could be useful for risk assessment in general-purpose AI:

- Safety by Design (SbD) is an approach that centres user safety in the design and development of products and services. For general-purpose AI products and services, this may take the form of restricting access to the model (e.g., through limiting the length of user interactions with the model).
- Safety analysis aims to understand the causal dependencies between the functionality of the individual components and the overall system so that component failures, which can lead to system-level hazards (e.g., aircraft crash or nuclear reactor core meltdown), can be anticipated and prevented to the extent possible.
- 'Safety of the intended function' (SOTIF) approaches require engineers to provide evidence that the system is safe when operating as intended.
- Some risk assessment methods, such as for the nuclear power sector, leverage mathematical models that are designed to quantify risk as a function of various design and engineering choices, accompanied by quantitative risk thresholds set by regulators (*682*).[5] A key advantage of this approach is that it lets a publicly accountable body define what risk is considered acceptable, in a way that is legible to the public and external experts.

However, translating best practices from these fields to general-purpose AI is difficult. Quantitative risk assessment methodologies for general-purpose AI are very nascent and it is not yet clear how quantitative safety guarantees could be obtained. Experience of other risk assessments in general-purpose AI suggests that many areas of concern may not be amenable to quantification (for example, bias and misinformation). If quantitative risk assessments are too uncertain to be relied on, they may still be an important complement to inform high-stakes decisions, clarify the assumptions used to assess risk levels and evaluate the appropriateness of other decision procedures (e.g. those tied to model capabilities). Further, "risk" and "safety" are contentious concepts – for instance, one might ask "safe to whom?" – which may require the involvement of diverse sets of experts and potentially impacted populations (*307*).

[5] For example, some regulatory commissions mandate that nuclear reactor operators produce probabilistic risk assessments, and ensures that the estimated risk of certain events are kept below specified thresholds. For example, operators are mandated to keep the estimated chance of radioactive release beyond the power plant below one in a million (*683*).

While there are currently no well-established safety engineering practices for general-purpose AI systems, the pipeline-aware approach to mitigating AI's harm takes inspiration from safety engineering and proposes scrutinising numerous design choices made through the general-purpose AI lifecycle, from ideation and problem formulation, to design, development, and deployment, both as individual components and in relation to one another (*684, 685*). Further work is needed to extend these ideas from traditional AI to generative AI.

**Safety cases:** Developers of safety-critical technologies such as aviation, medical devices, and defence software are required to make *safety cases*, which put the burden of proof on the developer to demonstrate that their product does not exceed maximum risk thresholds set by the regulator (*686, 687, 688, 689*). A safety case is a structured argument supported by evidence, where the developer identifies hazards, models risk scenarios, and evaluates the mitigations taken. Safety cases would be easier to make for general-purpose AI systems with limited capabilities since less capable models often pose less risk. Thereby, they are robust to scenarios of both slow and rapid progress in general-purpose AI capabilities (see 2.4 Capability progress in coming years). Safety cases leverage the technical expertise of the technology developer but still require that the regulator (or a suitable third party) has the technical expertise to evaluate safety cases. Safety cases often address only a subset of risks and threat models, leaving out important ones (*690, 691*). One mitigation to these limitations is to review safety cases alongside risk cases produced by a red team of third-party experts (*689*).

**The 'Swiss-cheese' model for general-purpose AI safety engineering:** The general-purpose, rapidly evolving, and inscrutable nature of highly capable models makes it increasingly difficult to develop, assess, and incentivise systematic risk management practices. Effectively managing the risks of highly capable general-purpose AI systems might therefore require the involvement of multiple stakeholder groups, including experts from multiple domains and impacted communities, to identify and assess high-priority risks. Further, it suggests that no single line of defence should be relied upon. Instead, multiple independent and overlapping layers of defence against those risks may be advisable, such that if one fails, others will still be effective. This is sometimes referred to as the *Swiss Cheese model of defence in depth (692)*.

**Current risk management among general-purpose AI developers:** Though not universally adhered to, it is common practice to test models for some dangerous capabilities ahead of release, including via red-teaming and benchmarking, and publishing those results in a 'model card' (*257*). Further, some developers have internal decision-making panels that deliberate on how to safely and responsibly release new systems. An increasingly common practice among these developers is to constrain decisions through voluntary pre-defined capabilities thresholds (*693*, 694**). Such thresholds determine that specific model capabilities must be met with specific mitigations that are meant to keep risks to an acceptable level (*682*). Such capabilities thresholds have the advantage of being observable in advance of capabilities being developed. However, more work is needed to assess whether adhering to some specific set of thresholds indeed does keep risk to an acceptable level and to assess the practicality of accurately specifying appropriate thresholds in advance.

# 5.2 Training more trustworthy models

## KEY INFORMATION

- There is progress in training general-purpose AI systems to function more safely, but there is currently no approach that can ensure that general-purpose AI systems will be harmless in all circumstances.
- Companies have proposed strategies to train general-purpose AI systems to be more helpful and harmless: however, the viability and reliability of these approaches for such advanced systems remains limited.
- Current techniques for aligning the behaviour of general-purpose AI systems with developer intentions rely heavily on data from humans such as human feedback. This makes them subject to human error and bias. Increasing the quantity and quality of this feedback is an avenue for improvement.
- Developers train models to be more robust to inputs that are designed to make them fail ('adversarial training'). Despite this, adversaries can typically find alternative inputs that reduce the effectiveness of safeguards with low to moderate effort.
- Limiting a general-purpose AI system's capabilities to a specific use case can help to reduce risks from unforeseen failures or malicious use.
- Researchers are beginning to learn to analyse the inner workings of general-purpose AI models. Progress in this area could help developers understand and edit general-purpose AI model functionality more reliably.
- Researchers are exploring how to obtain AI systems that are safe by design or provably safe, although many open problems remain to scale these methods to general-purpose AI systems.

## 5.2.1 Aligning general-purpose AI systems with developer intentions

'AI alignment' refers to the challenge of making general-purpose AI systems act in accordance with their developer's goals and interests (see 5.4 Technical approaches to fairness and representation in general-purpose AI systems for a discussion of the challenges to alignment posed by conflicting values of different stakeholders).

### 5.2.1.1 Two alignment challenges

There are two challenges involved in training aligned general-purpose AI systems: firstly, ensuring that they are trained with an objective that incentivises the intended goals; and secondly, ensuring that the outputs translate from their training contexts to the real world as intended, especially in high-stakes situations.

**It is challenging to precisely specify an objective for general-purpose AI systems in a way that does not unintentionally incentivise undesirable behaviours.** Currently, researchers do not know how to specify abstract human preferences and values in a way that can be used to train general-purpose AI systems. Moreover, given the complex socio-technical relationships embedded in general-purpose AI systems, it is not clear whether such specification is possible. General-purpose AI systems are generally trained to optimise for objectives that are imperfect proxies for the developer's true goals (*634*). For example, AI chatbots are often trained to produce text that will be rated positively by human evaluators, but user approval is an imperfect proxy for user benefit. Research has shown that

several widely-used chatbots sometimes match their stated views to a user's views regardless of truth (*238*). This is an ongoing challenge for general-purpose AI and similar AI systems (*489, 633, 634, 695**).

**Ensuring general-purpose AI systems learn behaviours that translate from their training contexts to real-world, high-stakes deployment contexts is also highly challenging.** Just as general-purpose AI systems are trained to optimise for imperfect proxy goals, the training context can also fail to adequately represent the real-world situations they will encounter after they are deployed. In such cases, general-purpose AI systems can still learn to take harmful actions even if they are trained with correct human-provided feedback (*636, 637, 638*). For example, some researchers have found that chatbots are more likely to take harmful actions in languages that are under-represented in their training data (*309*). Using more multilingual data and oversight may be able to more mitigate this type of failure.

### 5.2.1.2 Alignment techniques

To elicit the desired behaviours from state-of-the-art general-purpose AI systems, developers train models using human oversight. Improving performance in real-world contexts benefits from large amounts of data.

**State-of-the-art alignment techniques rely on feedback or demonstrations from humans and, as such, are constrained by human error and bias.** As discussed in 2.1. How does General-Purpose AI gain its capabilities?, developers fine-tune state-of-the-art general-purpose AI systems using a large amount of human involvement. In practice, this involves techniques that leverage human-generated examples of desired actions (*18*) or human-generated feedback on examples from models (*19, 20, 21*, 303*). This is done at scale, making it labour-intensive and expensive. However, human attention, comprehension, and trustworthiness are not perfect (*303*), which limits the quality of the resulting general-purpose AI systems (*696, 697*, 698*). Even slight imperfections in feedback from humans can be amplified when used to train highly capable systems with potentially serious consequences (see 4.1. Malicious use risks and 4.2.3. Loss of control).

**Improving the quality and quantity of human oversight can help to train more aligned models.** Some research has shown that using richer, more detailed forms of feedback from humans can provide a better oversight signal, but at the cost of increased time and effort for data collection (*699, 700, 701*). To gather larger datasets, leveraging general-purpose AI systems to partially automate the feedback process can greatly increase the volume of data (*158*, 702**). However, in practice, the amount of explicit human oversight used during finetuning is very small compared to the trillions of data points used in pre-training on internet data and may, therefore, be unable to fully remove harmful knowledge or capabilities from pre-training. Short of rethinking the way state-of-the-art general-purpose AI systems are trained, improving fine-tuning feedback data is unlikely to be a solution on its own.

**Maintaining uncertainty over goals can reduce risky actions.** Some researchers have proposed methods that involve incorporating uncertainty into the goals that general-purpose AI systems learn to pursue (*703, 704, 705, 706*, 707, 708*). By requiring general-purpose AI systems to act in a way that respects uncertainty about their objectives, these methods can reduce the risk of unexpected actions and encourage information-seeking or deference to humans in response to ambiguity. However, these methods have yet to be incorporated into state-of-the-art, highly capable AI.

**Some researchers are working toward safe-by-design approaches which might be able to provide quantitative safety guarantees.** Similar to the above approaches to estimate uncertainty about goals and predictions, it may be possible to design AI systems that are constructed to achieve quantified levels of safety (*626*). The advantage of mathematical guarantees and bounds is that they may provide safety assurances even outside of the domain where the AI has been trained and tested, in contrast with empirical trial-and-error methods that are currently the standard for designing deep learning systems. Currently, however, practically-useful, provable guarantees of safety are not possible

with general-purpose AI models and methods, and many open questions however remain in order to achieve those objectives for large-scale AI systems.

**It is unclear if and how humans might be able to oversee general-purpose AI systems with capabilities exceeding those of humans.** Future general-purpose AI models or systems that surpass the abilities of human experts across many or all domains would pose a particularly difficult challenge (see 4.2.3. Loss of control). Some research efforts, especially concentrated in the leading AI laboratories, study the extent to which humans might be able to oversee general-purpose AI with capabilities that exceed the human overseers generally or in a given domain (known as 'scalable oversight'). Some empirical research has studied the ability of less capable general-purpose AI systems to oversee more capable ones under the assumption that similar dynamics may exist for general-purpose AI exceeding human capabilities (*709*, 710**). Researchers have also proposed theoretical approaches that can, under certain assumptions, allow for stronger assurances than existing methods (*711, 712*, 713**). However, published research on these methods is highly preliminary.

## 5.2.2 Reducing the hallucination of falsehoods

**The hallucination of falsehoods is a challenge, but it can be reduced.** In AI, 'hallucination' refers to the propensity of general-purpose AI systems to output falsehoods and made-up content. For example, language models commonly hallucinate non-existent citations, biographies, or facts (*46, 47*, 48, 49, 50*), which could pose legal and ethical problems involving the spread of misinformation (*714*). It is possible but challenging to reduce general-purpose AI systems' tendency to hallucinate untrue outputs. Fine-tuning general-purpose AI models explicitly to make them more truthful – both in the accuracy of their answers and analysis of their competence – is one approach to tackling this challenge (*715**). Additionally, allowing general-purpose AI systems to access knowledge databases when they are asked to perform tasks helps to improve the reliability of language model generations (*716, 717*). Alternative approaches attempt to detect hallucinations rather than remove them from the model, and inform the user if the generated output is not to be trusted (*718*). However, reducing hallucination remains a very active area of research.

## 5.2.3 Improving robustness to failures

Sometimes, unfamiliar inputs a general-purpose AI system encounters in deployment can cause unexpected failures (*719*), and users or attackers can construct inputs that are specifically designed to make a system fail (*720*).

**Adversarial training helps improve robustness in state-of-the-art AI systems.** *Adversarial training* involves first, constructing 'attacks' designed to make a model act undesirably and second, training the system to handle these attacks appropriately. Attacks against AI systems can take many forms and can be either human- or algorithm-generated. Once an adversarial attack has been produced, training on these examples can proceed as usual. Adversarial training has become a principal method by which models are made more robust to failures (*2*, 3*, 22*, 204*, 207, 721*).

### 5.2.3.2 Open problems in robustness

While adversarial training is a valuable tool, it is not sufficient by itself (*219*).

**Making systems more robust to unforeseen classes of failure modes is a challenging open problem.** Adversarial training generally requires examples of a failure to fix it, (*722*, 723*). These limitations have resulted in ongoing games of 'cat and mouse' in which some developers continually update models in response to newly discovered vulnerabilities. A partial solution to this problem is to simply produce and train on more adversarial examples. Automated methods for generating attacks can help scale up adversarial training (*203, 210, 237*). However, the exponentially large number of possible inputs for general-purpose AI systems makes it intractable to thoroughly search for all types of attacks. One way to address this by applying adversarial training to general-purpose AI models' internal states, instead of inputs, has been proposed (*724*), but research on this remains preliminary. Mathematical proofs to certify a model's robustness would theoretically be a way to cover all possible attacks (*626*), but this is not possible with current models and methods.

**Adversarial training can sometimes harm a model's performance or robustness.** In vision models, there is often a trade-off between robustness to adversarial attacks and performance on non-adversarial data (*725, 726, 727*). Even when adversarial training is helpful to improve worst-case performance, it may not be used when it harms average case performance. Adversarial training can also sometimes make language models *less* robust to certain attacks that were not trained on (*722*, 724*). However, some refined approaches to adversarial training may be able to improve trade-offs between performance on clean and adversarial data (*728, 729, 730*).

## 5.2.4 Removing hazardous capabilities

**'Machine unlearning' can help to remove certain undesirable capabilities from general-purpose AI systems.** For example, removing certain capabilities that could aid malicious users in making explosives, bioweapons, chemical weapons, and cyberattacks would improve safety (*408*). Unlearning as a way of negating the influence of undesirable training data was originally proposed as a way to protect privacy and copyright (*586*) which is discussed in 5.5 Privacy methods for general-purpose AI systems. Unlearning methods to remove hazardous capabilities (*731, 732*) include methods based on fine-tuning (*733**) and editing the inner workings of models (*408*). Ideally, unlearning should make a model unable to exhibit the unwanted behaviour even when subject to knowledge-extraction attacks, novel situations (e.g. foreign languages), or small amounts of fine-tuning. However, unlearning methods can often fail to perform unlearning robustly and may introduce unwanted side effects (*734*) on desirable model knowledge.

## 5.2.5 Analysing and editing the inner workings of models

**Studying the inner workings of models can help to establish the presence or lack of specific capabilities.** One technique that researchers use is to analyse a general-purpose AI model's internal states to better understand what concepts they reason with and what knowledge they have (*278, 279, 735*). For example, these approaches have been used to study features related to fairness in visual classifiers (*736*) and what knowledge language models have (*737, 738**). However, methods for assessing a general-purpose AI model's internal representations are imprecise (*739, 740, 741*). They are also not currently competitively used over other types of evaluations for understanding a general-purpose AI model's capabilities (see 3. Methodology to assess and understand general-purpose AI systems).

**Understanding a model's internal computations might help to investigate whether they have learned trustworthy solutions.** 'Mechanistic interpretability' refers to studying the inner workings of state-of-the-art AI models. However, state-of-the-art neural networks are large and complex, and mechanistic interpretability has not yet been useful and competitive with other ways to analyse models for practical applications. Nonetheless, some researchers have provided thorough investigations of how very small neural networks perform very simple tasks (*628, 629, 630*). Some recent works have attempted more scalable techniques for designing human-interpretable models (*282, 631, 632*, 742*). This type of approach cannot be used to rule out the possibility of harmful or unexpected actions, but it could offer a useful lens into how models work that could be useful for understanding how safe a model is. Instead of trying to interpret internal computations of neural networks, one could use neural networks to generate interpretable and verifiable explanations that could yield quantitative safety guarantees (*626*), although how to do this efficiently remains an open problem.

**Understanding a model's internal workings can sometimes be used to guide edits to usefully change its behaviour.** Despite the difficulty of understanding models' inner workings, some

techniques can be used to guide specific edits to them. Compared to finetuning, these methods can sometimes be more compute- or data-efficient ways of modifying their functionality. Researchers have used a variety of methods for this, based on changes to their internal parameters (*743, 744, 745, 746, 747, 748, 749*), and neurons (*275, 282, 750*), or representations (*281, 751, 752, 753, 754*). However, these techniques are imperfect (*299*) and typically introduce unintended side effects on model behaviour (*755*). They remain an active area of research.

## 5.3 Monitoring and intervention

### KEY INFORMATION

- There are several techniques for identifying general-purpose AI system risks, inspecting general-purpose AI model actions, and evaluating performance once a general-purpose AI model has been deployed. These practices are often referred to as 'monitoring'. Meanwhile, 'interventions' refers to techniques that prevent harmful actions from general-purpose AI models.
- Techniques which are being developed to explain general-purpose AI actions could be used to detect and then intervene to block a risky action. However, the application of these techniques to general-purpose AI systems is still nascent.
- Techniques for detecting and watermarking general-purpose AI-generated content can help to avoid some harmful uses of generative general-purpose AI systems by unsophisticated users. However, these techniques are imperfect and can be circumvented by moderately skilled users.
- Techniques for identifying unusual behaviours from general-purpose AI systems can enable improved oversight and interventions.
- Having humans in the loop, and other checks before and during the deployment of general-purpose AI systems increase oversight and provide multiple layers of defence against failures. However, such measures can slow down general-purpose AI system outputs, may compromise privacy and could conflict with the economic incentives for companies that use general-purpose AI systems.

Monitoring during general-purpose AI system deployment refers to the ongoing identification of risks, inspection of model actions, and evaluation of performance. Interventions prevent potentially harmful outputs. Whereas 3. Methodology to assess and understand general-purpose AI systems discusses how systems are evaluated in order to allow for more informed decisions around their use, this section discusses how some techniques for monitoring and intervention can be built into AI systems themselves.

Different strategies that researchers are developing for general-purpose AI system monitoring and intervention are discussed below, including detecting AI-generated content, detecting risky situations, identifying harmful actions, explaining model actions, and intervening to override or block them.

### 5.3.1 Detecting general-purpose AI-generated content

Content generated by general-purpose AI systems – particularly deepfakes – could have widespread harmful effects (*756, 757*) (see 4.1. Malicious use risks). The ability to distinguish between genuine and general-purpose AI-generated content to prevent the malicious use of generative models.

**Some unreliable techniques exist for detecting general-purpose AI-generated content.** Just as different humans have unique artistic and writing styles, so do generative AI models. Some procedures have been developed to distinguish AI-generated text from human-generated text (*374, 375, 379, 380, 758*) and images (*759, 760*). Detection methods are typically based on specialised classifiers or assessing how likely it is that a given example was generated by a specific general-purpose AI model. However, existing methods are limited and are prone to false positives because general-purpose AI systems tend to memorise examples that appear in their training data, so common text snippets or images of famous objects may be falsely identified as being AI-generated. As general-purpose AI-generated content becomes even more realistic, it may be even more challenging to detect general-purpose AI-generated content.

**Watermarks make distinguishing AI-generated content easier, but they can be removed.** A 'watermark' refers to a subtle style or motif that can be inserted into a file which is difficult for a human to notice but easy for an algorithm to detect. Watermarks for images typically take the form of imperceptible patterns inserted into image pixels (*761*), while watermarks for text typically take the form of stylistic or word-choice biases (*382, 762*). Watermarks are useful, but they are an imperfect strategy for detecting AI-generated content because they can be removed (*374, 383*). However, this does not mean that they are not useful. As an analogy, fingerprints are easy to avoid or remove, but they are still very useful in forensic science.

**Watermarks can also be used to indicate genuine content.** In contrast to inserting watermarks into general-purpose AI-generated content, a contrasting approach is to put encrypted watermarks in non-AI-generated content (*763*). However, this would require changes in the hardware and software of physical recording devices, and it is not clear whether these methods could be bypassed by tampering.

## 5.3.2 Detecting anomalies and attacks

**Detecting anomalies and attacks on general-purpose AI systems allows precautions to be taken when they are identified.** Some methods have been developed that can help detect unusual inputs or behaviours from AI systems (*764, 765*). Other technical approaches aim to detect when the model outputs for a given input are uncertain, which can indicate that there is an attack or a risk of false outputs (*766*). Once detected, these examples can be sent to a fault-handling process or flagged for further investigation. It is also sometimes possible to detect and filter a significant proportion of malicious attacks before they are passed into a general-purpose AI model (*723, 767*) or detect potentially harmful outputs so that they can be blocked before they are sent to a user (*768, 769*).

## 5.3.3 Explaining model actions

**Techniques to explain why deployed general-purpose AI systems act the way they do are nascent and not widely applied yet, but there are some helpful methods.** The actions of general-purpose AI systems can be hard to understand. However, understanding the reasons why models act the way they do is important for evaluation and determining accountability for harms caused by general-purpose AI systems (*269, 770*). Unfortunately, simply asking general-purpose AI language models for explanations of their decisions tends to produce misleading answers (*771*). To increase the reliability of model explanations, researchers are working on improved prompting and training strategies (*772*, 773**). Other techniques for explaining general-purpose AI model actions (*774, 775*) have been shown to help with debugging (*207*). However, correctly explaining general-purpose AI model actions is a difficult problem because the size and complexity of general-purpose AI systems are beyond easy human understanding. State-of-the-art general-purpose AI systems are trained to produce outputs that are positively reinforced – not to do so for the desired reasons or in a self-consistent way.

## 5.3.4 Building safeguards into AI systems

As discussed in 5.1. Risk management and safety engineering, although there is no perfect safety measure, having multiple layers of protective measures and redundant safeguards in place increases the level of assurance. Once detected, interventions can identify and protect against potentially harmful actions from deployed general-purpose AI systems.

**Having a human in the loop allows for direct oversight and manual overrides.** Humans in the loop are expensive compared to automated systems. However, in high-stakes decision-making situations, they are essential. Instead of teaching general-purpose AI systems to act on behalf of a human, the human-AI cooperation paradigm aims to combine the skills and strengths of both general-purpose AI systems and humans and is seen as generally preferable in complex situations with potentially harmful ramifications (*703, 776, 777*, 778, 779*). However, having a human in the loop is not practical in many situations, when decision-making happens too quickly and cannot be slowed down (such as chat applications with millions of users), when the human does not have sufficient domain knowledge, or when human bias or error can exacerbate risks (*780*). As a result, humans in the loop can only be useful in certain situations.

**Automated processing and filtering methods can offer additional but generally imperfect layers of protection.** Some cyber-attacks against general-purpose AI systems often take the form of subtle, brittle patterns in their inputs. As a result, researchers have developed input pre-processing methods that can remove these patterns (*723, 781, 782, 783, 784, 785*). Sometimes attempted attacks can be detected and blocked before they are passed to a general-purpose AI model (*723, 767*), while harmful outputs may be detected before they are sent to a user (*768, 769*). These methods can significantly reduce risks from some failures, but can be vulnerable to attacks.

**Secure interfaces can be designed for general-purpose AI systems with potentially dangerous capabilities.** General-purpose AI systems that can act autonomously and open-endedly on the web or in the physical world pose elevated risks (see 4.4.1 Cross-cutting technical risk factors). For general-purpose AI systems with highly risky capabilities, limiting the ways in which they can directly influence people or objects is a proposed way to reduce potential risks (*625, 786*).

## 5.4 Technical approaches to fairness and representation in general-purpose AI systems

### KEY INFORMATION

- General-purpose AI models can capture and, at times, amplify biases in their training data. This contributes to unequal resource allocation, inadequate representation, and discriminatory decisions.
- Fairness lacks a universally agreed-upon definition with variations across cultural, social, and disciplinary contexts.
- From a technical perspective, the cause of bias is often the data, which may fail to adequately represent minorities of a target population. Bias can also stem from poor system design or the type of general-purpose AI technique used. These choices depend on the involvement of diverse perspectives throughout the general-purpose AI lifecycle.
- Mitigation of bias should be addressed throughout the lifecycle of the general-purpose AI system, including design, training, deployment, and usage.
- It is very challenging to entirely prevent bias occurring in current general-purpose AI systems because it requires systematic training data collection, ongoing evaluation, and effective identification of bias, trading off fairness with other objectives such as accuracy, and deciding what is useful knowledge and what is an undesirable bias that should not be reflected in the outputs.
- There are differing views about how feasible it is to achieve meaningful fairness in general-purpose AI systems. Some argue that it is impossible for a general-purpose AI system to be completely 'fair', while others think that from a practical perspective, near-complete fairness is achievable.

Fairness has no universally agreed-upon definition, and varies according to the cultural, social, and disciplinary contexts (*333, 787, 788, 789, 790*). Philosophically, fairness entails a deep ethical reflection on principles, values, and the distribution of resources and opportunities, while legal definitions may hinge on constitutional principles and case law. Achieving consensus on fairness proves to be elusive due to its multifaceted nature, necessitating engagement with diverse perspectives and context-specific factors.

The rise of AI applications brings algorithmic fairness to the forefront as a significant concern. Fairness in AI attempts to correct algorithmic bias in automated decision-making or content generation. AI fairness can be defined and measured in various ways (such as 'individual fairness'[6] versus 'group fairness'[7]) (*791*) which is proved appropriate depending on the context and the specific goals of the application (*333*), complicating the assessment of AI models. A classic example of this type of dilemma from the algorithmic decision-making literature is that the COMPAS software used to predict criminal recidivism is a well-known case of an unfair AI model. By some measures, COMPAS was found to be biased against African Americans while by other measures it was not (*788, 790*).

When an AI model is unfair, the actions that it takes are biased, harming individuals or communities. Bias in AI refers to unjustified favouritism towards or against entities based on their inherent or acquired characteristics (*788*). Bias is tightly intertwined with representation and fairness. The effectiveness of data-driven algorithms depends on the quality of the data they utilise. However, datasets often fail to adequately represent minorities, which can be reflected and amplified in the AI trained on these datasets (*227*). Biases in AI systems can lead to social harm, including unequal resource allocation and discriminatory decisions against marginalised groups (*792*) posing significant challenges across various domains (*793, 794, 795*).

### 5.4.1 Mitigation of bias and discrimination works throughout the stages of general-purpose AI development and deployment

Researchers deploy a variety of methods to mitigate or remove bias and improve fairness in general-purpose AI systems (*2*, 22**), including pre-processing, in-processing, and post-processing techniques (*796, 797*). Pre-processing techniques analyse and rectify data to remove inherent bias existing in datasets, while in-processing techniques design and employ learning algorithms to mitigate discrimination during the training phase of the system. Post-processing methods adjust general-purpose AI system outputs once deployed.

- Pre-processing bias mitigation - This includes targeting flaws in training data like harmful information or under-representation of certain demographic groups. Two prevalent techniques are data augmentation and data alteration. Data augmentation involves adding samples for under-represented groups to increase their representation in the datasets, either by creating modified copies of existing data or generating synthetic data (*798, 799, 800*). Data alteration modifies dataset samples based on predefined rules such as adding, removing, or masking attributes like gender and race from the data corpus (*227, 801*).

[6] Individual fairness emphasises similar treatment for similar individuals

[7] Group fairness, on the other hand, ensures forms of statistical parity (e.g. between positive outcomes, or errors) for members of different protected groups (e.g. gender or race)

- In-processing bias mitigation - Data manipulation alone often fails to ensure fairness in general-purpose AI systems (*802, 803*). Even with data that perfectly represents the population distribution, undesirable stereotypes and prejudices that are present in society can manifest (*803*). An additional strategy in pre-processing techniques is to integrate fairness into general-purpose AI systems by designing and utilising specialised learning algorithms to teach the general-purpose AI model to generate unbiased content (*804, 805, 806, 807, 808, 809, 810*). This can be achieved either by humans giving the general-purpose AI model feedback about what kind of output is and is not desirable (*2*, 22*, 702*, 811*) or by teaching the general-purpose AI model to follow human instructions (*3*, 812, 813, 814*). General-purpose AI models can also teach each other fairness, transferring information from a less biased general-purpose AI model (the 'teacher') to a second general-purpose AI model (the 'student') (*815*). A conventional method to mitigate bias in general-purpose AI systems is to train the general-purpose AI model with both unbiased and biased data samples, so that the model learns what to do and what not to do (*816, 817, 818*). Making under-represented attributes more salient is another pragmatic strategy toward model un-biasing (*819, 820**); yet it can potentially compromise privacy when trying to compensate for under-represented attributes (*821**).
- Post-processing mitigation of discrimination - When one cannot modify the full model directly (*822*). Another approach is to manipulate only the input or output of the general-purpose AI system toward fairness. Modifying general-purpose AI model inputs through prompts can help mitigate discrimination in outputs (*2*, 823*). In many cases, an external module such as a bias/safety classifier is responsible for detecting the unfair or unsafe output of the system (*812*). The external module can then rank outputs (*811*) and require the general-purpose AI system to regenerate the output to produce correct, non-discriminatory, content. If the topic is highly sensitive, the general-purpose AI system may be trained to refuse to reply to users seeking to prompt biased and discriminatory outputs (*820*, 822, 824*). These methods focus on manipulating the outputs of general-purpose AI systems to be fairer. These techniques are usually employed to complement other approaches.

There is no single technique that can ensure fairness in every situation or outcome. Consequently, leading AI companies utilise a combination of these approaches to iteratively improve fairness in their general-purpose AI systems (*20, 825**).

Importantly, increasing meaningful representation and participation can help to reduce the risk that under-represented groups are excluded. From a societal perspective, the AI alignment problem discussed in 5.2.1 Aligning general-purpose AI systems with developer intentions is ill-defined. It is not possible to make a general-purpose AI system represent the values of everyone in a diverse society where people sometimes disagree (*239, 307, 332*). Increased participation (*826*), representation (*827*), and dialogue (*332*) have been proposed as ways to reduce the risk that alignment to some people's interests will be harmful to others. However, some forms of participation can fail to be meaningful, and cannot fully solve challenges posed by disagreements between different people (*331*).

### 5.4.2 Is fairness in general-purpose AI systems achievable?

It is debated whether general-purpose AI systems can ever be completely 'fair'. There are arguments both for and against its feasibility. Mathematical results suggest that it may not be possible to satisfy all aspects of fairness simultaneously under reasonable assumptions (*828, 829, 830, 831*). This impossibility theorem of fairness is supported by results indicating the complexity of training unbiased general-purpose AI models (*832, 833*).

Many desirable properties involve trade-offs, such as the four-way trade-off among system fairness, accuracy, privacy, and efficiency (*834*). Studies suggest there is a trade-off between fairness and other values such as privacy (*821*, 834*) and predictive accuracy in general-purpose AI systems (*829, 835, 836*). A possible example is Google Gemini, which generated images of indigenous people and women of colour as US senators from the 1800s, and 'ethnically diverse' World War II-era German soldiers. These contents factually misrepresent the history, possibly as a result of an attempt to ensure racial diversity that failed to foresee and adjust to these specific use cases. To avoid prematurely prioritising specific aspects that inadvertently reflect the personal values of the stakeholders, technology can use quantitative and qualitative measures comprehensible to important technologists, helping them make well-informed decisions about trade-offs, which can then be enforced by the developed systems.

The counter argument is that, while theoretical limitations exist, practical solutions are attainable (*837, 838*). Some researchers suggest that fairness definitions can be reconciled with one another practically (*839, 840*), and it is possible to simultaneously satisfy multiple fairness criteria at least to a greater extent than they typically are (*841*). Empirical evidence challenges the idea that there is always a non-negligible trade-off between fairness and accuracy, indicating such trade-offs can often be resolved in practice. These studies suggest that reducing disparities in general-purpose AI system outputs may not necessarily entail significant drops in accuracy, or require complex methods (*838, 839, 840*).

Despite rigorous and comprehensive training and testing efforts, establishing general-purpose AI systems that are fair across all measures and across different cultural, social and scientific contexts remains challenging. No existing measure can entirely eliminate all potential risks of bias and unfairness which are inherent in the development of highly capable AI systems (*837*). Nevertheless, there exists the possibility of striving for continual refinement towards fairer systems.

### 5.4.3 Challenges in achieving fair general-purpose AI systems

Despite all the efforts to remove bias from general-purpose AI systems, major challenges have remained. Firstly, how fairness should be defined and measured is debated (*802, 842*). The line between useful and accurate world knowledge, and reinforcing harmful stereotypes, can be difficult to draw, and the perception of bias may vary depending on the situation (*227, 796*). Secondly, other aspects of ensuring general-purpose AI system safety can create or amplify bias: for instance, cleaning data to mitigate toxicity and privacy leakage can change the demographic distribution of the datasets, leading to more bias (*843*). Thirdly, some issues such as intersectional bias remain difficult to address (*844*); for instance, a general-purpose AI system might be fair to Asians and women separately, yet be biased toward Asian women. Finally, mitigation of bias requires ongoing effort throughout the development, deployment, and usage of the general-purpose AI systems. Bias in its various forms can emerge gradually, requiring specialised detection and mitigation techniques.

## 5.5 Privacy methods for general-purpose AI systems

#### KEY INFORMATION

- General-purpose AI systems present a number of risks to people's privacy, such as loss of data confidentiality, transparency and control over how data is used, and new forms of privacy abuse.
- Privacy protection is an active area of research and development. However, existing technical tools struggle to scale to large general-purpose AI models, and can fail to provide users with meaningful control.

As described in 4.3.5. Risks to privacy, advanced general-purpose AI systems pose risks to people's privacy, such as loss of data confidentiality, transparency and control over how data is used, and new forms of privacy abuse. Existing technology and policy only partially address these threats.

**Current privacy-enhancing technologies do not scale to large general-purpose AI Models.** While various privacy techniques can be applied to AI models to protect individual privacy while still allowing for useful insights to be derived from data (*845, 846*), these techniques can significantly impair model accuracy, are hard to scale to large models, and may not be suitable for all use cases, in particular for general-purpose AI models trained on text (*847*). For domains with highly sensitive data (e.g., medical or financial), it may be possible to attain strong privacy guarantees by adapting powerful general-purpose AI models that are first pre-trained on publicly available data from the internet (*848, 849*), but such techniques have rarely been applied in production, to date. Another solution is using synthetic data to avoid using sensitive data in general-purpose AI systems training pipelines. However, researchers have demonstrated that there is an important utility/privacy trade-off. If the synthetic data utility is high, then they may carry as much information as the original data and enable mostly the same attacks (*850, 851, 852*).
The confidentiality and data centralisation issues raised by general-purpose AI systems could, in principle, be addressed using secure computation solutions such as cryptographic approaches (*853*), federated learning (*854*), and hardware protections (*855*). Existing techniques, however, have not been scaled to the largest and most capable models being trained today. These solutions also all impose costs which might be prohibitive at scale, although researchers are trying to find ways to reduce these costs. Advances in hardware protections for accelerators, a class of specialised hardware designed to accelerate AI applications such as artificial neural networks and machine vision, could in future provide a practical avenue for training and running general-purpose AI models without accessing sensitive data.

**Measures to address the lack of data transparency and control in other areas could be applied to general-purpose AI systems.** Developing better mechanisms for individuals to control and trace their data would promote transparency and accountability in general-purpose AI systems. This includes providing user-friendly interfaces for managing data permissions, implementing secure data provenance systems to track how data is used and shared, and establishing clear processes for individuals to access, view, correct, and delete their data (*856*). The technological means to provide these controls exist and are successfully deployed in other areas (for example, user-controlled dashboards allowing users to decide how various websites or companies collect or use their personal data). It is possible that such and other approaches could be expanded to general-purpose AI systems. It may also be possible to redistribute the wealth derived from personal data in a more traceable and equitable manner, for example by using economic tools for data valuation (*857*). However, providing people with transparency and control over how their *public* data is used (i.e. information found readily on the Web) is much more challenging. While individual service providers such as social media platforms could prohibit their data from being used by external general-purpose AI systems, the control is ultimately not in the hands of the end-user and only covers a small portion of the data on the Web.

Another challenge is to provide meaningful controls for the use of *derived* data, or data that is not identified but allows inference about a person. Such cases will likely be common in AI systems that use personal data. More research is needed to explore ways to reduce the risks of unauthorised data use and sharing.

**Some forms of privacy abuse are hard to prevent through technical means.** In part, due to the lack of data transparency and control, new forms of privacy abuse stemming from general-purpose AI, such as non-consensual deepfakes or stalking, are hard to prevent via technical means. Some legal frameworks aim to hold creators and distributors accountable for malicious use (*858*), and to provide remedies for individuals whose privacy has been violated. Some recent regulations also call for AI systems to be developed and deployed in a manner that respects privacy principles, such as by enforcing data minimisation and purpose limitation (*859, 860*), yet how to achieve these properties, or the extent to which they are achievable is questionable (*861*).

# 6 Conclusion

This interim International Scientific Report on the Safety of Advanced AI finds that the future trajectory of general-purpose AI is remarkably uncertain. A wide range of possible outcomes appears possible even in the near future, including both very positive and very negative outcomes, as well as anything in between. Among the most promising prospects for general-purpose AI are its potential for education, medical applications, research advances in a wide range of fields, and increased productivity leading to more prosperity. If managed properly, general-purpose AI systems could substantially improve the lives of people worldwide.

But to reap the benefits of this transformative technology safely, researchers and policymakers need to identify and take informed action to mitigate the risks that come with it. Malicious use of general-purpose AI as well as malfunctioning general-purpose AI are already causing harm today, for instance through deepfakes, scams and biased outputs. Depending on the rate of progress of future general-purpose AI capabilities, the technical methods that developers and regulators employ to mitigate risks, the decisions of governments and societies in relation to general-purpose AI, and the degree of successful global coordination, it is also possible that further risks could emerge. The worst outcomes could see the emergence of risks like large-scale unemployment, general-purpose AI-enabled terrorism, or even humanity losing control over general-purpose AI systems. There is no consensus among experts about how likely these risks are and when they might occur.

This report also examines the factors that make addressing these risks difficult. Despite rapid advances in capabilities, researchers currently cannot generate human-understandable accounts of how general-purpose AI models and systems arrive at outputs and decisions. This makes it difficult to evaluate or predict what they are capable of, how reliable they are, and obtain assurances on the risks they might pose.

There are technical approaches to addressing the risks from general-purpose AI: methods for reducing model bias, improving our understanding of the inner workings of general-purpose AI models, assessing their capabilities and potential risks, and making them less likely to respond to user requests that could cause harm. There are also complementary techniques for monitoring and mitigating harmful actions from general-purpose AI systems. However, no existing techniques currently provide quantitative guarantees about the safety of advanced general-purpose AI models or systems.

Nothing about the future of AI is inevitable. How general-purpose AI gets developed and by whom, which problems it gets designed to solve, whether we will be able to reap general-purpose AI's full economic potential, who benefits from it, and the types of risks we expose ourselves to — these and many other questions depend on the choices that societies and governments make today and in the future to shape the development of general-purpose AI. Since the impact of general-purpose AI on many aspects of our lives is likely to be profound, and since progress might continue to be rapid, the precautionary principle implies an urgent need to broker consensus and to put resources into understanding and addressing these risks. Constructive scientific and public discussion will be essential for societies and policymakers to make the right choices.

For the first time in history, this interim report brought together expert representatives nominated by 30 countries, and the EU and the UN, and several other world-leading experts to provide a shared scientific, evidence-based foundation for these vital discussions. We continue to disagree on several questions, minor and major, around the capabilities, risks, and risk mitigations for general-purpose AI. But we consider this project essential for improving our collective understanding of general-purpose AI and its potential risks, and for moving closer towards consensus and effective risk mitigation to ensure people can enjoy general-purpose AI's benefits safely. The stakes are high. We look forward to continuing this effort.

# Chair's note on the interim report

This interim report is the product of extremely rapid collaboration between a diverse and large group of AI experts, including an Expert Advisory Panel nominated by 30 countries as well as the EU and the UN. The field of AI research is moving at pace, and on many important questions the field is far from having a consensus view. Against this backdrop, I am particularly impressed by what the 75 international experts contributing their diverse perspectives to this report have achieved in such a short time. Writing a report that discusses the capabilities, risks and potential risks of general purpose AI in a balanced way has been especially important to me. I am very grateful to the experts contributing to the report for the collaborative spirit with which they approached this important project.

The short amount of time available for writing this interim report also means that several difficult procedural decisions and decisions about the scope of the report had to be made, leaving many important issues unaddressed or only covered briefly. My goal for the next publication, which will build on this interim report, is for the contributing experts to jointly identify the most important areas of improvement to work on for the next report.

This might, for example, include some of the following aspects:

- We aim to further improve how the report evaluates and synthesises the scientific evidence on capabilities and risks. I am satisfied with how we have done this for the interim report given the short time available, but for the next report improvements can and should be expected. In particular, I expect the next report to:
  - Take into account an even larger body of scientific work to provide an even more comprehensive discussion of the literature;
  - In some cases, be more explicit about how compelling particular pieces of research are based on their methodology;
  - Do an even better job of synthesising the evidence in order to provide more nuanced and concise assessments of the state of the science on particular questions.
- For the interim publication, we restricted input to the report to the writing group, senior advisors, and expert input we needed on particular sections, including from a very limited set of civil society organisations. We intentionally excluded input from private sector representatives to avoid conflicts of interest. However, companies have been the main driver of recent advances in AI capabilities and are also active contributors to research, which we cite, on risk assessment and mitigation. Similarly, there is an active and diverse ecosystem of civil society organisations whose work has been crucial for the field's understanding of AI risks and risk mitigation. Therefore, now that the interim report has been published, we would like to broaden the input to the report. We are calling for submissions of evidence from companies and civil society to inform the development of the next, full report. We will set out more details on how to do this in due course. In the meantime, do email secretariat.AIStateofScience@dsit.gov.uk if you have evidence to submit.
- Ensuring that the report is comprehensive in its discussion of the risks of advanced AI has been a top priority for the writing team and senior advisors, but it has also been clear to us from the beginning that the short time available for producing the interim report would not be enough to address all issues with the depth they deserve. An example of a topic that I would like the next report to discuss in more detail is the global 'AI Divide' that prevents people around the world enjoying any benefits of AI equally. Another topic that the next report will discuss in greater detail than in the present report is the impact of AI development and deployment on the environment. In aiming to make the report even more comprehensive, we will remain conscious of the potential trade-offs between breadth, depth, and the coverage of evidence, and will aim for a balance that we consider most useful to policymakers.

- We have restricted the scope of this interim report to general-purpose AI. I am confident that this was the right decision, but it is also clear that the variety of types of AI models and systems pose a challenge for defining any subgroup of AI cleanly. For the next report, whatever its scope, we will aim to delineate in even more detail what types of AI are in and out of scope of the report.

I am extremely grateful to all the experts who contributed to this interim report and look forward to working on the next publication.

# Differing views

The Chair and Secretariat worked to reach consensus between members of the Expert Advisory Panel on the content of reports. As per the report's principles & procedures, a near-final version of the interim report was shared with the Expert Advisory Panel. Where differing views remained with the Panel, Panel members were offered the option for these to be noted. Below are the views noted.

## 1. Ciarán Seoighe (Ireland)

Noted concern that the general tone of the report is excessively negative. Noted that while the report does note that the future of AI is not predetermined, the language used could create the impression that the outlook for humanity is bleak no matter what steps are taken, and that consequently, the report's impact on policymakers could be undermined.

## 2. Ciarán Seoighe (Ireland)

Requested that future iterations of the report should include:

- Assessment of the societal impacts of generative AI in domains such as the information ecosystem, education, and democratic processes from both a technical and socio-technical perspective. It is important also to incorporate the range of regulatory approaches that could mitigate societal risk along with technical solutions.
- The importance of human/fundamental rights assessments and their role in mitigating risks before AI systems are developed and deployed. Noted this to be a significant omission in the current version of the report.

# Glossary

The explanations below should all be taken for the use of the term in the context of AI or general-purpose AI.

**Adaptivity:** The ability to identify patterns, reason, and make decisions in contexts and ways not directly envisioned by human programmers or outside the context of a system's training data.

**AI agents / agent / autonomous agent:** AI systems that are capable of accomplishing multi-step tasks in pursuit of a high-level goal with little or no human oversight. AI agents may do things like browsing the internet, sending emails, or sending instructions to physical equipment.

**AI deployers:** Any individual or organisation that supplies or uses an AI system to provide a product or service. Deployment can be 'internal', where a system is only used by the developers, or 'external', allowing the public or other non-developer entities to use it.

**AI developers:** Organisations or individuals who design, build, train, adapt, or combine AI models and applications.

**AI end user:** Any intended or actual individual or organisation that uses or consumes an AI-based product or service as it is deployed.

**AI lifecycle:** All events and processes that relate to an AI system's lifespan, from inception to decommissioning, including its design, research, training, development, deployment, integration, operation, maintenance, sale, use, and governance.

**AI risks:** The combination of the probability of an occurrence of harm arising from the development or deployment of AI models or systems, and the severity of that harm.

**Algorithmic Transparency:** The degree to which the factors informing general-purpose AI output, e.g. recommendations or decisions, are knowable by various stakeholders. Such factors might include the inner workings of the AI model, how it has been trained, what data it is trained on, what features of the input affected its output, and what decisions it would have made under different circumstances.

**Alignment:** The process of ensuring an AI system's goals and behaviours are in line with its developer's values and intentions.

**Application Programming Interface (API):** A set of rules and protocols that enables integration and communication between AI systems and other software applications.

**Artificial General Intelligence (AGI):** A potential future AI system that equals or surpasses human performance on all or almost all cognitive tasks. A number of AI companies have publicly stated their aim to build AGI. However, the term AGI has no universally precisely agreed definition.

**Autonomy / autonomous:** Capable of operating, taking actions, or making decisions without the express intent or oversight of a human.

**Biological design tools (BDTs):** In this report, biological design tools (BDTs) refers to AI systems trained on biological data that can help design new proteins or other biological agents such as enzymes.

**Black-box:** A system deployed with restrictions such that a user cannot access or analyse its inner workings. See also "White-box" below.

**Capabilities:** The range of tasks or functions that an AI system can perform and the proficiency with which it can perform them.

**Cloud labs:** Remotely controlled automatised biochemical laboratories.

**Cognitive tasks:** Tasks involving a combination of information processing, memory, information recall, planning, reasoning, organisation, problem solving, learning, and goal-oriented decision-making.

**Compute:** Computational resources, required in large amounts to train and run general-purpose AI models. Mostly provided through clusters of Graphics Processing Units (GPUs).

**Cross-lingual differences:** Discrepancies in how a general-purpose AI model or system might respond to the same input in different languages.

**Deep Learning:** A set of methods for AI development that leverages very large amounts of data and compute.

**Deployment:** The process of releasing an AI system into a real-world environment, such as a consumer-facing AI system.

**Disinformation:** Deliberately false information generated or spread with the intent to deceive or mislead.

**Ecosystem Audit:** A broad evaluation of an AI system and its surrounding ecosystem. Ecosystem audits might consider AI models, their training data, the circumstances of their deployment, and surrounding operational practice.

**Evaluations:** Systematic assessments of an AI system's performance, capabilities, or potential impacts. Evaluations can include benchmarking, red-teaming, and audits.

**FLOPS:** 'Floating point operations per second' – a measure of the computing power of a computer.

**Foundation models:** Machine learning models trained on very large amounts of data that can be adapted to a wide range of tasks.

**Frontier AI:** For the AI Safety Summit at Bletchley Park, frontier AIs were defined as models that can perform a wide variety of tasks and match or exceed the capabilities present in today's most advanced models.

**GPU (Graphics Processing Unit):** A piece of computer hardware assembled from semiconductors widely used as the central source of computational power for general-purpose AI. GPUs were originally designed for graphics rendering applications.

**Guardrails:** Pre-defined safety constraints or boundaries set up in an attempt to ensure an AI system operates within desired parameters and avoids unintended or harmful outcomes.

**Heuristic:** A rule-of-thumb, strategy, or a simplified principle that, in the context of computer science, has been developed to solve problems more efficiently when classic methods are too slow or fail to find an exact solution.

**Input (to an AI system):** The data or prompt fed into an AI system, often text or an image, which the AI system processes before producing an output.

**Large Language Model (LLMs):** Machine learning models trained on large datasets that can recognise, understand, and generate text and other content.

**Massive Multitask Language Understanding (MMLU):** A widely used benchmark AI research that assesses a general-purpose AI model's performance across a broad range of tasks and subject areas.

**Misgeneralisation:** When an AI system trained to perform well in one context fails to perform well in a new context. For instance, if an AI trained mostly on pictures of white cats labels a black cat as a 'dog', it is misgeneralising from its training data.

**Misinformation:** Incorrect or misleading information, potentially generated and spread without harmful intent.

**Modalities:** The types and nature of data that an AI model can process, such as text, image, sound, or videos. Models might be unimodal, i.e. only able to process one type of data, or multimodal, i.e. able to process multiple types of data.

**Model Card:** A document providing important information on a general-purpose AI model, such as its purpose, its performance on evaluations and benchmarks, and safety features.

**Narrow AI:** An AI system that only performs well on a single task or narrow set of tasks, like sentiment analysis or playing Chess.

**Open-ended domains:** Scenarios or environments that have a very large set of possible states and inputs to an AI system – so that developers cannot anticipate all types of contexts of use and thus cannot test the AI's behaviour in all possible situations.

**Pre-training:** The first stage of developing a modern general-purpose AI model, in which models learn from large amounts of data. Pre-training is the part of general-purpose AI training that requires the most data and computational resources.

**Prompt:** An input to an AI system, often a text-based question or query, that the system processes before it produces a response.

**Red-teaming:** A method to evaluate the safety and robustness of systems by attempting to design inputs that make them fail. This is often done by developing 'adversarial attacks' or challenging conditions. Red-teaming tries to reveal worst-case behaviours or malicious use opportunities.

**Risk factors:** Elements or conditions that can increase downstream risks. For example, weak guardrails constitute a risk factor that could enable an actor to malicious use an AI system to perform a cyber attack (downstream risk).

**Safety and security:** The protection, wellbeing, and autonomy of civil society and the population. In this publication, safety is often used to describe prevention of or protection against AI-related harms. AI security refers to protecting AI systems from technical interference such as cyber-attacks or leaks of the code and weights of the AI model.

**Scaffold:** Additional software that aids in processing inputs and outputs of an AI model while leaving the model itself unchanged. For example, a scaffold allows GPT-4 to power the autonomous AI agent AutoGPT. The scaffold prompts GPT-4 to break down a high-level task into sub-tasks, assign sub-tasks to other copies of itself, save important information to memory, and browse the internet.

**Semiconductors:** Fundamental material components of modern computer hardware, such as GPUs.

**Synthetic data:** Data, e.g., text, images, etc., that has been generated artificially, for instance by general-purpose AI models. Synthetic data might be used for training general-purpose AI models, such as in cases of scarcity of high-quality natural data.

**System integration:** The process of combining different software elements into one cohesive system assembled to perform some function. For instance, system integration might combine a general-purpose AI model, a content filter, a user interface, and various other components into a chatbot application.

**Transfer learning:** A machine learning technique in which a model's completed training on one task or subject area is used as a starting point for training or using the model on another subject area.

**Transformer architecture:** A deep-learning architecture at the heart of most modern general-purpose AI models. The transformer architecture has proven particularly efficient at converting increasingly large amounts of training data and computational power into better model performance.

**Weights:** Parameters in a model that are akin to adjustable dials in the algorithm. Training a model means adjusting its parameters to help it make accurate predictions or decisions based on input data, ensuring it learns from patterns it has seen.

**White-box:** A system deployed without restrictions such that a user can access or analyse its inner workings. See also “Black-box” above.

# References

* denotes that the reference was either published by an AI company or at least 50% of the authors of a preprint article have an AI company as their affiliation.

## How to cite this report:

Bengio, Y., Mindermann, S., Privitera, D., Besiroglu, T., Bommasani, R., Casper, S., Choi, Y., Goldfarb, D., Heidari, H., Khalatbari, L., Longpre, S., Mavroudis, V., Mazeika, M., Ng, K. Y., Okolo, C. T., Raji, D., Skeadas, T., Tramèr, F., Fox, B., de Leon Ferreira de Carvalho, A. C. P., Nemer, M., Pezoa Rivera, R., Zeng, Y., Heikkilä, J., Avrin, G., Krüger, A., Ravindran, B., Riza, H., Seoighe, C., Katzir, Z., Monti, A., Kitano, H., Kerema, M., López Portillo, J. R., Sheikh, H., Jolly, G., Ajala, O., Ligot, D., Lee, K. M., Hatip, A. H., Rugege, C., Albalawi, F., Wong, D., Oliver, N., Busch, C., Molchanovskyi, O., Alserkal, M., Khan, S. M., McLean, A., Gill, A., Adekanmbi, B., Christiano, P., Dalrymple, D., Dietterich, T. G., Felten, E., Fung, P., Gourinchas, P.-O., Jennings, N., Krause, A., Liang, P., Ludermir, T., Marda, V., Margetts, H., McDermid, J. A., Narayanan, A., Nelson, A., Oh, A., Ramchurn, G., Russell, S., Schaake, M., Song, D., Soto, A., Tiedrich, L., Varoquaux, G., Yao, A., & Zhang, Y.-Q. (2024). International Scientific Report on the Safety of Advanced AI: Interim Report. International Scientific Report on the Safety of Advanced AI, Report 1 (interim report). https://www.gov.uk/government/publications/international-scientific-report-on-the-safety-of-advanced-ai

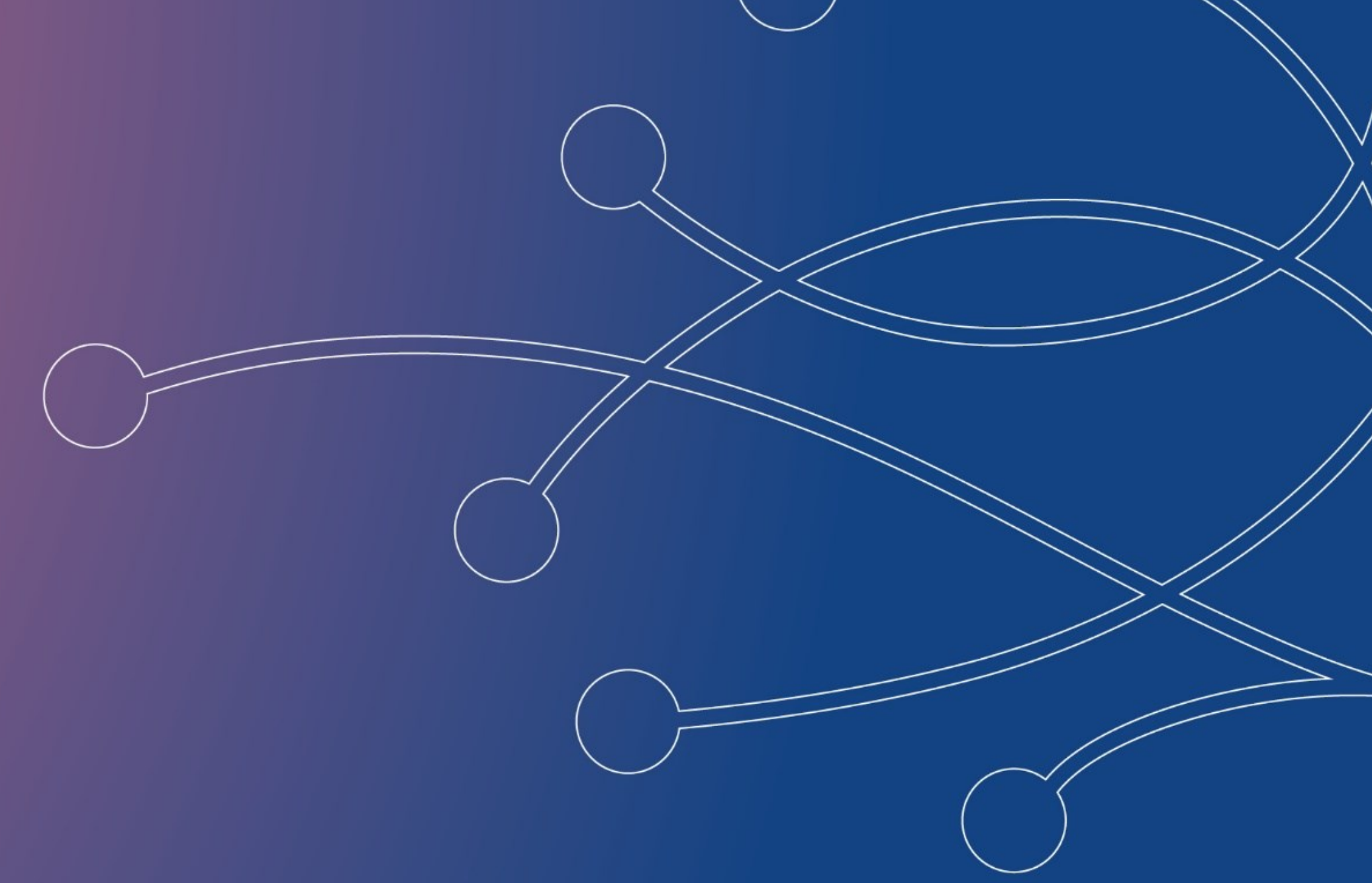

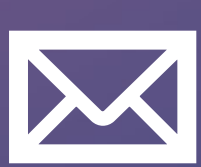

**Any enquiries regarding this publication should be sent to: secretariat.AIStateofScience@dsit.gov.uk**

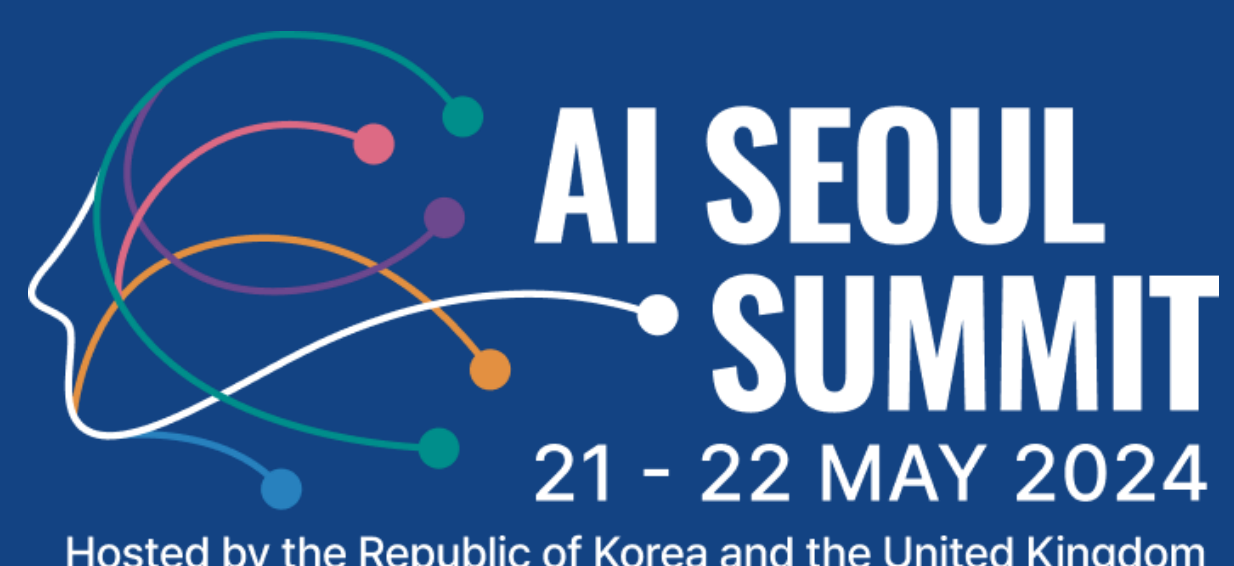